\documentclass[proof]{WileyASNA-v1}

\articletype{Article Type}%
\usepackage{apacite}
\received{2018 Mar 9}
\accepted{2018 Apr 8}

\begin{document}

\sloppy

\title{Reconstructed sunspot positions in the Maunder Minimum based on the correspondence of Gottfried Kirch}

\author[1]{R. Neuh\"auser*}

\author[2]{R. Arlt}

\author[1,3]{S. Richter}

\authormark{Neuh\"auser et al.: New sunspot positions in the Maunder Minimum from Kirch}

\address[1]{\orgdiv{Astrophysikalisches Institut}, \orgname{Universit\"at Jena}, \orgaddress{\state{Schillerg\"asschen 2-3, 07745 Jena}, \country{Germany}}}

\address[2]{\orgdiv{Leibniz-Institut f\"ur Astrophysik}, \orgname{Potsdam}, \orgaddress{\state{An der Sternwarte 16, 14482 Potsdam}, \country{Germany}}}

\address[3]{\orgdiv{Franckh-Kosmos Verlags-GmbH \& Co. KG}, \orgname{Pfizerstrasse 5-7}, \orgaddress{\state{70184 Stuttgart}, \country{Germany}}}

\corres{*Ralph Neuh\"auser. \email{rne@astro.uni-jena.de}}

\abstract{
We present reconstructed sunspot positions based on observations reported in letters
between Gottfried Kirch and other contemporary astronomers from AD 1680 to 1709,
i.e. in the last decades of the Maunder Minimum.
The letters to and from Gottfried Kirch in Latin and German language were compiled and edited by Herbst (2006).
The letters (and observations) from Kirch are mostly by Gottfried Kirch, but some also by his 2nd wife Maria M. Kirch
(married 1692) and their son Christfried Kirch (born 1694).
Using excerpts from the letters, some with drawings,
we found some 35 sunspot groups (often for several days in a row or with interruptions)
by Kirch and/or his letter partners (in three cases, only the month is given: 1704 Jan, Feb, 1707 Mar, otherwise
always the exact dates) -- usually one group at a time.
We also found 17 explicit spotless days, several of them new (previously without any known observations).
We could constrain the heliographic latitude by Bayesian inference for 19 sunspot groups -- five of them completely new
(one group 1680 May 20-22 from Kirch and Ihle, one to two groups 1680 Jun 15-23 for Kirch, 
one group 1684 May 6 from Ihle, and one group 1688 Dec 14-15 from Kirch),
while the others mostly agree (within $2 \sigma$) with previously published values for those dates by others.
With these data, we then amend the butterfly diagram for the Maunder Minimum.
By comparison of our data with the sunspot group catalog in Hoyt \& Schatten (1998), 
we noticed a number of discrepancies, e.g. that dates for British observers in the Maunder Minimum
(Flamsteed, Caswell, Derham, Stannyan, Gray, and Sharp) as listed in Hoyt \& Schatten (1998)
are their original Julian dates, not converted to the Gregorian calendar (10-11 day offset in Hoyt \& Schatten).
Most of these modifications also apply to the modified sunspot group catalog in Vaquero et al. (2016).
We also present two aurorae observed by the Kirchs in 1707 and 1716.
}


\maketitle

\section{Introduction}
\label{introduction}

The reconstruction of solar activity helps to understand the physical processes 
in the Sun and predicting future activity and space weather. 
Given that our Sun may have entered another Grand Minimum with the current weak
Schwabe cycle, it is most important to study in more detail the Maunder Minimum (MM).

Ribes \& Nesme-Ribes (1993, henceforth RNR93) used the collected records at the Paris observatory 
of sunspots during the MM to arrange them in a diagram 
showing the heliographic latitudes over the years from AD 1671 to 
1719 -- the butterfly diagram.
The usual symmetric, double wing-shaped distribution of sunspots was all but 
missing during the MM and spots were mainly observed in the southern hemisphere. 

The MM was first noted by Sp\"orer (1889), 
then amplified by Maunder (1890) and Eddy (1976);
it has received much attention since then (e.g., RNR93, 
Usoskin et al. 2007, Vaquero et al. 2011, Vaquero 2012, Vaquero \& Trigo 2014, 2015, 
Clette et al. 2014, Zolotova \& Ponyavin 2015, Usoskin et al. 2015, 
and Carrasco \& Vaquero 2016; the latter also used
a letter exchange (the one of Flamsted) to study sunspots,
as we do here with the letters to and from G. Kirch).
The Maunder Minimum
is usually dated from 1645 to 1715.

Here, we present observations by the astronomer Gott\-fried Kirch from Germany
for the time from AD 1680 to 1709. During this period of the MM, 
the sunspots were found in very low numbers only and almost all in the southern hemisphere,
but solar activity recovered in the first decade of the 18th
century -- the last Schwabe cycle of the MM.

Sect. \ref{kirch} gives an introduction to the astronomer and calendar-maker Gottfried Kirch, 
calendar issues relevant for the 17th and early 18th century,
and some information on how they observed and recorded sunspots at that time.
In Sect. 3, we then present the correspondence between G. Kirch and his contemporaries
regarding sunspot observations -- always together with our reconstruction of the
sunspot location on the solar disk and a comparison to what was previously known
for those spots from other observers.
At the end of Sect. 3, we present Table 1 with all our spot latitudes --
and Table 2 with all discrepancies between the data from the Kirch letters and 
the tables in Hoyt \& Schatten (1998, henceforth HS98).
In Sect. 4, we summarize our findings and present an updated butterfly diagram for the Maunder Minimum.
In tables in the appendix, 
we list additional texts 
about solar observations
from the Kirch letters, which did not result in spot detections, 
but do, for example, indicate explicitely spotless days.
We would like to stress that a significant part of the observations and letters 
since 1692 are actually from Gottfried Kirch's 2nd wife Maria M. Kirch; in the edition of the
corresponence of G. Kirch by Herbst (2006), only those letters (or copies or concepts)
written by Maria (or Christfried) Kirch are included, 
which appear to be written {\it for} or {\it in the name of} G. Kirch -- 
and also letters directed to Maria Kirch were excluded.

\section{Gottfried Kirch and his time}
\label{kirch}

We introduce here G. Kirch (2.1), then also comment briefly on calendar issues (2.2),
and also discuss briefly the observational method in the 17th century (2.3).

\subsection{Astronomer and calendar maker Gottfried Kirch}

Biographical information on Gottfried Kirch and his family can be 
found in, e.g., Herbst (2006):
Gottfried Kirch was born AD 1639 Dec 8 on the Julian calendar (i.e., Dec 18 Gregorian) 
in Joachimsthal, Germany. After he went to school in Guben, Germany, 
he worked as a schoolmaster in Langgr\"un near Schleiz and Neundorf near Lobenstein, 
Germany, from 1663.
In 1666, he published a calendar for the first time.

G. Kirch studied astronomy with Erhard Weigel in Jena, Germany, 
and Johannes Hevelius in Gdansk, Poland. 
After a yearlong stay in K\"onigsberg (today Kaliningrad in Russia), 
G. Kirch moved back to Lobenstein in 1675. 
There he had the idea of founding an astronomical society, 
which never got realised. Only one year later, he moved to Leipzig and 
worked as an independent scholar, living in the famous Collegium Paulinum.
After many years of teaching (except 1680/81) in Leipzig, 
G. Kirch went back to his home town Guben in 1692, because of hostilities against his pietistic belief.

Eight years later, he was appointed Royal Astronomer of the 
``Brandenburgische Societ\"at der Wissenschaften'' (Brandenburgian academy 
of science) and therefore moved to Berlin, where he 
died on 1710 Jul 25 (Greg.).

His discoveries and inventions are 
numerous:
in 1679, 
G. Kirch designed a micrometer which enabled its user to measure separations 
on the sky very precisely, it is still used today in a slightly modified way.
One year later, he discovered the ``Great Comet of 1680'' (C/1680 V1) through a telescope.
He discovered the star clusters M11 and M5 in 1681 and 1702 and the variability of $\chi$ Cygni in 1686.

G. Kirch published some scholarly journals and articles of high importance, 
e.g. the {\it Himmels-Zeitung} (sunspot observations from this journal
are shown below in Sects. 3.1 \& 3.2) or in ``Acta Eruditorum''. 
His calendars received
a lot of
attention. G. Kirch is considered to be one of the first 
exponents of the early Enlightenment in Germany. This is especially interesting, 
because he was a convinced proponent of the Pietism as well.

After his first wife Maria Lang (married since 1667) died in 1690, 
he married Maria Margaretha Winckelmann (1670-1720) in 1692. She assisted G. Kirch and discovered 
a comet in 1702 by herself; she is also known for her weather predictions.
Gottfried Kirch wrote about her in letter no. 547 to Adam Adamandus Kochanski SJ
(Warsaw, Poland): {\it meine Ehefrau ... hat gro\ss e Lust zur Sternkunst, 
und ist mir, wie sonsten, also auch hierinnen
eine gute Geh\"ulffinn} (Herbst 2006, p. 194-195), i.e. {\it my wife [Maria Margaretha] ... 
has a lot of fun in astronomy, so that she is here, as elsewhere, a good assistant}.

Some of the sunspot observations by G. Kirch were done together with his wife 
Maria Margaretha and his
son Christfried Kirch (1694-1740), who was an astronomer himself, 
or even without G. Kirch (HS98 list G., M.M. and C. Kirch).
We should consider much of the observations and letters from 1692-1710
as collaborative work by both Gottfried and Maria M. Kirch.

\subsection{Calendar issues}

The time period studied here is after
the Gregorian calendar reform, which replaced the previous Julian calendar:
1582 Oct 4 was immediately followed by Oct 15, i.e. the ten days Oct 5-14 were left out,
while the sequence of weekdays was uninterrupted.
This reform was initiated by Pope Gregory XIII (AD 1502-1585) modifying slightly
the previous calendar by Gaius Julius Caesar (100-44 BC).
The implementation of the reform was slow and took place at different
times depending on region and religion (protestant or catholic),
e.g. in most protestant German states, the reform was implemented by
jumping from 1700 Feb 18 to 1700 Mar 1 (see von den Brincken 2000).

G. Kirch himself, being protestant and working in a protestant part of Germany,
used the old Julian calendar until early 1700, 
similar to most 
-- but not all -- of his colleagues
in Germany, with whom he exchanged letter.

HS98 presumably have transformed all dates in their catalogue to the Gregorian calendar (new style).
However, by comparison of some of the English observers with the other observers, in particular
from France and Italy, 
who all already used the 
Gregorian calendar, we will see below that the
date ranges for English observers in HS98 (for the life time of Kirch) are mostly still 
Julian (except Harriot), i.e not shifted by 10 days from Julian to Gregorian.
In England, the Gregorian reform was implemented as late as 1752 by jumping from
1752 Dec 2 to 14 (see von den Brincken 2000) -- the Julian calendar countries did have a leap day in 1700,
but not the Gregorian calendar countries, so that the jump had to be that strong after 1700 Feb.
This applies to the 
observer Flamsteed (Carrasco \& Vaquero 2016), and also to
Derham, Stannyan, Gray, and Sharp.
Vaquero (2007) was the first to notice such calendar problems in HS98.

We list calendar dates either in Gregorian new style or give the date for
both the Julian and Gregorian style in the form {\it year month x/y} with
x being the Julian date (day) and y being the Gregorian date (day),
e.g. 1582 Oct 5/15;
this does not indicate a date range, but the two different dates
in the Julian and the Gregorian calendar
for the very same day.

\subsection{Observational methods of the 17th century}
\label{methods}

Like every astronomer of the 17th century, G. Kirch used simple, 
several foot long telescopes, called ``Tubus'' (plural ``Tubi'') to observe the sky.

In a letter to Georg Samuel D\"orffel in 1683, 
G. Kirch describes precisely his equipment and explains the effect that every part of an instrument has
(our English translation is given below).
\begin{quote}
Ein iedweder Tubus hat eigentlich 2 Gl\"aser, das Ocular, so dem Auge am
n\"achsten, und das Objectiv, so dem Objecto am n\"achsten.
Das Objectiv ist stets Convex, entweder auff beyden Seiten, oder auff einer
convex, und auff der andern plan. Das Ocular aber kann entweder auff einer
Seite concav und auff der andern plan seyn, oder auff beyden concav
beyderley Art stellen die objecta auffrecht. Oder aber das Ocular kann auff
beyden Seiten convex seyn, oder auff einer convex und auff der andern plan,
beyderley stellen die objecta verkehrt.
Will man durch convexe Ocularia die objecta auffrecht stellen, so mu\ss{} man 3
Ocularia haben.
\end{quote}

Staying as close as possible to the original German text, we translate to English as follows:
\begin{quote}
Every telescope consists of two lenses, the eyepiece, which is closest to the eye, 
and the objective, closest to the object.
The objective is always convex, either on both sides or convex on the one and flat on the other side. 
The eyepiece can be either concave on one side and flat on the other or concave on both sides. 
In both cases the object is shown upright.
Alternatively the eyepiece is either convex on both sides or convex on one and flat on the other, 
which shows the object the wrong way around.
If one wants to show the object the right way up through convex eyepieces, three eyepieces are needed.
\end{quote}

The most common combination was the second one, where the convex / flat-convex objective 
and convex / flat-convex eyepiece show the object flipped in E-W and N-S. 
The reason for the acceptance of the rotated image can be found in the same letter,  
where G. Kirch describes that three lenses would reduce the field of view immensely.

For solar observations G. Kirch and his contemporaries used every sort of Camera Obscura, 
whether a small box or the whole wall as a screen, together with a telescope to
project the sunlight into the dark chamber (camera helioscopica).
They also attached a blank piece of paper behind the eyepiece of their telescopes to project the Sun onto it.

Drawings consist either of a plain circle with a cross through the middle and a spot, 
where the sunspot was seen, or of a large circle divided into six smaller circles, 
called ``digiti'' (lat. digitus: finger). Such a digitus was also called zoll or inch.
Since the eastern and western half-circles were counted separately, the measurements run from 0 to 12 inches.
For many of the observations discussed below, the position of the spot is given
as a digitus or inch (one twelfth of a solar diameter), and we assume an 
error bar of $\pm 0.4$ inches (unless otherwise stated), i.e. a bit smaller than the
typical resolution of half an inch with which the measurements were given.
E.g., Ihle wrote to Kirch (Sect. 3.11): {\it ... 1/5 of one inch,
that would be 1/60 of the whole [solar] diameter ...}.

As specified by early telescopic sunspot observers in the 17th century like 
Malapert (see e.g. Neuh\"auser \& Neuh\"auser 2016), solar observations
were mostly obtained at noon time, when celestial equator and horizon coincide.
In about half the reports below, where the hour is given, the observations indeed
happened roughly around noon time (not counting those during a solar eclipse).

By about AD 1700, the nature of sunspots as dark and cold regions on the photosphere
due to magnetic activity was not yet known.
Gottfried Kirch himself expressed his opinion in one of his letters 
(letter no. 891 to G.W. Leibniz on 1708 Dec 1, quoted from Herbst 2006)
as follows:
\begin{quote}
Weil nun die Maculen, wie ich g\"antzlich dar vor halte,
ein Rauch seyn, von einem neuen, in der Sonnen entstandenen, Brande;
Wie solches auch die Faculen, (welche gemeiniglich nur am Rande der Sonnen
zu sehen seyn) bezeugen: Als k\"onte man solches billich ein Himmlisches Feuer-Werck nennen.
\end{quote}
We translate this to English as follows:
\begin{quote}
Because the spots, as I fully support, are a smoke from a new fire forming in the Sun,
as also the faculae attest it (which are commonly seen only on the Sun's limb);
one could well call this a heavenly firework.
\end{quote}

We would like to note that there are almost no dated aurora observations
in the letter exchange to and from G. Kirch (with one exception in AD 1707,
see below and Sect. 3.21). 

In his hand-written aurora catalog
(see Schr\"oder 1996 for an edition and de Mairan 1754, pp. 499/500 and 515/516 for an extraction),
Gottfried Kirch's son Christfried Kirch listed
eight entries for the period AD 1645-1715, namely for AD 1657 Jan 3, 1660 Nov 9, 
1672 Jan 24, 1682 Oct 28 (probably Julian dates), as well as
1707 Mar 6, Oct 21, Oct 29, Nov 27 (probably Gregorian dates);
the earliest of those is listed by de Mairan (1754, p. 515-516) in his extraction
of the C. Kirch aurora catalog for 1657 Jan 13 (Gregorian),
and the last four are given for the same dates (p. 500).
For 1657-1682, none of them are from their own observations, but C. Kirch is quoting and citing others:
for the first three, night-time in not mentioned; the last five (AD 1682 Oct 28 and 1707) 
fall into the time period studied here (AD 1680-1709); for 1682, C. Kirch just wrote
{\it 1682 - 28 Oct. Hevel. Ann. Climat. p. 135}; the full text is also quoted in
Link (1964), including {\it evening ninth hour ... luminous rays ... to the south-west ... 
like tails of 3-4 comets};
this event is dated Nov 7 (Gregorian) in Fritz (1873), 
and it listed for both dates (Oct 28 and Nov 7) in K\v{r}ivsk\'y \& Pejml (1988):
while both should presumably be Gregorian according to K\v{r}ivsk\'y \& Pejml (1988), 
only the latter can be Gregorian, while the earlier is a misduplication.
Of the discriminative aurora criteria in Neuh\"auser \& Neuh\"auser (2015a),
only night-time is fulfilled (while it would not be impossible to see an aurorae
in the SW). 
For the four events in 1707, no details are given in C. Kirch's catalog except date, observer name
(always Kirch himself), and/or location, see Sect. 3.21 below for some more 
discussion for 1707 March (simultaneous with sunspots).
For a full evaluation of aurora suggestions during
the Maunder Minimum, we refer to Neuh\"auser \& Neuh\"auser (in prep.).

\section{Sunspots by Gottfried Kirch and colleagues}
\label{spots}

In the following we will present all reports from the Kirch correspondence (Herbst 2006), 
in which sunspots are described. 
In addition, we will show one more early observation of a sunspot by Kirch
(1680 May 20--29) as published by him in his {\it Himmels-Zeitung} (Sect. 3.1).
We will briefly quote the most important sections 
from the letters, translated by us to English as close as possible to the original German or Latin.
We then reconstruct or constrain the heliographic latitude(s) of the spot(s) observed
using the solar B$_{0}$- and P-angles: angle B$_{0}$ is the heliographic latitude of the central point 
of the solar disk ranging from $-7.25^{\circ}$ to $+7.25^{\circ}$,
and angle P is the position angle between the geocentric north pole and the solar rotational 
north pole measured east from geocentric north and ranging from $-26.77^{\circ}$ to $+26.77^{\circ}$.
We are using the ephemeris provided by the JPL Horizons service
\footnote{http://ssd.jpl.nasa.gov/horizons.cgi} which gives
$P$ from the true-of-date direction to the geocentric north.

If some form of positional information for the sunspots
is given for more than one day, we obtain probability
distributions for the heliographic latitude through
Bayesian inference. The average and $68\%$ confidence intervals
are derived from these distributions. The method is
a simplified version of the one described in Arlt \& Fr\"ohlich
(2012). Since the information provided is very scarce,
we need to reduce the number of unknowns drastically.
Instead of inferring the differential rotation, we use
the one by Balthasar et al. (1986), which was also
derived from spots. Since almost all measurements are
radial distances from the solar limb/centre, the position
angle of the solar disk drops out of the computations.
The only unknowns are the longitude and latitude of the spot.

All reconstructed spot latitudes are listed in Table 1.
The date ranges given in the subsection headings are the Gregorian date ranges of the
observations of spots including spotless days as far as mentioned.

\subsection{1680 May 20--29}
\label{22.05.1680}

The first observations by Kirch are actually from the printed
{\it Himmels-Zeitung} (sky journal) in which he describes sunspot
observations in May and June 1680 (Kirch 1681). He reports
first for May 22, then for May 23:
\begin{quote}
\dots versuchte ich auch den zehensch\"uhigen Tubum, fand durch
denselben in der Sonnen selbst alle beyde Maculn a und b / auch
die gar d\"unne Macul c / welche an b zu kleben schiene (Besihe
die Figur C.)

Diese Maculen hat Herr Johann Abraham Ihle/Churf\"urstlicher
Brandenb. Factor in Leipzig / mein sehr wehrter Freund und
grosser G\"onner/schon am 10. May durch einen f\"unffsch\"uhigen
Tubum observiret / und befunden / da\ss{} damals die Macul
a etwan 2. Zoll oder etwas mehr / und die Macul b ungefehr
1. Zoll / oder ein wenig dr\"uber / vom Ostrande
in der Sonnen gewesen.

Den 13. (23.) Maji / fr\"uhe um 7. Uhr waren a und b durch
den zehensch\"uhigen Tubum sehr gut auf der Scheibe in der
finstern Kammer zu sehen / mich deuchtete sie w\"aren
mercklich n\"aher beysammen / als am n\"achst vorhergehenden
12. (22.) May.
\end{quote}

We translate this to English as follows (square brackets from us):
\begin{quote}
\dots I also tried the 10-foot telescope, and found both
spots a and b through it, even the very weak spot c, which
seemed to glue to b (see figure C). \\
These spots were already observed by Johann Abraham Ihle,
Brandenburgian Electoral Factor in Leipzig, my very
valued friend and great favourer, on May 10 [Greg.: May 20] through
a five-foot telescope, who found that the spot a was
located at about 2 inch or a bit more, and the spot b
was located at about 1 inch or a bit more from the
eastern limb of the Sun. \\
On May 13 ([Greg.:] 23) at 7 in the morning, a and b were readily
visible through the 10-foot telescope in the dark chamber,
I thought they would be notably closer together than on
the day before, May 12 (22).
\end{quote}
We show Kirch's drawing (his figure C) here in Fig. 1.

Kirch further writes that spots a and b were still visible on May 24
(Greg.), but not c. Spot a was visible also on May 27 and 28 (Greg.), 
close to the western limb of the Sun, while b and c were not
visible. On May 29 at 9 am, the Sun was spotless.

\begin{figure}
\includegraphics[width=0.485\textwidth]{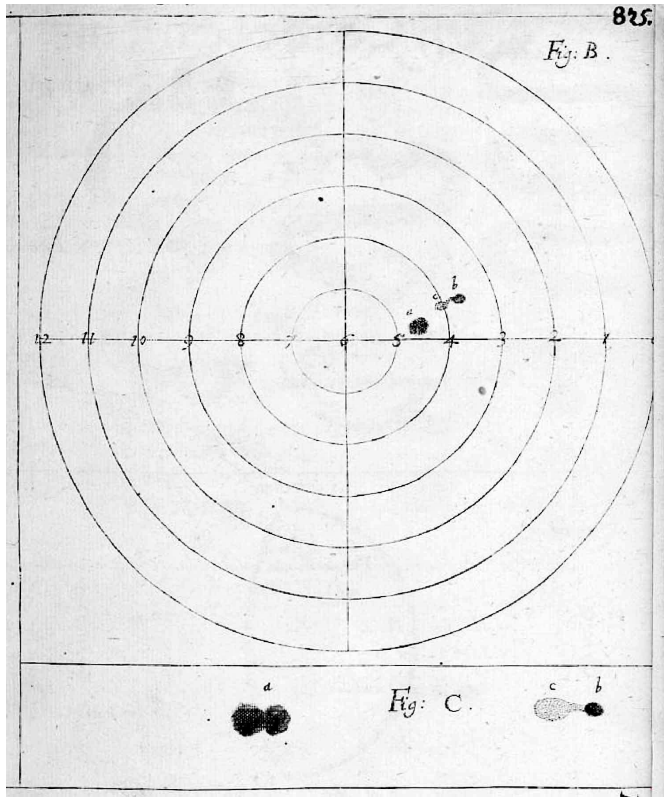}
\caption{Original sunspot drawing by Kirch on 1680 May 22 (Greg.) as
observed with a 10-foot-telescope and using a projection screen.}
\label{1680_05_22_drawing}
\end{figure}

We have two options to obtain the spot positions. We
can measure the positions directly from Kirch's figure 
assuming that west is to the left, see Fig.~\ref{1680_05_22_drawing}.

The second option consists of using the separations of the spots
from the solar limb from Ihle's and Kirch's observations and
matching the two assuming the solar differential rotation profile.

\begin{itemize}
\item May 20: Spot a: 3.9 inch, Spot b: 4.9 inch (from Ihle's text)
\item May 22: Spot a: 1.42 inch, Spot b: 2.31 inch (from drawing)
\end{itemize}

See also Figs. 1 and 2.
For spot a, we obtain a double-peaked probability distribution for the solar latitude
with peaks at $\sim -11.6 \pm 6.0^{\circ}$ and $\sim 10.0 \pm 6.0^{\circ}$.
For the spots called b and c, we obtain $-0.6 \pm 13.2^{\circ}$.
Given the drawings, this is also a good approximation for the latitude of spot a.
The large uncertainties indicate that the positional estimates by Ihle are not very precise.

Spots a, b, and c together form most likely one group given today's definition,
where spot a is leading (and larger) and where the trailing spot
is smaller and fragmented (b and c), which is typical for groups.

\begin{figure}
\includegraphics[width=0.485\textwidth]{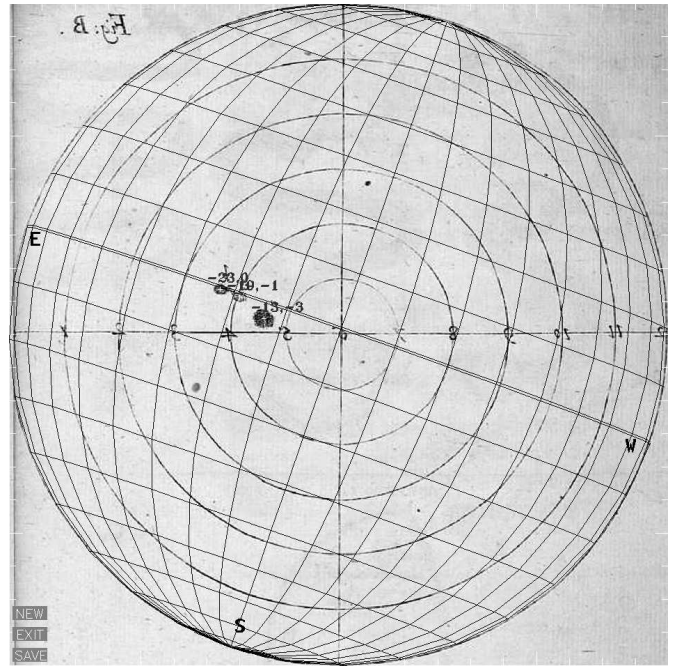}
\caption{Mirrored version with measured sunspot positions. The
tilt angle of the Sun was $B_0=-1.20^{\circ}$.}
\label{1680_05_22_tip}
\end{figure}

For Kirch himself, HS98 report one group for
1680 May 22-27 and a spotless Sun for May 29.
This is slightly different from Kirch's records,
where he said that spot a was also still seen on May 28.
HS98 report for Ihle on 2 sunspot groups for May 20.
By comparison to Kirch's account, this is most certainly
also the same one group (with 2 major spots) as observed by Kirch,
so Ihle saw only one group on May 20.

HS98 list only one more observation by someone else for 1680 May 20-29,
namely
Cassini with one group seen May 20-30,
which is almost consistent with Kirch, 
who did not see any spots any more on May 29.
Sp\"orer (1889) listed a spot from 1680 May 20-30
and remarked that the same spot was observed by Ihle and Kirch.

\subsection{1680 June 15--29}
\label{juni.2680}

In the journal article in {\it Himmels-Zeitung}, G. Kirch writes about
the June 1680 sighting of two sunspot groups:
\begin{quote}
Den 5.(15.) Junii war eine Macul um 3. Uhr Nachmittage
in der Sonnen 1 1/4 Zoll vom Ostrande. Den 6.(16.) Jun. um
3. Uhr n. war gedachte Macul um einen Zoll f\"order ger\"ucket
und 2 1/4 Zoll in der Sonnen. Diese Macul war viel kleiner als die
vorigen, welche sich im May sehen liessen, sie war auch nicht so dicht und
schwartz. Den 8.(18.) Junii um 7. Uhr vormittage war noch eine
Macul in die Sonne ger\"ucket, und etwan 3/4 eines Zolls drinnen, sie war
etwas kleiner als die erste. Den 9.(19.) Junii um 8. Uhr v. befand sich
diese letzte Macul 1 1/2 Zoll in der Sonnen.
Den 10.(20.) Junii um halbweg 5. Uhr n. war die letzte Macul
etwan 3. Zoll vom Ostrande, und auch so weit vom Centro der Sonnen,
und gr\"osser als die erste, welche das Mittel der Sonnen \"uberschritten,
und 2. Zoll von ihrem Centro stunde: beyde waren mehr als ein
Viertel des Sonnen-C\"orpers von einander. Den 12.(22.) Junii
war die erste Macul, fr\"uh um halbweg 5. noch 2. Zoll vom Westrande,
und den n\"achstfolgenden Tag um 10. Uhr v. 1 1/3 Zoll. Forthin konte
man wegen tr\"uben Wetters nichts gewisses mehr observiren. Den
19.(29.) Junii war die Sonne wiederum gantz rein ohne Maculen.
\end{quote}

The paragraph can be translated (first always the Julian date,
then the Gregorian in brackets) to English as follows:
\begin{quote}
On the Jun 5 (15) at 3 pm, there was a spot in the Sun, 1.25 inches
from the eastern limb. On Jun 6 (16) at 3 pm, the same spot had
advanced by one inch and was at 2.25 inches in the Sun. This spot
was much smaller than the one that appeared in May, it was also
less dense and black. On Jun 8 (18) at 7 am, another spot had moved
into the Sun, at about 0.75 inches, and was somewhat smaller than the
first one. On Jun 9 (19) at 8 am, the second spot was 1.5 inches in
the Sun.
On Jun 10 (20) at 4:30 pm, the second spot was about 3 inches from
the eastern limb, and as much from the centre of the Sun. It was
larger than the first one, which had passed the middle of the Sun
and stood at 2 inches from the centre. Both were separated by
more than a quarter of the solar body. On Jun 12 (22) at 4:30 am,
the first spot was 2 inches from the western limb, and 1.33 inches
on the following day, at 10 am. After that, dull weather did not
allow observations. On Jun 19 (29), the Sun was yet again entirely
spotless.
\end{quote}

The description contains the following positional information for
spot~1, measured from the disk centre:
\begin{itemize}
\item 1680 Jun 15, 3:00 pm: 4.75 inches
\item 1680 Jun 16, 3:00 pm: 3.75 inches
\item 1680 Jun 20, 4:30 pm: 2 inches
\item 1680 Jun 22, 4:30 am: 4 inches
\item 1680 Jun 23, 10:00 am: 4.67 inches
\end{itemize}
and for spot~2:
\begin{itemize}
\item 1680 Jun 18, 7:00 am: 5.25 inches
\item 1680 Jun 19, 8:00 am: 4.5 inches
\item 1680 Jun 20, 4:30 pm: 3 inches
\end{itemize}

\begin{figure}
\begin{center}
\includegraphics[width=0.485\textwidth]{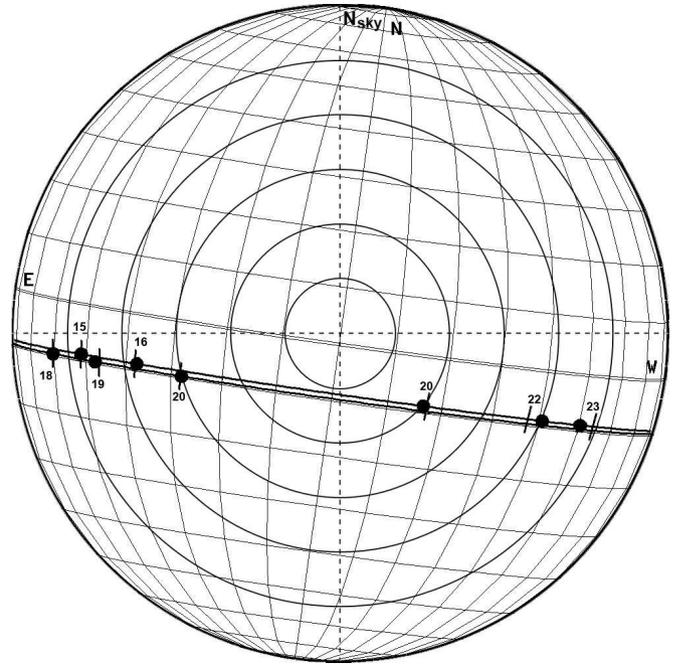}
\end{center}
\caption{Reconstructed positions for the sunspots seen on 1680 Jun~15--23 (Greg.),
the corresponding days in June are indicated in the graph. The Sun
had a tilt between $B_0 = 1.68^{\circ}$ and $B_0 = 2.57^{\circ}$ on these days.}
\label{1680_june}
\end{figure}

The first spot delivers an optimum solution at $b=-8.8 \pm 7.0^{\circ}$.
The second spot was situated at $b=-9.6 \pm 6.8^{\circ}$, according
to the best solution. Fig.~\ref{1680_june} shows the two most likely
paths of the spots across the solar surface. There are less likely
solutions on the northern hemisphere for both spots which we omitted
here.
Interestingly, the difference in heliographic longitude between
the two groups is about 45--50$^{\circ}$, which is much less
than the ``quarter of the solar body'' stated by Kirch. Given
the distances from the limbs, there is very little freedom in
the heliographic longitudes in our reconstructions; the uncertainties
remain in the latitudes. It is therefore unclear, how Kirch came
to the above mentioned statement.

HS98 report on one group on Jun~15, 16, and 18--22 seen by Kirch.
This differs from Kirch's publication in that Kirch saw one group
also on Jun~23, but did not report on Jun~21, and he actually saw
2~groups on Jun~18--20. The spotless day on Jun~29 agrees with HS98.
Other observers were Cassini (one group on Jun~13), 
Picard (spotless on Jun~20), Eimmart (spotless on Jun~27), 
and Siverus (spotless Jun~23 to Aug~3),
and presumably also Flamsteed (spotless on Jun~18, 23, 25).

Flamsteeds observing days were taken from Flamsteed (1725),
which contains a large number of altitude measurements of
the Sun and no mention of sunspots. Together with his
statement {\it These appearances \dots have been so rare of late
that this is the only one I have seen in his face since December 1676}
(Flamsteed 1684),
HS98 inferred a group number of zero for all these
observing days (Hoyt \& Schatten 1995). We do not know, however, how
carefully Flamsteed checked for spots when making his positional
observations at the quadrant,
see also Carrasco \& Vaquero (2016).

\subsection{1681 July 21-31}
\label{25.07.1681}

In letter numbered 94 by Herbst (2006) of 1681 July 16/26 (Jul./Greg.), 
G. Kirch reports to Gottfried Schultz the following:
\begin{quote}
Itzt befindet sich eine Macul in der Sonnen: 
Den 13/23 Julii haben wir sie zum ersten mahl durch einen 10- und 5 sch\"uhigen Tubum gesehen. 
Den 11/21 Jul. ist sie noch nicht drinnen gewesen: 
Sie war l\"anglicht, etwan ein paar Zoll vom Eintrits Rande, 
nur ungefehr, denn es war bald tr\"ube. 
Den 14/24 Jul. vor Mittage um 08:45 Uhr war die Macul 2,5 Zoll in der Sonnen: 
Den 15/25 Julii um 9:30h v. war sie 3,5 Zoll drinnen. 
Heute habe ich sie, wegen des tr\"uben Wetters noch nicht sehen k\"onnen: [...] 
Die Macul ist nicht gar gro\ss{}, 
kan durch einen 4 sch\"uhigen Tubum gar schwerlich gesehen werden. [...] 
Den 24 Jul/3 Aug sch\"atze ich werde sie aus der $\odot$en treten.
\end{quote}

This we translate directly to (square brackets from us):
\begin{quote}
Now there is a spot in the Sun: 
On July 13/23 [Jul./Greg.] we saw it [spot] for the first time through a 10- and 5-foot telescope. 
On July 11/21 it was not yet on it: 
It was longish, a few inches from the ingress limb, just approximately, because it soon got dull. 
On July 14/24 before noon around 8:45h the spot was 2.5 inches in the sun's disc: 
On July 15/25 at 9:30h a.m. it was 3.5 inches in it [i.e. 3.5 inches from the eastern limb]. 
Because of the bad weather I could not observe it today: [...] 
The spot was not large, barely visible through a 4-foot telescope. 
[...] I suppose it will leave the Sun's disc on July 24/Aug 3.
\end{quote}

Another mention of that spot can be found in letter 96
from 1681 July 27 / Aug 6 (Herbst 2006),
where G. Kirch writes to Gottfried Schultz:

\begin{quote}
Die $\odot$en Macul war den 16. Jul Sonnabends gegen Abend schon ziemlich schwach,
konte den 18/28 Jul. noch gesehn werden. Donnerstags war sie auch durch den
13 sch\"uhigen Tubum nicht mehr zu finden.
\end{quote}

\begin{quote}
The solar spot was already small on Saturday evening, July 16/26
and still visible on July 18/28.
On Thursday [July 31 Greg.] it was not visible any more, even through a 13-foot telescope.
\end{quote}

In summary, G. Kirch saw the group July 23-25, but not yet on Jul 21 (spotless),
and then also on July 26 \& 28, but not any more on Jul 31 (spotless), all dates Gregorian.

\begin{figure}
\begin{center}
\includegraphics[width=0.485\textwidth]{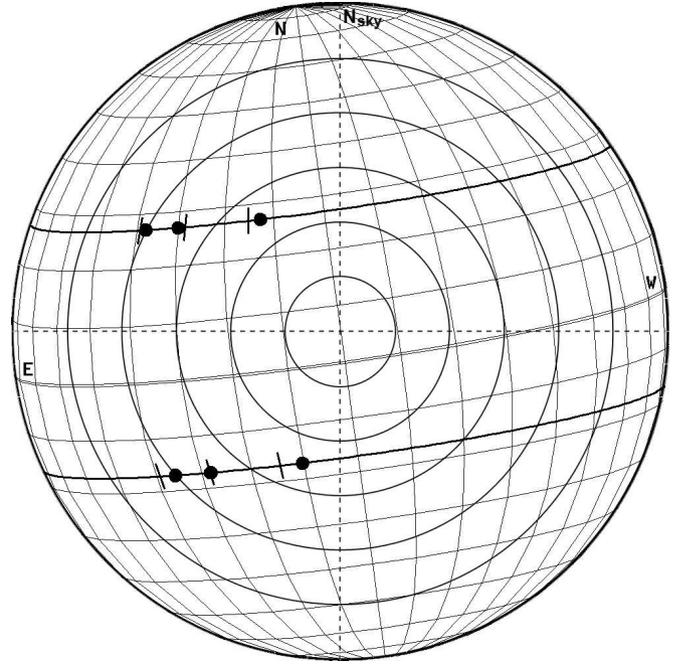}
\end{center}
\caption{Reconstructed positions for the sunspot seen on 1681 Jul~23, 24, and 25 (Greg.).
The Sun had a tilt of $B_0=5.49^{\circ}$, $5.55^{\circ}$, and $5.63^{\circ}$ on these days, respectively.}
\label{1681_07_23_1201_tip}
\end{figure}

With only the separations from the center which are explicitly given, we obtain:
\begin{itemize}
\item Jul 23: {\it few inches from limb} $\simeq 4$ inch from center
\item Jul 24: {\it 2.5 inches from limb} = 3.5 inch from center
\item Jul 25: {\it 3.5 inches from limb} = 2.5 inch from center
\end{itemize}
Omitting the inferred near-limb cases, we obtain two solutions:
$b_1=-17\pm9^{\circ}$ and $b_2=27\pm9^{\circ}$, with the former having a slightly
higher peak probability density. The resulting latitudes, together
with the observations, are shown in Fig.~\ref{1681_07_23_1201_tip}.

\begin{figure}
\begin{center}
\includegraphics[width=0.485\textwidth]{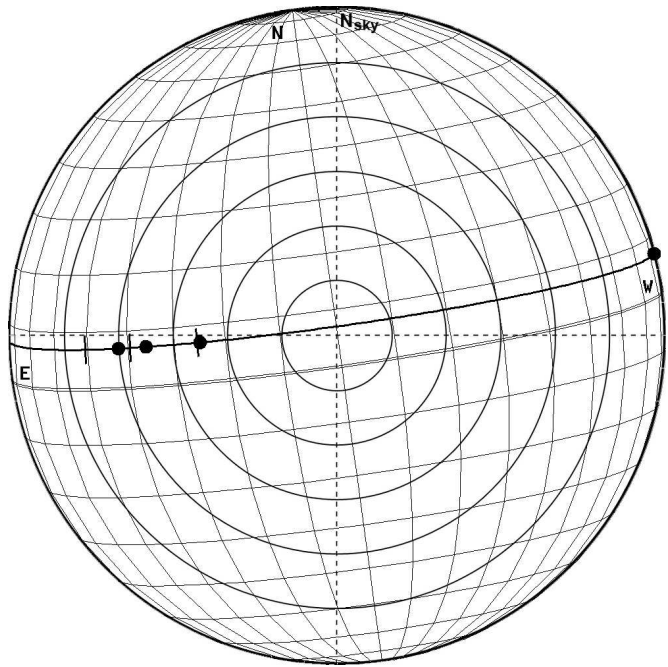}
\end{center}
\caption{Reconstructed positions for the sunspot seen on 1681 Jul~23, 24, 25,
and as expected for Aug~3 (Greg.) at the limb, when the tilt was $B_0=6.01^{\circ}$.}
\label{1681_07_23_1200_tip}
\end{figure}

Since the rotation of the Sun was very well known at Kirch's
time, the expected exit from the solar disk is another observational constraint
on the location of the sunspot. Kirch must have had the longitude of the spot
in order to make such an estimate. We therefore assumed it to be exactly on the
solar limb on Aug~3 (Fig.~\ref{1681_07_23_1200_tip}).

We then obtain a single solution with $b=+7\pm9^{\circ}$. The error
margin is the 68\% confidence interval of the probability density distribution. 

If we move the limb case to July~31, as the Sun was reported as
spotless on that day, the possible latitudes actually increase.
This is because at given separations from the center (4 inch, 3.5 inch, and 2.5 inch
on the eastern side of the solar disk),
the spots must have been closer to the central meridian, i.e.
at higher latitudes. The result would be $-22^{\circ}$ and $+34^{\circ}$ with
the first latitude having the larger likelihood.
We cannot arrive at a final unique result.

Let us compare these dates with other observers and catalogues:
Neither the Sp\"orer catalogue (1889) nor the diagram by RNR93 show matching entries for these dates
(for RNR93, we used here and below the tabular compilation by Vaquero et al. 2015). 
The only observer listed in HS98 is G. Kirch himself with an observation 
of one sunspot group 1681 July 24-27 (Greg.), which differs in dates.
For the correspondence about this particular spot between G. Kirch and Schultz,
see also Herbst (2005).

The data in HS98 for Flamsteed for a spotless Sun for certain days in July were
probably not meant as such by Flamsteed (1725) himself, see Sect. 3.2.
Picard (from Paris, hence quite certainly Gregorian) 
reported a spotless Sun for July 21, 22, 24, 26, \& 28 (HS98),
Varin spotless for July 21, 22, 24, 26 \& 28 (HS98),
which appear to be consistent with each other,
but inconsistent with the reported spot by Kirch.
Did Kirch saw a spot that was too small for Picard and Varin?
Furthermore, HS98 report that Siverus from Hamburg would have reported a spotless
Sun on each and every day from 2 Jan 1677 until 31 Dec 1690 with few exceptions
(when other detect spots), so that these data in HS98 probably are based on
a generic statement by Siverus that is mis- or over-interpreted by HS98
(like for Marius, see Neuh\"auser \& Neuh\"auser 2016);
Vaquero et al. (2016) already suggested to remove all these generic zeros from Siverus.

\subsection{1684 May 5-6}
\label{16.05.1684}

\begin{figure}
\begin{center}
\includegraphics[width=0.485\textwidth]{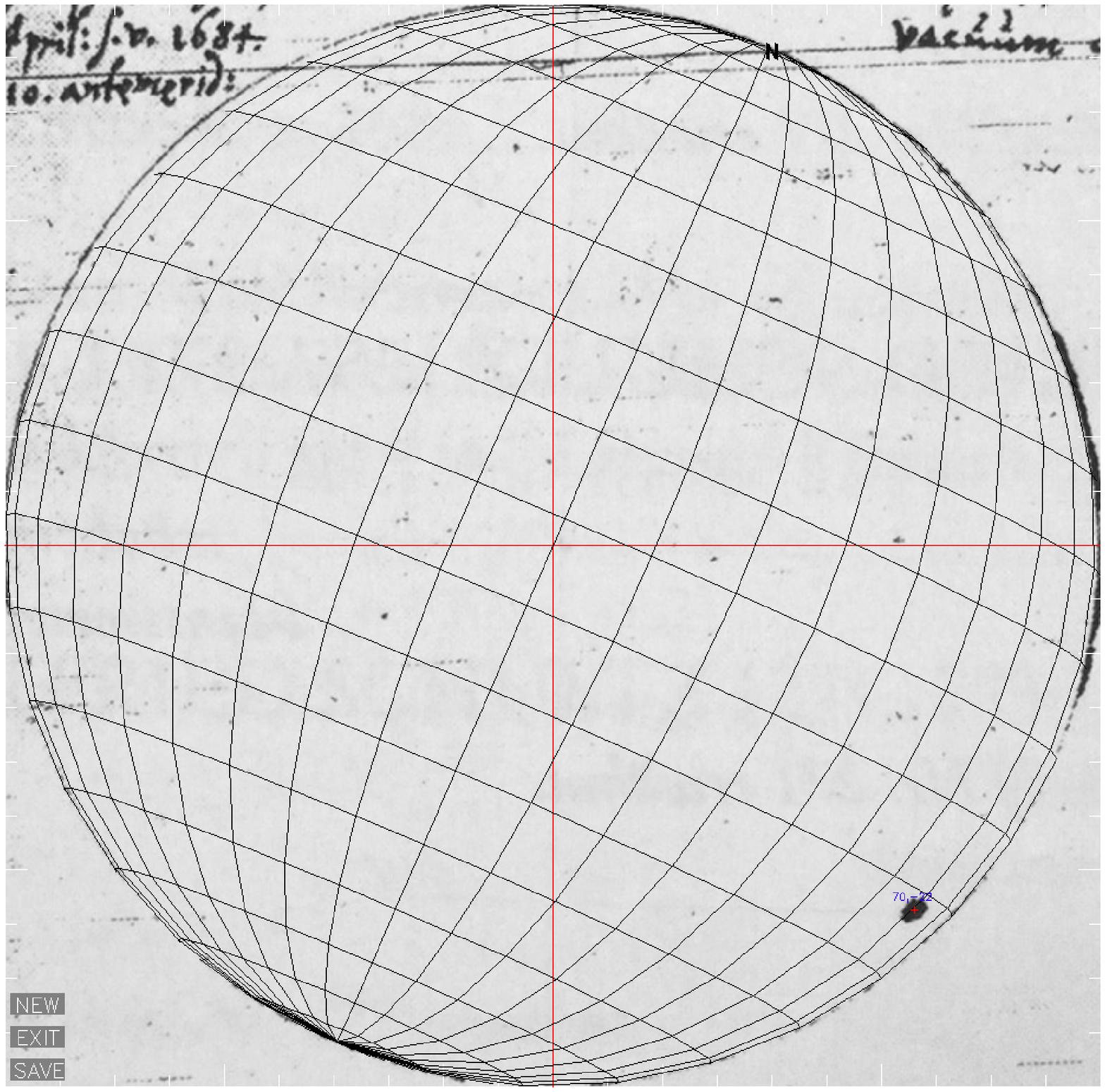}
\includegraphics[width=0.485\textwidth]{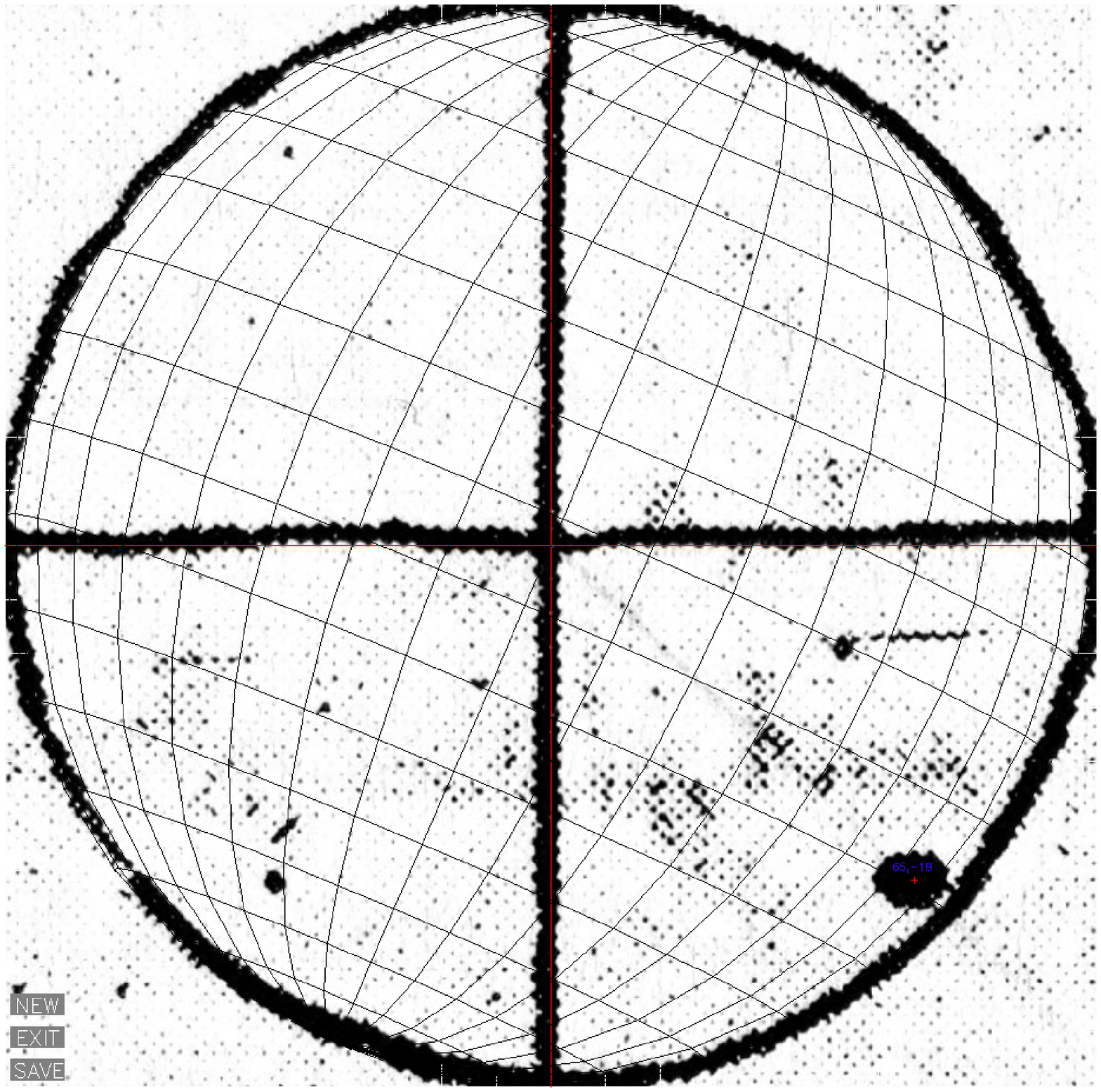}
\end{center}
\caption{Reconstructed positions for the sunspot reported for 1684 May~6
(Greg.) by Ihle (top) and Kirch (bottom); the coordinate system is added by us.}
\label{1684_05_06_1000_grid}
\end{figure}

The letter from G. Kirch to Johannes Hevelius from 1684 Apr 26/May 6, number 267 in Herbst (2006), 
is the first description of a sunspot in the correspondence containing a small sketch by G. Kirch.
In the library of the Observatory of Paris, where this letter is located, an additional page
is attached to this letter, supposedly another drawing by Kirch of the same spot.
However, the writing on that additional page is apparently by the hand of J.A. Ihle,
another well-known sunspot observer, and this drawing (lower panel of Fig.~\ref{1684_05_06_1000_grid})
might be from his letter to
Hevelius dated 1685 June 10 (Greg.), as suggested by Herbst (2006).

The mentioned description by G. Kirch reads as follows:
\begin{quote}
Itzt da ich schlie\ss{}en will, k\"ombt Hr. Ihle, und berichtet mich, da\ss{} eine Macula in der Sonnen, 
welche Er gestern nicht gesehen, um 10 vormittags soll sie etwan gestanden seyn, wie beystehende Figur anzeiget.
\end{quote}

Staying as close to the original wording as possible, this translates into:
\begin{quote}
Now, coming to an end, Mr. Ihle is coming and reports that a spot [is] in the Sun, 
which he has not seen yesterday. 
Around 10 o'clock before noon it [the spot] was said to be standing like the attached figure shows.
\end{quote}

The two drawings may be assumed to be aligned with an equatorial
coordinate system. The resulting latitudes are then $-21.9^{\circ}$ (Ihle)
and $-19.2^{\circ}$ (Kirch). See Fig.~\ref{1684_05_06_1000_grid} for the
original drawings with our reconstructed coordinates. The position angle of the solar rotation axis for
this date is $23.2^{\circ}$ clockwise.
The two positions from the sketches average to a latitude of $-20.6 \pm 3.5^{\circ}$.
The error is the $68\%$ confidence interval taken from a
$t$-distribution for poorly sampled averages.

Here, it is important to note that Ihle reported on May 6
that no spot was seen on 1684 May 5 (Greg.), 
and also on none of the days before in that year  --
as reported on his figure caption:
\begin{quote}
1684 Avril 26 [Julian]:
Typus notabilis maculae Solaris, Lipsiae visae, die 26 April: s.v. 1684
[Julian]. Hora 10. antemeridiana.
Die Praecedente, et anno amplius, Solem plane vacuum offendimus.
\end{quote}
i.e. in English:
\begin{quote}
Drawing of a remarkable sunspot, seen in Leipzig [Germany] on 1684 May 6 [Greg.],
in the tenth hour before noon.
On the day before as in the whole year so far, the Sun was without spots.
\end{quote}

The latter is a generic statement which may not hold for each and every day,
as it is well possible that Ihle could not observe on some days,
e.g. due to clouds.
Since HS98 do not list Ihle for 1684 (not even zeros), 
we can conclude that HS98 did not use this letter nor the
drawing of Ihle as source(s).

HS98 list a number of observers for this period: 
La Hire (1684 May 1-8, 1-2 spot groups), Flamsteed (Apr 25-May 8, 1 group), 
Cassini (May 5-17, 1 group), Clausen (May 4, 5, and 7, 1 group),
Eimmat (spotless on May 4, 5, 7, and 8),
and Ettmuller (Apr 26, 28-30, May 3, 5, 7, and 8, 1 group),
as well as G. Kirch (Apr 26-May 7, 1 group).

The data given for G. Kirch in HS98 are not consistent with the statement in the letter from G. Kirch
to Hevelius (nor the information from Ihle), 
so that we can conclude again that HS98 did not use this letter 
nor the Ihle drawing as source(s) --
or that HS98 did not correct Kirch's Julian date to a Gregorian date
(already Sp\"orer 1889 wrote that Kirch observed this spot since 26 April old style).

If the data given for Flamsteed (Apr 25-May 8, 1 group) in HS98 would be
Julian, then he would have detected a spot May 5-18 (Gregorian),
which is well consistent with Cassini (May 5-17, 1 group), who quite certainly
used the Gregorian calendar. Still, La Hire in Paris saw one or two groups 
only May 1-8, but not May 10-12 and 14-18.
Eimmart in Nuremberg would also have reported a spotless Sun for May 5 (like Ihle),
but did not observe on May 6 (HS98).

The spot seen by Kirch and Ihle near the western limb on May 6 definitely rotated
out of view within one to few days, and did not last until May 17.
This may indicate that there were two different spot groups in May 1684.

In the diagram by RNR93 we found no spots for 1684 May
(but $-11^\circ$, $-11.6^\circ$, $-12.4$, and $-10.5^\circ$ for 1684 Jun, Aug, and Sep).
The Sp\"orer catalogue lists a spot observation May 5-17 with a heliographic latitude of $-11^\circ$ 
for Cassini (Lalande, Mem. 1778, p. 401).
Neither the latitudes in Sp\"orer nor RNR93 for spring and summer 1684
seem to be consistent with the drawing by Ihle, so that Ihle's spot could be another spot.
On the other hand, we have no error bars for the data in Sp\"orer's catalogue nor in
RNR93, so that it is hard to conclude whether or not they are consistent with Ihle.

Given that the spot drawn by Ihle and G. Kirch will obviously rotate out of view
within one or very few days after May 6, and given that 
Cassini (and maybe Flamsteed) observed a spot until May 17,
listed also by Sp\"orer for Cassini at $b = -11^\circ$,
we conclude that the spot drawn by Ihle and G. Kirch might be
a different spot compared to the one listed by Sp\"orer,
and that the one listed by the latter authors 
may have been too small for Ihle.
Hence, we list this extra spot in Table 1. 

\subsection{1684 June 28-July 7}
\label{05.07.1684}

This spot was first mentioned by G. Kirch in a letter 
to Georg Samuel D\"orffel on (Gregorian) 1684 July 6 (no. 275, Herbst 2006):
\begin{quote}
Itzt ist eine Macul in der Sonnen, sie ist nun schon wie ich davor halte zum dritten mahl drinnen, 
und fein gro\ss{}, es ist m\"oglich da\ss{} sie wol gar zum 4ten mahle drein komme.
am 18 jun. war ihr dritter Eintritt, heute um 6 nachmittags stehet sie gleich noch 2 Zoll vom Austritts Rande.
\end{quote}

\begin{quote}
Now is a spot in the Sun, I guess it is its third passage. 
It is large and might possibly be visible a fourth time. 
On June 18/28 was its third entrance, today [July 6 (Greg.)] 
at 6 in the afternoon it is two inches from the egress limb.
\end{quote}

A few days later 
(July 12/22),
G. Schultz wrote to G. Kirch about his observation on July 5-7 using Gregorian dates
(no. 277, Herbst 2006):
\begin{quote}
Da Ich sie denn noch gutt genug befunden habe, und zugleich, 
nach wuntsch eine Maculam in quadrante occidentali partis inferioris Disci Solaris, 
ohngefehr 3 zoll vom centro gefunden, welche Ich auch folgenden 6 und 7 Julij, 
so lange das Wetter gutt gewesen, mit fernerer ann\"aherung zum margine occidentali, darinnen gesehen.
\end{quote}

\begin{quote}
Since I approved them [lenses] I, as I wished, found a sunspot in the lower, 
western quadrant of the Sun's disc, around 3 inches from the centre. 
As long as the weather was good, I could see it also 6 and 7 July, approaching the western edge.
\end{quote}

Herbst (2005) discusses the correspondence between G. Kirch and G. Schultz 
between 1681 and 1692 including this spot.

We summarize that Schultz saw the spot on July 5, 6, and 7,
and that G. Kirch saw it on July 6.

We obtain the positions
\begin{itemize}
\item 1684 July 5 (assuming noon): 3 inch from disk-centre
\item 1684 July 6 (6 pm): 4 inch from disk-centre
\end{itemize}

\begin{figure}
\begin{center}
\includegraphics[width=0.485\textwidth]{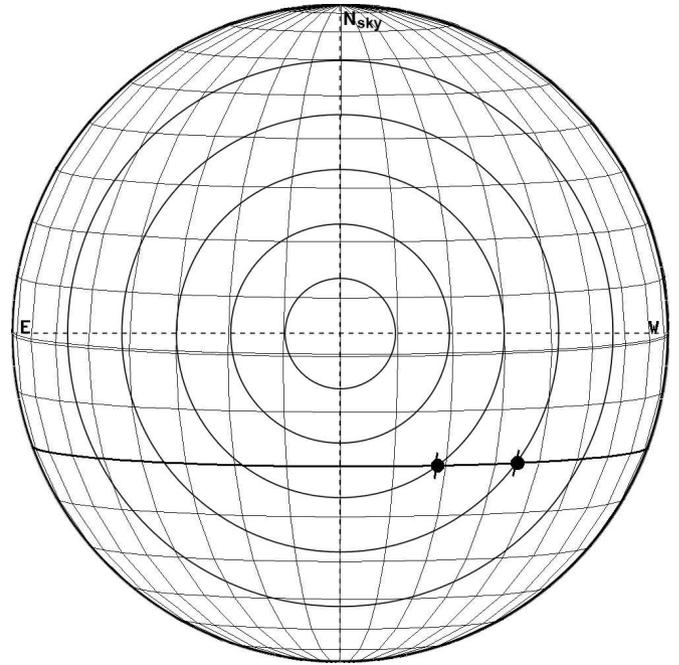}
\end{center}
\caption{Reconstructed positions for the sunspot seen on 1684 Jul~5 and~6
(Greg.), $B_0=3.88^{\circ}$ and $4.00^{\circ}$, respectively.}
\label{1684_07_05_1200_tip}
\end{figure}

Additionally, we know that the spot was in the south-western quadrant of the Sun
({\it Schultz: lower, western quadrant}). The statement of the ingress of the spot on June~28
was probably not an actual observation (but calculated) and was not used. (Including the limb position
on June~28 would lead to an unlikely latitude of $-35^{\circ}$).
The solution is shown in Fig.~\ref{1684_07_05_1200_tip}
where the best-fit latitude of $-20.4 \pm 13.6^{\circ}$ is shown.

This spot can be found in Sp\"orer's catalogue for Cassini (Lalande, Mem. 1776, p.474). 
For the time of observation (June 28 - July 9) he reports a heliographic latitude 
of $-10.8^\circ$, consistent with our result.
RNR93 shows no matching observation
(but $-11^\circ$, $-11.6^\circ$, $-12.4$, and $-10.5^\circ$ for 1684 Jun, Aug, and Sep).

HS98 list a number of astronomers, who observed one group each:
La Hire June 28-July 1 and July 3-10, 
Flamsteed June 24 and 27 (but spotless for June 28 and July 2),
Caswell June 26 (but spotless for June 30), 
Cassini June 27-July 9, 
Eimmart June 22-28 (and spotless June 29), 
Hevelius June 28-July 9 (and spotless July 10 and 12), and Gulielmini July 6 and 8,
as well as G. Kirch for June 18-31 (and spotless July 1-15),
the latter being clearly inconsistent with the letter from G. Kirch
(HS98 may have assumed that Kirch's statement for June 18 is not a calculation,
but a real observation, and they did not convert Kirch's Julian date to Gregorian).
Vaquero et al. (2016) already suggested to remove the Kirch value for 1684 June 31
(but did not comment on the other dates in HS98 this year for Kirch).
If the spot was seen until about July 10, it cannot have been
seen as early as June 18. 
The 1684 data from Caswell from London as listed in HS98 may well be his Julian dates,
not yet corrected to Gregorian by HS98, so that they would then be consistent with
the other observers.
The same could apply here to Eimmart from Nuremberg, who is listed for 1 group from 1684 June 22-28
(and spotless in June 29), which would be consistent with, e.g. La Hire, Cassini, and Hevelius
(who certainly used the Gregorian calendar), if we would shift Eimmart's dates by 10 days
(he lived in a protestant area, where the old calendar style was still in use until 1700 Feb).
The observer Schultz is not listed at all in HS98.

From the fact that G. Kirch assumed in his letter that the spot
is seen for a third time, we can conclude that he also saw or knew
about a spot seen about one and two months earlier.
A spot seen by him two months earlier (early May 1684) was discussed
in the previous subsection ($b = -20.6 \pm 3.5^{\circ}$,
i.e. similar to our conclusion for this section).
Kirch's assumption that the spot is the same as the one observed on May~6 
would only give a match with the data in this subsection, 
if the spot was in the eastern hemisphere on May~6.
HS98 list one group for G. Kirch also for May 29 to June 6
(and also one group for Clausen from May 21 to June 3),
which are not mentioned in the records we study here.

\subsection{1686 Apr 25-May 1}
\label{25.04.1686}

On 1686 April 17/27 G. Kirch mentioned another sunspot seen on April 15/25. 
In this letter to Gottfried Teubner, number 319 in Herbst (2006), 
his description turns out to be fairly short.
\begin{quote}
Vorgestern fand ich sie zum ersten mal. Um 5:30h nachmittags war sie etwan 1 Zoll vom 
Centro der $\odot$en. Heute um 7 vormittags war sie 2 Zoll davon.
\end{quote}

Nonetheless one can still obtain the separation from the centre as well as the date:
\begin{quote}
Two days ago [Apr 25 (Greg.)] I saw it [the spot] for the first time. 
Around 5:30h in the afternoon it was about 1 inch from the centre of the Sun. 
Today [Apr 27 (Greg.)] 7 o'clock a.m. it was 2 inches from it.
\end{quote}

On the same day G. Kirch wrote to Georg Christoph Eimmart as well, using the same words. He only adds:
\begin{quote}
Sie ist etwan 12'' oder 15'' gro\ss.
\end{quote}

\begin{quote}
It is about 12'' or 15'' large.
\end{quote}

H. Kirch sent his complete observation to John Flamsteed on 1686 Oct 31 
(letter no. 335, Herbst 2006). The original text written in Latin:
\begin{quote}
Die 15 Aprilis, Hor. 10 1/2 ante meridiem Maculam in Sole vidi, qvae distabat a centro Solis, 
H. 5 1/2 post meridiem adhuc 1 Dig. Die 17 Apr. H. 6 post meridiem 2 1/2 Dig. A Limbo Solis vero die 19 Apr. H. 
6 post meridiem 1 1/2 Dig. et die 21 Apr. H. 7 ante meridiem 1/3 Dig.
\end{quote}

We translate this to English as follows based on the German translation in Herbst (2006):
\begin{quote}
On Apr 15 [Apr 25 (Greg.)] around 10:30h a.m. I found a spot in the Sun. 
At 5:30h p.m. it was one inch from the solar centre. 
On Apr 17 [Apr 27 (Greg.)] around 6h p.m. 2.5 inches. 
But on Apr 19 [Apr 29 (Greg.)] around 6h p.m. 
1.5 inches from the solar limb and on Apr 21 [May 1 (Greg.)] at 7 o'clock 1/3 inches.
\end{quote}

So the positions adopted are
\begin{itemize}
\item 1686 Apr 25, 5:30 pm: 1.0 inch from disk centre
\item 1686 Apr 27, 7 am: 2.0 inch
\item 1686 Apr 27, 6 pm: 2.5 inch
\item 1686 Apr 29, 6 pm 4.5 inch
\item 1686 May 1, 7 am: 5.67 inch
\end{itemize}

\begin{figure}
\begin{center}
\includegraphics[width=0.485\textwidth]{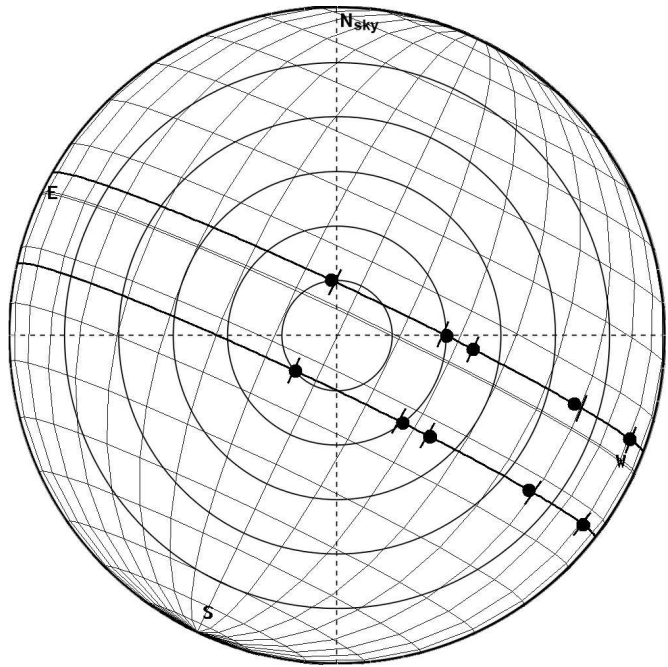}
\end{center}
\caption{Reconstructed positions for the one sunspot seen on 1686 Apr~25--May~1.
(Greg.), $B_0=-4.24^{\circ}$, $-4.09^{\circ}$, $-4.04^{\circ}$, $-3.83^{\circ}$, and $-3.68^{\circ}$, respectively,
being either north or south.}
\label{1686_04_25_1730_tip}
\end{figure}

We obtain two possible solutions: $-13\pm5^{\circ}$ and $+4\pm5^{\circ}$.
Fig.~\ref{1686_04_25_1730_tip} shows the corresponding
tracks of the spot on those latitudes. 

Sp\"orer gives a heliographic latitude of $-15^\circ$ for the observation between Apr 23-May 1
by La Hire (Mem. X, p. 708 and Mem. 1778, p. 402);
in 1686, RNR93 shows no spots before July
(heliographic latitudes of $-1.8^\circ$, $-14.8^\circ$, and $-15.3^\circ$ for July,
most of these values are consistent with one of our solutions being $-13\pm5^{\circ}$.

HS98 list La Hire for one group Apr 26 and 28 to May 1,
and Cassini for one group on Apr 29.
HS98 also list G. Kirch for one group Apr 25 to May 1.

\subsection{1688 Dec 14 to 1689 Jan 19}
\label{14.12.1688}

This spot is mentioned for the first time in letter number 594 (Herbst 2006), 
for which addressee and date are missing in the letter itself.
Given that Ulrich Junius wrote in his letter (no. 593 in Herbst 2006) that he
sent it (his own letter) to Kirch on 1698 Jan 22 / Feb 1 together with a letter
from Ihle (in the same envelope), Herbst (2006) argues that the undated letter
no. 594 is from Ihle from the same date.
In letter no. 594, its author writes to G. Kirch
(with Julian dates):
\begin{quote}
Die sehr denckw\"urdige Erscheinung de\ss{} Sterns in Collo Ceti, 
wie auch derer 2 macularum in Sole, wolle doch der Herr k\"unfftig in einen Kalender bringen.
\dots die sonne habe ich \dots aber gantz ledig befunden am 20. und 21. Decemb: 6. 7. 19 Januar:
zwischen solchen tagen aber ist sie niemahls sichtbar, oder sonst zum observiren
gelegenheit gewesen
\end{quote}

\begin{quote}
The very remarkable appearance of the star in the neck of Cetus, 
as well as the two sunspots might be published in one of your future calendars.
\dots the sun I saw \dots fully clear on Dec 20 and 21 and Jan 6, 7, 19 [Jul.]:
between those days, it was never visible, or otherwise there was no possibility to observe.
\end{quote}
The date of this sunspot observation is not given 
(see below for the exact dating of those spotless days).
The mentioned star in the neck of Cetus ({\it in Collo Ceti}) is
the pulsating variable star Mira (o Ceti).

Herbst (2006) cites an extract from G. Kirch's ephemeris from 1690, where it reads:
\begin{quote}
Die 4 Decemb. [1688] H 11 a.m. 2. Maculae 1 dig, infra centrum Solis apparebant. 
Die 5. Dec. 2 dig. a centro Solis. D. 8 Dec. Maculas in Sole invenire non amplius potui. 
Die sequente, coelo defaecatissimo, Sol purus cernebatur.
\end{quote}

\begin{quote}
On Dec 4 [1688 Dec 14 (Greg.)] at 11 o'clock a.m. 2 sunspots 1 inch beneath the Sun's 
centre appeared. On Dec 5 [Dec 15 (Greg.)], 2 inches from the solar centre. 
On Dec 8 (Dec 18 (Greg.)] I could not find the spots any more. 
The following day [Dec 19 (Greg.)] the sky was clear and a spotless Sun was seen.
\end{quote}

Herbst (2006) suggests that the two spots mentioned in the letter from Ihle to Kirch
are the spots seen by G. Kirch on 1688 Dec 14 and 15 --
and that the spotless days given by Ihle would then be 1688 Dec and 1689 Jan
(instead of 1697 Dec and 1698 Jan).
But what is so {\it very remarkable} with those two spots, that Ihle mentioned them
together with the {\it new star}? Maybe their mere existence given the
dearth of spots those years.

\begin{figure}
\begin{center}
\includegraphics[width=0.485\textwidth]{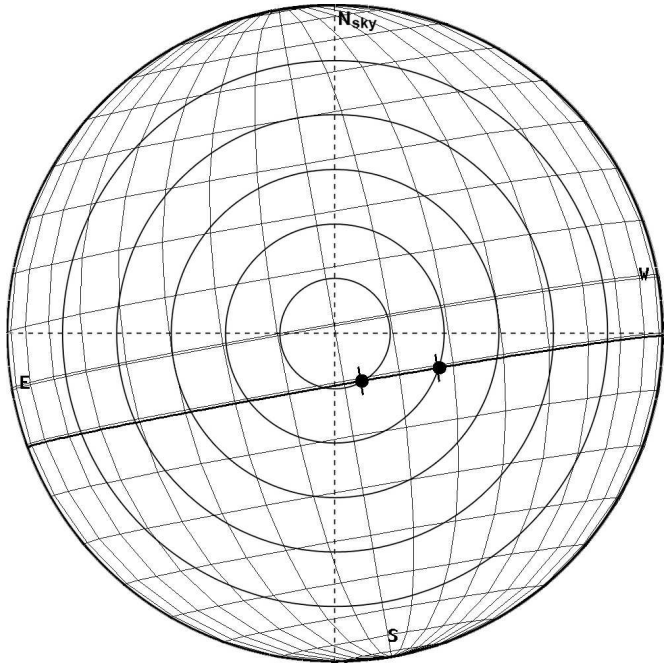}
\end{center}
\caption{Reconstructed positions for the sunspot seen on 1688 Dec~14--15.
(Greg.), $B_0=-1.48^{\circ}$ and $-1.61^{\circ}$, respectively.}
\label{1688_12_14_1100_tip}
\end{figure}

We adopt the positions
\begin{itemize}
\item 1688 Dec 14, 11 am: 1 inch
\item 1688 Dec 15: 2 inch (time not given, we assume noon)
\end{itemize}
and obtain two solutions at $-10.5 \pm 6.0^{\circ}$ and $+7.6 \pm 6.0^{\circ}$,
of which only the southern one can be correct, because Kirch stated
the spot was {\it beneath the Sun's centre}.
Fig.~\ref{1688_12_14_1100_tip} shows the corresponding matches
with the observations. The error may be underestimated here, since
there is an uncertainty in the time difference between the two
observations. 

Neither Sp\"orer's catalogue nor RNR93 provided a matching entry,
so that this is a new latitude for the butterfly diagram.

HS98 do not list any observations from Kirch for 1688.
We will consult the ephemeries of G. Kirch in a later work,
as this would have been beyond the scope of this present paper.
HS98 listed an observation by Arnold from Berlin, 
who saw two sunspot groups during the beginning of December (Dec 2/3). 
Here it is possible that HS98 did not convert a Julian date to the Gregorian,
because the old style was still in use in Berlin, where also Kirch worked.

For 1688 Dec 18 \& 19 (Greg.), G. Kirch specified that the Sun was spotless.
According to HS98, La Hire saw a spotless Sun also on Dec 17 \& 18 --
Flamstead also on Dec 19, but that is probably Julian, i.e. Dec 29 (Greg.).

In addition, according to Ihle and the dating by Herbst (2006),
we have spotless days on [1688] Dec 20 and 21 and [1689] Jan 6, 7, 19 [Julian],
i.e. on Gregorian dates 1688 Dec 30 and 31 as well as 1689 Jan 16, 17, and 29. 
For 1689 Jan 16, no other
observers are listed by HS98 (except generic zeros), but for Jan 17 and 29, HS98 list La Hire
as observer, who also saw a spotless sun; there were many generic
zeros for Jan 1689 in HS98 for Siverus, Dechales, and Cassini, which
may be incorrect.
For 1688 Dec 31, one other observer is listed, La Hire, who also reported a spotless sun;
for Dec 30, no other observer is listed by HS98, except Siverus, who would
have reported a spotless Sun for
1688 Dec 7-31 (and also most other days that year), which appears inconsistent
with Kirch; HS98 probably used a generic statement from Siverus,
which in fact does not apply to each and every day (similar as done incorrectly
for Marius 1617 and 1618, see Neuh\"auser \& Neuh\"auser 2016);
Vaquero et al. (2016) also suggested to remove those generic zeros from Siverus for 1688.
Hence, for the dating of those five spotless days to 1688/9, three could 
be confirmed (all by La Hire).

Alternatively, the author of the letter (where no date nor author is given,
Ihle according to Herbst 2006) 
could have meant Dec 1697 and Jan 1698 for spotless days in his
letter dated by Herbst (2006) to 1698 Feb 1 (Greg.).
Then, we would have spotless days on 1697 Dec and 21 and [1698] Jan 6, 7, 19 [Julian],
i.e. on Gregorian dates 1697 Dec 30 and 31 as well as 1698 Jan 16, 17, and 29.
We compare those days also to HS98:
for 1697 Dec 30 and 31, we have no other observations
(except Angerholm with incorrect generic zeros);
for 1698 Jan 16, 17, 29 (Greg.), we have La Hire and Cassini with a spotless day on Jan 17
(and Angerholm with incorrect generic zeros),
i.e. in total two confirmations for this dating.

It is therefore hard to decide, whether the dating of the observations 
by Ihle (?) to 1688/9 by Herbst (2006) is correct -- they could also have been
at the turn from 1697 to 1698.

One could also try to date the letter using the light curve of Mira,
whose brightening was mentioned:
one of its maxima was observed on 1698 Jan 29 (Prager 1934), 
which might be too late for the mentioning in the letter
(which was dated 1698 Feb 1 by Herbst 2006).
However, with a period of 332 days (Hoffleit 1997) and
ten periods earlier, we would arrive on 1688 Dec 27 (Greg.),
which would be fully consistent with a dating of the observations 
(of Sun and Mira) to 1688/9.

\subsection{1700 June 10-13}
\label{1700}

In the 750th of the collected letters (Herbst 2006), 
G. Kirch received a letter from his colleague Johann Philipp Wurzelbaur from Nuremberg. 
Note that this letter is from 1701 Jan 15; 
dates since 1700 Mar 1 given in the letters are now Gregorian dates;
the protestant German countries (including Nuremberg, where Wurzelbaur observed)
implemented the Gregorian calendar on AD 1700 Mar 1.
We can assume that the dates given in their letters after AD 1700 Mar 1 are Gregorian
(as confirmed by comparison with French observers in the next sub-section).
Some other non-catholic countries like England changed the calendar even later.
%

Wurzelbaur writes about the lack of sunspots since 1695 and his further sighting in June 1700. 
He continues to the last date those spots were seen and writes in German:
\begin{quote}
\dots auch nach A$^\circ$: [Anno] 1695 im Ma\"yen, 
unerachtet fast t\"aglicher durchsuchung disci Solaris keine ersehen k\"onnen, 
bi\ss{} im Juni des j\"ungstabgewichenen Jahres, da ich einige sehr 
schwache von 10 bis 13 des Monats erblicket, \dots
\end{quote}

We translate this to English as follows:
\begin{quote}
\dots 
not even after 1695 May, in spite of nearly daily examination of the Sun's disc, 
until June last year, when I observed some faint [spots] from 10th to 13th of the month.
\dots
\end{quote}

Wurzelbaur's generic statement that he did not notice any spots from May 1695
until 1700 June 9, {\it in spite of nearly daily examination} should
again be seen with caution, as he most certainly could not observe each and every day --
in the next section, he gives an example of bad weather.
However, the active day fraction was probably very low from 1695 May until 1700 May.

A comparison with HS98, RNR93, and the Sp\"orer catalogue shows
no entry in any of them for 1700 June.
However, HS98 do list a few reports about spotlessness: La Hire for June 10, 12, 13,
Eimmart for June 11 \& 12, Stancarius for June 8-15, and Agerholm for 1695 May 31 to 1700 Oct 31
(HS98: spotless each and every day, but probably a mis-interpreted generic statement,
see Vaquero et al. 2016);
for Stancarius, the observations are found in Manfredi (1736), but these are
not meant to be observations of sunspots ({\it observationes meridianae}).
A spotless Sun as reported by La Hire, Eimmart, and Stancarius would be inconsistent
with the positive detection by Wurzelbaur -- but Wurzelbaur mentioned that the
spots were {\it faint}, i.e. weak, so that others may not have been able to detect them.
Wurzelbaur's observations for June 1700 are not listed in HS98.

\subsection{1700 Nov 6-13}
\label{1700-nov}

Wurzelbaur then continued in letter no. 750, the previous citation, as follows (Herbst 2006):
\begin{quote}
\dots letztverschienen 6 Novemb aber, da nachmittag die Winde das Gew\"olcke in etwas zertheilet, 
fande ich 2 ansehnliche flecken in der Sonne, welche folgenden tages 
gegen die Mitte der $\odot$scheiben fortger\"ucket waren. Den 9:$^{ten}$ 
aber da Vormittags etwas hell Wetter gewesen habe ich durch meinen 18sch\"uhigen tubum 
und vorgesezte gr\"une Gl\"aser in die Sonne geschauet, und \"uber 2 grose noch einen 
kleinen Flecken angetroffen, die 2 gr\"osere waren von ganz differenter Art, 
die untere war sehr schwarz mit einem etwas hellen schein umgeben und vergleichete 
sich einer Grube oder See umb welche ein damm aufgeworfen, die Obere aber so zwar 
mercklich gr\"osser, war, wie auch die kleineste wie ein d\"unner und in etliche 
\"Aste ausgebreiteter Rauch anzusehen: in solcher gestalt praesentirten sie sich 
auch folgenden tages, den 11:$^{ten}$ aber war die kleinere verschwunden, 
den 12:$^{ten}$ war tr\"ub Wetter, den 13:$^{ten}$ umb halb 9 Vormittag mehr 
nicht als ein ganz schmahle macula jedoch annoch mit dem hellen schein umbgeben 
nechst am Rande, nachmittag aber das geringste vestigium davon nicht mehr anzutreffen.
\end{quote}

We translate as follows:
\begin{quote}
\dots they were lastly seen on [1700] Nov 6 in the afternoon, 
as the clouds dispersed. There I saw two considerable spots in the Sun,
which the next day moved to the centre of the Sun's disk. On [Nov] 9, 
when the weather was bright before noon, I looked through my 18-foot telescope 
with prefixed green glasses and discovered above the two larger spots another small spot. 
The two larger one [spots] had a different appearance, the lower one was very black, 
and surrounded by a bright shine, like a pit or lake, surrounded by a dam, 
the upper one was notable larger and, like the smallest one, fainter and like ramified smoke. 
In that way they were observed the next day. On [Nov] 11 the smallest one 
was vanished, on [Nov] 12 the weather was dull, on [Nov] 13 around 9h before noon 
only a slender spot, however surrounded by a bright shine, was seen near the limb. 
In the afternoon not the smallest sign of it could be seen.
\end{quote}

\begin{figure}[h]
\begin{center}
\includegraphics[width=70mm,height=70mm]{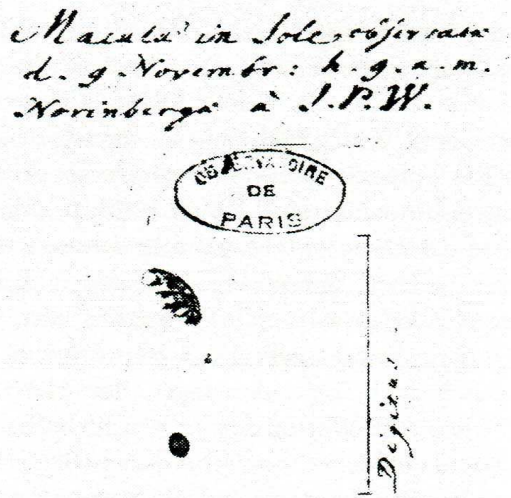}
\end{center}
\caption{Attached sketch by J.P. Wurzelbaur from 1700 Nov 9, 
original caption in Latin: \textit{Maculae in Sole observatae d. 9 Novembr: 
h. 9. ante meridiem Norinbergae a J.P.W.}; translation following Herbst (2006): 
``Sunspots in the Sun, observed on [1700] Nov 9 [at] 9h a.m. in Nuremberg by J.P. W[urzelbaur]''.
To the right of the drawing, the caption says {\it digitus} indicating the scale,
i.e. 1/12 of the solar diameter being 2.7 arc min in the sky or, when placed in the disk centre,
a heliocentric distance of 9.6 degrees.
}
\label{1700_11_07_wurzelbaur}
\end{figure}

Wurzelbaur described the spots quite well with a dark inner area (umbra),
a {\it dam} around them (penumbra), being {\it surrounded by a bright shine} (faculae).
By using the wording {\it smoke}, he may point to the seeing effect.
He also attached a drawing
showing the spots without any sort of placement on the Sun's 
disc (Fig. \ref{1700_11_07_wurzelbaur}). 

Wurzelbaur's statement that the spot had {\it moved to the centre of the Sun's disk}
should probably be interpreted as a progression in longitude towards the central meridian, 
rather than as a location {\it in\/} the centre.
We can obtain a heliographic latitude of $\sim 0^\circ$ for 1700 Nov 7.

We may adopt the following positions:
\begin{itemize}
\item 1700 Nov 7 (assuming noon): $\sim 0$ inch
\item 1700 Nov 13, 8:30 am: $\sim 5.9$ inch
\end{itemize}
Both separations are very rough estimates. Still they deliver a
reasonable solution for the latitude (mostly fixed to low
latitude because of the 0 inch on the first day) which is
$3.2\pm7.2^{\circ}$. The error is now based on a measurement
uncertainty of $\pm 0.8$ inch, which is twice as much as for the other
observations with apparently more precise measurements.

\begin{figure}
\begin{center}
\includegraphics[width=0.485\textwidth]{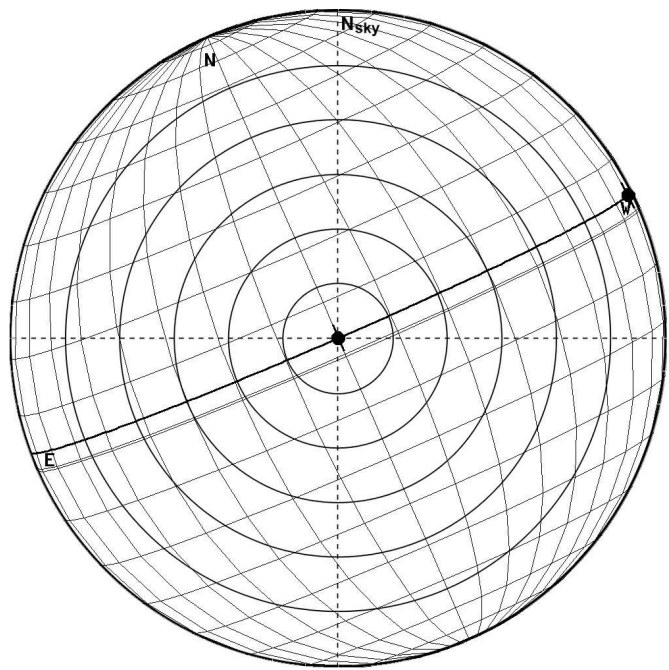}
\end{center}
\caption{Reconstructed positions for the sunspot seen on 1700 Nov~7 and~13
with $B_0=3.25^{\circ}$ and $2.58^{\circ}$, respectively.}
\label{1700_11_06_1500_tip}
\end{figure}

In Sp\"orer's catalogue one can find an observation from Nov 7-13 with 
a heliographic latitude of -9.5$^\circ$, namely by La Hire, Cassini and Wurzelbaur, whereas in RNR93, there is no data point. 
If the spot listed by Sp\"orer at -9.5$^\circ$ is the same as observed by Wurzelbaur,
then his statement {\it in the centre} has to be understood as rough (or meaning only the longitude).

In HS98 one can find three entries: 
La Hire, who saw two sunspot groups on 1700 Nov 9 and one group on Nov 10-12,
Cassini and Wurzelbaur himself are each listed from Nov 7-13 
with the observation of one sunspot group, i.e. 6 Nov is missing.
Stancarius would have reported a spotless Sun for Nov 8 and 13 (HS98),
Eimmart for Nov 11 \& 13 (the spots became very faint at the end according
to Wurzelbaur, so that some observers may not have detected them).

We summarize that Wurzelbaur saw two large spots Nov 6-11 (see figure for Nov 9)
plus one smaller spot nearby on Nov 10 \& 11,
and one of the larger spots were still visible on Nov 13 
(but disappeared during that day) --
all three spots probably forming one group.
The dates are consistent with Sp\"orer, La Hire, and Cassini
and, hence, they were probably indeed Gregorian as given by Wurzelbaur.

\subsection{1700 Dec 31}
\label{1700-dec}

Another observation in 1700 Dec can be found in letter 751 (Herbst 2006) 
from G. Kirch to Samuel Reyher, dated to 1701 Jan 17. 
G. Kirch mentions briefly:
\begin{quote}
1700. den 31 Dec. 2 Maculen in der $\odot$.
\end{quote}
We translate this to English:
\begin{quote}
1700 Dec 31, 2 spots in the Sun.
\end{quote}

G. Sp\"orer's catalogue gives a heliographic latitude of $-3^\circ$ for an observation on Dec 30
from La Hire. 
In RNR93 we found no spots around the turn of the year 1700/1701 (first in 1701 March).

The date given by Kirch (Dec 31) compares well with dated observations by La Hire and Cassini,
who always used the Gregorian calender, so that the date by Kirch is also Gregorian here:
HS98 lists three observations: 
La Hire (1700 Dec 30) and G. Kirch (1700 Dec 31) are each listed 
with the observation of two sunspot groups, 
as well as their observation of one group on 1701 Jan 2
(we do not know the source of HS98 for one spot group on Jan 2 by G. Kirch). 
Cassini is listed with the observation of one group on 1701 Jan 1 and 2.
Stancarius is given by HS98 to have seen a spotless Sun on 1701 Jan 2.
It is not clear from the letter by Kirch, whether he saw two groups
or two spots in one group.

\subsection{1701 Oct and Nov 3-10}
\label{03.11.1701}

The letter from Johann Abraham Ihle (no. 761, Herbst 2006) to G. Kirch 
contains another generic statement about low sunspot numbers.
Writing about how he could barely see any spots over the last 20 years, 
he then describes his new sightings:
\begin{quote}
Nach dem ich nun \"uber 20 Jahr sehr wenig maculas solares vermercket, 
auch von andren Orthen, da\ss{} dergleichen gesehen worden, nicht viel vernommen, 
habe ich mich offtmahls sehr verwundert, 
woher doch solche grosse infreqventia entstehen m\"oge [...] 
an j\"ungst verwichenem letzten Octobris war die Sonne noch gantz rein. 
Am 3 Novemb: recht im mittage erblickte ich endlich eine zwischen 
dem zergehenden Gew\"olcke, und zwar, wie ichs aestimirte, 
in initio 4 digiti, sie war fein rund und bunt, bedeckte etwan 1/5 unius digiti, 
das w\"are 1/60 totius diametri, und zwar unter einem diametro Solis, Horizonti parallelo. 
folgende tage habe ich nichts thun k\"onnen. Am 7 Nov: judicirte ich sie im 10 digito, 
schon abnehmend. Am 9 Nov: Mittags im 11 digito, etwas \"uber dem diametro Solis, 
Horizonti parallelo, sehr dilut und geringe, folgenden tag konte ich davon 
gar nichts mehr sp\"uhren, da doch die Sonne klar genug war, doch eine kurtze zeit.
\end{quote}

\begin{quote}
Now, after I noticed only very few sunspots for 20 years,
I also did not hear anything different from other places, 
I often wondered much, how such large irregularities arise [...] 
in the recent October, the Sun was completely clear. 
On Nov 3: right at noon, I finally saw one [spot] between the dispersing clouds,
namely, as I estimated, at first in the fourth digitus [inch],
it was nicely round and colourful and covered about 1/5 of one inch, 
that would be 1/60 of the whole [solar] diameter, 
namely beneath the Sun's diameter, parallel to the horizon. 
The following day, I could not do anything. 
On Nov 7: I estimated it to the 10th inch, already diminishing. 
On Nov 9: at noon in the 11th inch, slightly above the Sun's diameter, parallel to the horizon, 
very faint and small. The next day, I could not see it, 
even though the Sun was clear at least for a short while.
\end{quote}
Hence, Ihle saw one spot on 1701 Nov 3, 7, and 9.
He also specified that the Sun was spotless in October and on Nov 10.

\begin{figure}
\begin{center}
\includegraphics[width=0.485\textwidth]{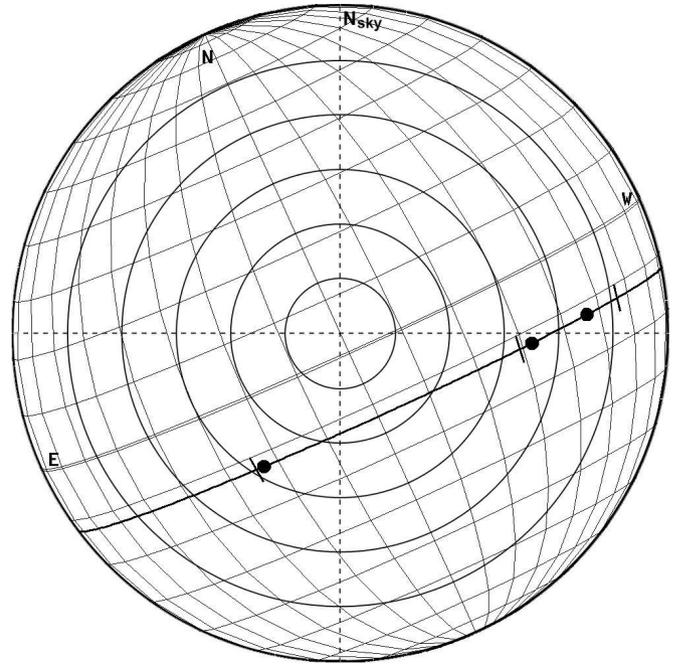}
\end{center}
\caption{Reconstructed positions for the sunspot seen on 1701 Nov~3, 7, and 9. The
corresponding $B_0$ are $3.73^{\circ}$, $3.28^{\circ}$, and $3.05^{\circ}$.}
\label{1701_11_03_1200_tip}
\end{figure}

The assumed separations from the centre are
\begin{itemize}
\item 1701 Nov 3, 12h noon: 2.8 inch [in initio 4 digiti]
\item 1701 Nov 7 (assuming noon): 3.5 inch
\item 1701 Nov 9, 12h noon: 4.5 inch
\end{itemize}
We assumed that ``in the 10th inch'' means that the spot
has not reached the 4-inch line counted from the center,
but was in the band between the 3-inch line and the 4-inch
line. The resulting solutions are $-12.4\pm8^{\circ}$ and
$+22.0\pm7^{\circ}$. Since the ecliptic is quite inclined against the
celestial equator in early November, a position in the
northern hemisphere of the Sun can actually lie 
{\it beneath the Sun's diameter}. However, this is not possible
for the derived northern latitude of $22^\circ$, and we
conclude that the southern latitude of $-12.4^{\circ}$ is the correct one.
The corresponding tracks across the solar
disk are given in Fig.~\ref{1701_11_03_1200_tip}.

According to Sp\"orer's catalogue, spots observed Oct 31-Nov 10 by Cassini
had a heliographic latitude of $-12 ^\circ$, consistent with our estimate 
and the date range from Ihle.
The RNR93 diagram does not contain any data for Oct or Nov,
but $-11.8^\circ$ for Dec.

HS98 list La Hire, who observed one sunspot group Nov 1 to 6
as well as 9 and 10, as well as Cassini and Jartoux both for Oct 31-Nov 11 and Nov 1-13.
Those dates include Nov 10, for which Ihle reported a spotless Sun,
but maybe these observers could still detect the spot which was already lost by
or too small for Ihle.
Eimmart is given to have seen a spotless Sun on Nov 2, 3, and 12 (HS98),
also, Stancarius would also have reported a spotless Sun for Nov 1-4, 9 and 12 (HS98) --
both partly inconsistent with the detections by Ihle, La Hire, and Cassini.
Ihle himself is not listed by HS98 for 1701.

The detection of a spot by Cassini on Oct 31 appears inconsistent with Ihle saying that
the Sun was spotless all October, but in fact Ihle may have had some days with overcast,
so that we just says that he never detected a spot (whenever he observed);
La Hire, Eimmart, and Stancarius also did not detect a spot in October, but had several bad weather days (HS98).
Ihle is also not listed for Oct 1701 in HS98.

Ihle also mentioned that he and others {\it noticed only very few sunspots for 20 years},
i.e. from about 1681 to 1701 Nov.

\subsection{1702 Dec 17 to 1703 Jan 1}
\label{24.12.1702}

The descriptions of the sunspot from the end of the year 1702  
also give an answer to the question whether the telescope's image was rotated.
Letters 781 by Ihle and 786 by Wurzelbaur (Herbst 2006) to G. Kirch give the hint
that the astronomers very well knew how to compensate the rotation, at least in their descriptions.
Ihle writes on 1702 Dec 24:
\begin{quote}
Am 17 Decembr: Mittags befunde ich die Sonne gantz rein, folgende tage waren tr\"ube, 
am 22 Decemb: nach Mittage 3 uhr war eine feine de\"utliche macula, schon im XI digito, 
zu sehen, gestern war es wieder tr\"ube, he\"ute, den 24 Dec: scheinets noch ungewi\ss{}, da\ss{} 
die Sonne recht klar werden m\"ogte. \dots\\
Gleich ietzo XI uhr k\"ommt die Sonne noch herfur; aber ich bin gantz confus: 
Am 22 Dec: meinte ich, sie st\"unde im XI digito; he\"ute aber mu\ss{} 
ich dencken da\ss{} es nicht der XI. sondern der 2. m\"usse gewesen seyn, 
die zeit leidet ietzo ein mehrers nicht zu gedencken.
\end{quote}

\begin{quote}
On Dec 17: around noon I found the Sun fully clear, 
the following days were dull. 
On Dec 22: 3 o'clock in the afternoon, there was a distinct spot visible, already in the 11th inch.
Yesterday it was dull again. Today, on Dec 24: seems to be uncertain whether the weather [Sun]
will be good [clear] enough. \dots\\
Right now, at 11 o'clock, the Sun comes out; but I am utterly confused: 
on 22 Dec: I thought it stood in the 11th inch; 
but today, I must think that it was not the 11th [inch], 
but rather the second. But I cannot observe any longer now due to a lack of time.
\end{quote}

\begin{figure}[h]
\begin{center}
\includegraphics[width=80mm,height=40mm]{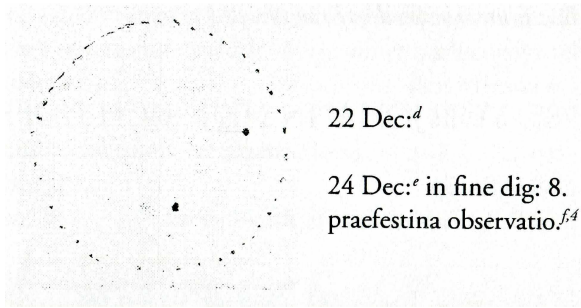}
\end{center}
\caption{Ihle's drawing attached to the letter and Herbst's adopted caption, 
translated to English by us as follows: 
{\it Dec 22 (upper), Dec 24: at the end of the  8th inch, very hurried observation (lower);
the footnotes in the drawing as given by Herbst (2006) are as follows (translated to English by us):
(d) Kirch remarks {\it after midday} at the margin, (e) Kirch remarks {\it Hor. XI.}, i.e. 11th hour,
above the drawing, (f) Kirch remarks to have received the letter ({\it Praes. 29 Dec. 1702}) on AD 1702 Dec 29,
(4) original text in Latin.}
Please note that the upper spot (Dec 22) drawn to the upper right quadrant should be mirrored and
was located in the upper left quadrant, namely in the 2nd inch from the left, 
instead of in the 11th inch (which is the 2nd inch from the right), Ihle: {\it I was utterly confused}.
The spot drawn in the bottom for Dec 24 is placed correctly to the end of the 8th inch.
}
\label{1702_12_24_ihle_own}
\end{figure}

Attached to that letter is the small drawing (Fig. \ref{1702_12_24_ihle_own}) 
by Ihle and the caption: one can see two spots in one drawing, which are meant
as observations of one spot on two different dates -- with corresponding dates in the caption.

\begin{figure}
\begin{center}
\includegraphics[width=0.485\textwidth]{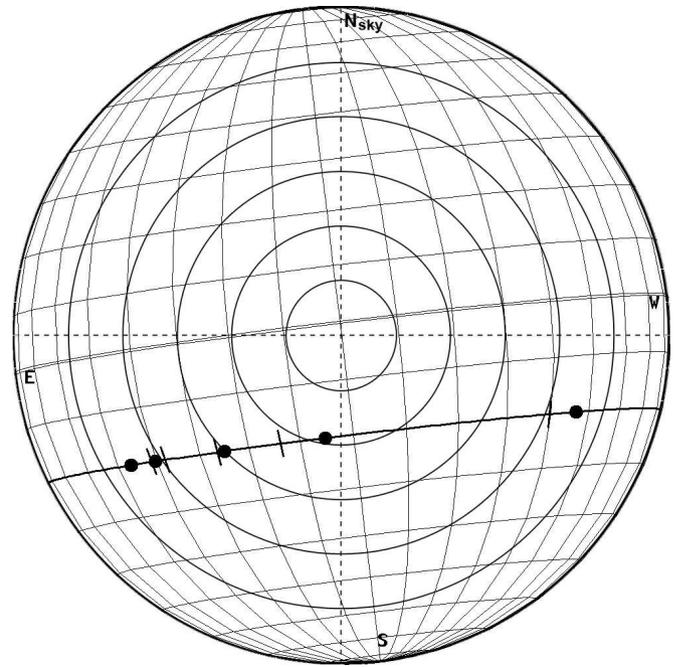}
\end{center}
\caption{Reconstructed positions for the sunspot seen on 1702 Dec~22, 23, and 28
by Wurzelbaur. The corresponding $B_0$ are $-2.27^{\circ}$, $-2.40^{\circ}$, and $-3.00^{\circ}$.}
\label{1702_12_22_0903_tip}
\end{figure}

The caption to the figure drawn by Ihle includes the Latin text 
{\it 24 Dec: in fine dig: 8.}
This was translated by Herbst (2006) to German as follows: {\it 24. Dez.: am Ende 8 Fingerbreiten},
i.e. in English {\it Dec 24, at the end (in the) 8(th) inch}. While this is a possible translation of the
abbreviated Latin words, it would imply that the spot was observed last on Dec 24 (within the 8th digitus) and then
dissolved. However, a spot was seen again on Dec 28 by Wurzelbaur. Therefore, the alternative
(and intrinsically correct) translation of the abbreviated Latin words is {\it at the end of the 8th inch}.
The meaning of the colon (:) after the Latin {\it dig} ({\it dig:}) is not similar to today's normal
meaning of a colon (that something important and relevant is to follow now), 
but the colon was used by Ihle to signal that the word was abbreviated,
namely {\it dig:} instead of {\it digiti}. He also used the colon to signal abbreviations
in the rest of his letter, e.g. {\it 22 Dec:} for what we would abbreviate as {\it 22 Dec.},
i.e. he used a colon instead of a full stop (.) after the abbreviated word, see Fig.~\ref{1702_12_22_0903_tip}.

In comparison to Ihle's text, Wurzelbaur writes on 1703 Feb 16 the following
(letter no. 786):
\begin{quote}
Am 22:$^{ten}$ Decembr: fande ich nach etlichen tr\"uben t\"agen umb 9 Uhr Vormittag 
eine ziemlich grosse und schwarze maculam in der Sonne, 
welche albereit auf 1,5 dig: in den discum hineinger\"ucket war, 
die habe ich auch den 23:$^{ten}$ auf 3 digg: dort-, 
nach etlichen tr\"uben tagen aber, nemlich den 28:$^{ten}$ 
nur noch 1,5dig. vom westlichen Rande, den 31:$^{ten}$ Decembr: 
aber und primo Januarii 1703 nichtes mehr davon angetroffen.
\end{quote}

\begin{quote}
On Dec 22: after many dull days, around 9 a.m., I found a rather large and black spot in the Sun, 
which was already 1.5 inches, moved into the Sun's disc, 
on Dec 23, on the third inch. 
There, after many dull days, on Dec 28, only 1.5 inches from the western edge. 
But on Dec 31: and on 1703 Jan 1, nothing of it was seen.
\end{quote}

While Ihle saw the spot on Dec 22 and 24, but Dec 17 was spotless;
Wurzelbauer detected it on Dec 22, 23, and 28, but gives a spotless Sun for Dec 31 and Jan 1.

By first assuming that Ihle was in a hurry and therefore did not follow his arising confusion, 
his first description and drawing match each other. 
Even the lower point on the sketch of his observation on Dec 24 
shows an appropriate continuity and velocity of the spot.

Reading by contrast Wurzelbaur's text, 
written nearly two months after the actual incident and well revised, 
we identify this description as a correct one. 
Apart from this, Ihle made us understand that the astronomers (otherwise)
corrected the image, by re-mirroring it in their descriptions.
By mirroring Ihle's drawing and data to the left (eastern) hemisphere of the Sun, 
both descriptions match perfectly.

Though we do not know, if Ihle's sketch is horizontally mirrored as well, 
we can assume Wurzelbaur's text to be correct.

So we obtain the following positions:
\begin{itemize}
\item 1702 Dec 22, 9 am: 4.5 inch (in the 2nd inch)
\item 1702 Dec 23 (assuming noon): 3.0 inch
\item 1702 Dec 28 (assuming noon): 4.5 inch
\item 1702 Dec 22, 3 pm: 4.1 inch (as measured from sketch)
\item 1702 Dec 24, 11 am: 1.9 inch (interpreting `end of 8th inch')
\end{itemize}
There are two solutions possible: one at $-20.0 \pm 6.8^{\circ}$ and one at $13.9 \pm 6.4^{\circ}$.
The corresponding reconstruction is given in Fig.~\ref{1702_12_22_0903_tip}.
The discrepancy to Ihle's sketch can be explained by (i)
the fact that Ihle corrected his Dec 22 position towards
the eastern limb, and (ii) the fact that it was seen in the
afternoon and the Sun was rotated clock-wise in the sky. We can
also see that the best location for the Dec 24 spot was
slightly left of the vertical diameter instead of slightly to
the right as indicated by Ihle. Perhaps Ihle missed to note
that also this observation needed to be mirrored.

Sp\"orer gives a heliographic latitude of -11$^\circ$ for an observation on Dec 22-31
by Cassini and La Hire, 
which can well be consistent with the spot seen by Ihle and Wurzelbaur
(but the latter did not notice a spot on Dec 31).
In RNR93, there is no matching data point. 

HS98 list La Hire (1702 Dec 22, 24, 26, 27, 29, \& 30), 
Cassini (Dec 22, 24, \& 25), and Manfredi (Dec 29),
all for one group each, also consistent with Ihle and Wurzelbaur,
who are not listed in HS98 for these dates.

HS98 list Eimmart for spotlessness for 1702 Dec 16, 22, 23, and 26
and Stancarius for Dec 18, 19, 25-29 and 31,
and also La Hire, Cassini, and Eimmart for spotlessness for 1703 Jan 1.
Eimmart and Stancarius often report spotlessness, when other can still detect a spot
(for Stancarius, the observations are found in Manfredi (1736), but these are
not meant to be observations of sunspots ({\it observationes meridianae}).

\subsection{1703 May 24-June 2}
\label{25.05.1703}

\begin{figure}[h]
\begin{center}
\includegraphics[width=60mm,height=60mm]{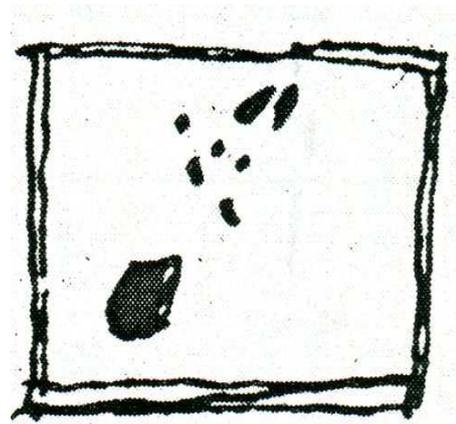}
\end{center}
\caption{Sketch from Johann Heinrich Hoffmann to G. Kirch in letter 791 (Herbst 2006), 
showing the observed sunspot group on 1703 May 27.}
\label{1703_05_27_hoffmann}
\end{figure}

The spots of 1703 were first mentioned in letter no. 791, 1703 May 28, 
by Johann Heinrich Hoffmann (Herbst 2006). Next to a detailed description, 
Hoffmann gives a little sketch as well (Fig. \ref{1703_05_27_hoffmann}).
\begin{quote}
die gro\ss{}e Maculam habe heit wieder geme\ss{}en, 
scheint als ob sie schon \"uber das centrum $\odot$ weg, 
oder zum wenigsten in dem Medio $\odot$ sey, 
die zwey kleine aber zertheilen sich und werden viel andere kleinere daraus 
welche man aber kaum erkenne kan. 
gestern Mittag fand ich sie in solchem situ gegen einander; da sich noch 5 kleinere zeigeten.
\end{quote}

This translates to:
\begin{quote}
Today [1703 May 28] I measured the large spot again. 
It seems like it has already passed the Sun's centre 
or it might at least be in [near] the middle of the Sun. 
The two smaller spots divide into many other, smaller spots, 
which are almost invisible. Yesterday [May 27] at noon 
I found them in that position to each other, when 5 smaller [spots] showed up.
\end{quote}

In letter no. 794, J.A. Ihle writes to G. Kirch on 1703 June 12 (Herbst 2006) 
that the group was visible probably already one day earlier (May 26):
\begin{quote}
Die erste erscheinung j\"ungst gedachter maculae $\odot$ am 26 Maji 
in digito primo war sehr geringe, ja gantz zweifelhafftig, 
folgende tage aber gar fein de\"utlich und rund; 
ihre letzte Erscheinung aber im 12 digito sehr nahe am rande den 4 Jun: si recte memini.
\end{quote} 

\begin{quote}
The first appearance of the solar spots on [1703] May 26 in the first inch 
was a very faint, in fact even dubious one, but the following days it was very distinct and round; 
its last appearance was on June 4 in the 12th inch very close to the edge: 
if I remember correctly.
\end{quote}

G. Kirch himself describes the spots in a letter to Gottfried Wilhelm Leibniz 
from 1703 August 13 (no. 797, Herbst 2006).
After talking about the printing of his newest calendar, G. Kirch writes the following:
\begin{quote}
Unterde\ss{}en berichte, da\ss{} ich dem 25 Maj. eine gro\ss{} Macul in der Sonnen gefunden. 
Sie war kurtz vor dem Untergange der Sonnen 32 partes micrometri vom Eintrits-Rande: 
thut 6'. 47'' oder 2,5 zoll. 
Der Diameter Solis war 152 partes, thut 32'. 14''. 
Diese Macul hatte noch eine schwache neben sich, 
welche sich aber nach etlichen Tagen verlohr. 
Die gro\ss{}e schwartze hingegen, r\"uckete nach und nach durch die Sonnenscheibe, 
wie gew\"onlich, und ward am 2 Junii zu letzt, gantz nahe am Austrits Rande gesehen. 
\end{quote}

\begin{quote}
Meanwhile I report that I found a large spot in the Sun on May 25. 
Shortly before sunset it was 32 partes micrometri (p.m.) from the ingress limb: 
equal to 6' [arc min] 47'' [arc sec] or 2.5 inches. 
The Sun's diameter was 152 p.m., means 32' 14''. 
Next to this spot was another, weak spot, 
which vanished after a few days. 
The large, black one moved on through the Sun's disc as usual 
and was seen for the last time on June 2, very close to the egress limb.
\end{quote}

Note that Kirch now uses {\it partes micrometri} as unit for angular separations
(or sizes) on the solar disk (it means something like micrometer parts); 
this unit is often abbreviated with {\it p.m.} One p.m. corresponds to 8.6 arc sec.

In summary, Hoffmann saw the spots on May 27 \& 28,
Ihle detected it first on May 26 and last on Jun 4,
G. Kirch observed it from May 25 until Jun 2.
Hoffmann's observation of the group in the centre of the solar disc (1703 May 28) 
results in a heliographic latitude of $\sim 0^\circ$.

\begin{figure}
\begin{center}
\includegraphics[width=0.485\textwidth]{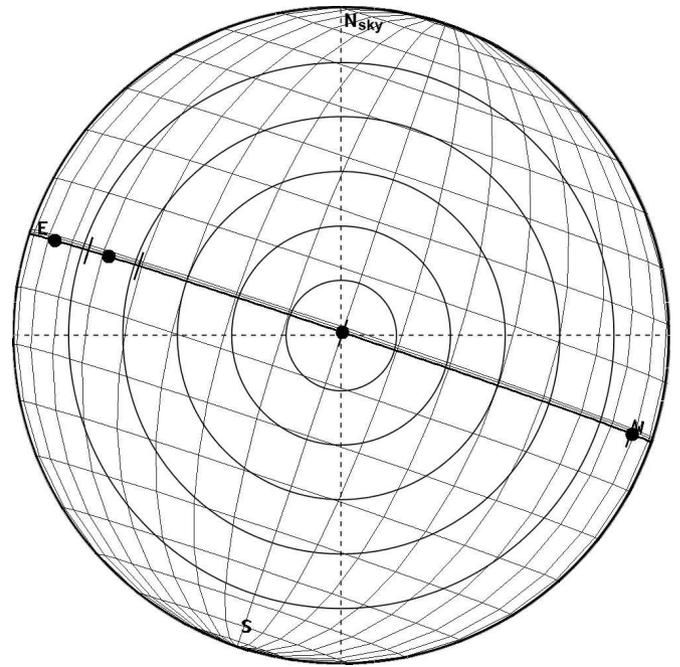}
\end{center}
\caption{Reconstructed positions for the sunspot seen on 1703 May~24--Jun~2 by
Kirch, Ihle, and Hoffmann. The tilt angles of the Sun were $B_0=-1.19^{\circ}$ (May~24),
$-1.04^{\circ}$ (May~25), $-0.71^{\circ}$ (May~28), and $-0.10^{\circ}$ (Jun~2).}
\label{1703_05_24_1200_tip}
\end{figure}

Preliminarily, we may just compile what has been said
(always assuming noon time for the observation):
\begin{itemize}
\item 1703 May 25, 7 pm: 3.47 inch (Kirch)
\item 1703 May 26: 5.5 inch  (Ihle)
\item 1703 May 28: 0 inch    (Hoffmann)
\item 1703 Jun 2: 5.6 inch  (Kirch)
\item 1703 Jun 4: 5.6 inch  (Ihle)
\end{itemize}
However, we see immediately, that the positions by Ihle do not fit into a reasonable passage 
of a spot through the solar disk. Firstly, while on May~25, Kirch saw the spot at considerable 
separation from the limb, Ihle reported it to be close to the limb  a day later.  
Secondly, while Kirch saw the spot near the western limb on Jun 2, Ihle saw it there, two days later. 
If we assume that Ihle's report is wrong by 2~days, we obtain:
\begin{itemize}
\item 1703 May 24: 5.5 inch  (Ihle)
\item 1703 May 25, 7 pm: 3.47 inch (Kirch)
\item 1703 May 28: 0 inch    (Hoffmann)
\item 1703 Jun 2: 5.6 inch  (Kirch)
\item 1703 Jun 2: 5.6 inch  (Ihle)
\end{itemize}
These locations result in a sharp probability density distribution with an average latitude of $-0.7\pm3.5^{\circ}$. 
The corresponding reconstruction is shown in Fig.~\ref{1703_05_24_1200_tip}.
The sharp density distribution may point to just one sunspot group.
(Leaving out the observation by Hoffmann leads to a much broader distribution in latitude
with nearly the same average but an uncertainty of $\pm18^{\circ}$.)

Sp\"orer's catalogue lists an observation from May 24 to June 3 
with a heliographic latitude of $-2^\circ$ by Cassini and La Hire, 
well consistent with our solution (Fig.~\ref{1703_05_24_1200_tip});
this could also be the spot given by Hoffmann to lie close to the center of the Sun.
RNR93 does not show a data point for this time 
(but $-1.7^{\circ}$ for Aug).

HS98 list the observation by La Hire May 25-Jun 1 \& 3, 
who observed 2 sunspot groups on May 27 and the other days just one group.
It is well possible that what HS98 list as 2 groups for La Hire for May 27
are just the two spots seen by Hoffmann and G. Kirch at the end of May,
who both mention that the second spot weakened and disappeared after a few days.
According to HS98, Manfredi (May 26-June 2), Eimmart (May 22 to June 2),
Cassini (May 24 to June 3), Hoffmann (May 26-June 3), and Stannyan (May 15-23 and June 3) 
all would have observed one group --
noone observed until June 4, so that Ihle is probably wrong here by two days.
The date range for Stannyan from England (as given in HS98) may well be Julian,
not corrected to Gregorian by HS98 (as in the next subsection, a difference of 11 days by this time),
because La Hire, Manfredi, and Eimmart all consistently report spotless
days from May 15-21, and a spot afterwards.
The date range for Hoffmann given in HS98 is not supported by the letter
from Hoffmann to G. Kirch,
Kirch himself is given with one group from 1703 May 25 to June 2 in HS98,
i.e. as in his letter cited above.
Ihle is not listed in HS98 for this year.

\subsection{1703 June 18-29}
\label{1703-june}

After G. Kirch's description of the spots 1703 May 25 to June 2
in his letter to Leibniz,
1703 August 13 (no. 797, Herbst 2006),
G. Kirch speculates that the very same spot appeared
one solar rotation period later:
\begin{quote}
\dots am 2 Junii zu letzt, gantz nahe am Austrits Rande gesehen.
Diese Macul kam zum andern mal wieder in die Sonne am 18 Jun. 
Den 29 Jun. war sie wieder nahe am Austrits-Rande.
\end{quote}

\begin{quote}
\dots on June 2, very close to the egress limb.
This spot came again into the Sun on Jun 18. 
On Jun 29 it was again near the egress limb.
\end{quote}

At a (low-latitude) synodic rotation period of
27~d, the spot moves at 13.3 degrees/d. If the separations
from the ingress and egress limbs were equal, the
central-meridian separations should have been $\pm73.2^{\circ}$.
Near the equator, that corresponds to a normalized
separation from the solar centre of 0.957 R$_{\odot}$ (solar radii) or, in the
language of the historic observations, 0.26 inches from the limb.
There is a very wide range of latitudes possible for
this situation; we do not derive a certain latitude
at this point, but note that the above mentioned longitudes are fully compatible
with representing the same spot as the one of June~2, within less than $4^{\circ}$.

Sp\"orer's catalogue lists a spot with heliographic latitude of $-2^\circ$ 
for June 18-30 for the observer Cassini and La Hire;
this is the same latitude as one month earlier,
which may have motivated G. Kirch to conclude that it was
the same spot that was seen one solar rotation earlier. 
RNR93 contained no data for June (but $-1.7^{\circ}$ for Aug).

HS98 list spot groups as follows, always one group:
La Hire June 20, 21, 23-27,
Manfredi June 15, 22, 23, 26-28,
Eimmart June 21-23, 27, 
Cassini June 18-30,
Blanchini June 20-29,
Hoffman June 18-30,
Gray June 15, 16, 18, 27, 28 (and spotless June 26),
R{\o}mer June 21,
Stannyan June 3, 7-12,
and G. Kirch June 18, 20, 21, 25, 27 (one group each),
as well as Derham for one group June 13, three groups June 18 \& 19, two groups June 28, and three groups June 29 to July 1.
The date range for G. Kirch given by HS98 is slightly different from Kirch's letter cited above,
where he says June 18 to 29.
The date range given by HS98 for Derham is not consistent with the other reports (starting too early), 
neither if Gregorian nor if uncorrected Julian;
the date range given for Stannyan from England may well be Julian, 
not corrected to Gregorian by HS98 (as in the previous subsection, 
there was an 11 day shift between Julian and Gregorian since March 1700).

\subsection{1703 July 8-15 (and early Aug)}
\label{08.07.1703}

G. Kirch describes another observation in letter no. 797 to G.W. Leibniz dated 1703 Aug 13:
\begin{quote}
Den 8 Julii fanden wir eine neue Macul in der Sonnen, 
am 10 und 11 Jul. waren es ihrer drey, 
und am 15 Jul. konte man nur noch eine, nahe am Austrits-Rande sehen.
\end{quote}

\begin{quote}
On July 8 we found a new spot in the Sun, 
on July 10 and 11 there were three [spots], 
and on July 15 one could only see one [spot], near the egress limb.
\end{quote}

In letter no. 800 from G. Kirch to Olaus R{\o}mer (dated 1703 Oct 25, Herbst 2006), 
G. Kirch uses a similar wording as in letter 797:
\begin{quote}
Den 8 Jul fand sich eine andere Macul, etwan im andern zoll vom Eintrits-Rande. 
Den 10 und 11 Jul. waren ihrer 3. 
Den 15 Jul. konten wir (zwar durch ziemlich dicke Lufft) 
nur noch eine Macul, am Austrits-Rande sehen. 
Am Eintrits-Rande aber war die erste Macul nicht wiederkommen.
\end{quote}

\begin{quote}
On [1703] July 8 another sunspot was found, in the other inch from the ingress limb. 
On July 10 and 11 there were three [spots]. 
On July 15 we only could see one spot (through very thick air) 
at the egress limb. The first spot did not come back on the ingress limb.
\end{quote}

The last sentence in letter 800 is additional information compared to letter 797,
namely that G. Kirch did not see this spot one rotation period later,
i.e. early August, and probably no spots at all.
HS98 list R{\o}mer to have seen one group on Aug 3.

The spot mentioned by G. Kirch for 1703 July 8-15 is also 
described by J.H. Hoffmann in letter 795 
dated 1703 Jul 8 (Herbst 2006) to G. Kirch:
\begin{quote}
Als ich diesen Morgen noch vor 6 Uhr die $\odot$ in meine kammer bekahm, 
und der Himmel noch rein war, betrachtete ich die $\odot$ mit dem gro\ss{}en Tubo, 
fand in derselben unweit von dem Eintrits rande wieder eine feine Maculam 
doch nicht so starck als die vorige, dem augen ma\ss{} nach war sie 
wenigstens 2 zoll \`{a} margine $\odot$ \dots Diese Macula hat wieder verschiedene Faculas um sich.
\end{quote}

\begin{quote}
As the Sun shone into my room before 6 o'clock this morning [July 8] 
and as the sky was still clear, 
I observed the Sun through the large telescope. 
I saw a spot close to the ingress limb again, 
but it was not as strong as the last [spot]. 
Judging by eye it was at least two inches from the limb 
of the Sun \dots This spot has different faculae around it.
\end{quote}

We obtain:
\begin{itemize}
\item Jul 8: 2 inch from E limb ({\it in the other inch from the ingress limb})
\item Jul 15: 1 inch from W limb ({\it at the egress limb})
\end{itemize}
These constraints yield a very broad solution of $7 \pm 21^\circ$.

Sp\"orer gives a heliographic latitude of $-19^\circ$ for July 7-16 
for La Hire, which can well be the spot mentioned by Hoffmann
(and also seen by G. Kirch).
The RNR93 diagram shows a spot one month later in 1703 Aug 
with a heliographic latitude of $-1.7^\circ$,
i.e. different than in Sp\"orer, but also consistent with Hoffmann. 

HS98 list La Hire's observation of one group on July 8, 2 groups July 9-13, then one group July 15 \& 16.
Other observers of 1 group in HS98 are 
Manfredi (July 10 and 15), 
Eimmart (July 9, 13, 15), 
Cassini (July 8-16, spotless on July 7 and 17), 
R{\o}mer (July 5), Stannyan (July 17, then spotless), and
Derham (June 28-July 1 2-3 spots, July 4-6 \& 9 one spot each),
the latter are again most certainly Julian dates (11 day offset since 1700).
The observer Sharp (from Horton, England) is listed in HS98 to have reported
a spotless Sun from July 8-24, which would be inconsistent with many
other reports -- either he could not detect the spot(s) or the dates are wrong;
if the dates by Sharp and Stannyan are Julian (still in use in England at the time),
the Sun would have been spotless since July 19 (Gregorian), 
and indeed La Hire, Manfredi, and Eimmart reported a
spotless Sun on some days from July 19 to Aug 10 -- an exception being one
group on July 25 by Eimmart, and R{\o}mer one group on Aug 3).
G. Kirch is listed in HS98 with an observation of 3 sunspot groups 
on July 9, but one group on July 8 and 15, while G. Kirch himself writes in the letter 
that he saw 1 spot on July 8, 3 on July 10 \& 11, and one on July 15,
probably all in one group.

\subsection{1704 Jan, Feb, Mar, Apr}
\label{1704}

Further spots, even though without positional information, 
are mentioned in letter no. 810 from 1704 May 17 (Herbst 2006), 
where J.P. Wurzelbaur lists a variety of observations:
\begin{quote}
Die maculae solares welche vor 20 Jahren wohl rare gewesen, 
stellen sich nun \"offters ein: massen 
derer seither anfang dieses Jahres zum vierdten mahl al\ss{} Januario, 
Februrario, vom 21 bi\ss{} 24 Martii und 23 und 24 Aprilis eingefunden,
welche beedesmahlige letztere von geringer wehrung gewesen und aus deren, 
wie auch bey etlichen Jahren her erschienener vieler andrer, ver\"anderungen 
und derselben genauer betrachtung die Natur des Sonne C\"orpers fast unschwehr 
zu errathen seyn will \dots
\end{quote}

\begin{quote}
Sunspots, which apparently were rare 20 years ago, are visible more often nowadays: 
Since the beginning of the year [1704] we measured them four times,
in January, February, March 21-24, and April 23 and 24. 
The last two were weaker and, like the many others years ago,
changes and detailed observations helped us almost without difficulty
to understand the nature of the solar body \dots
\end{quote}

The same months and dates are listed in G. Sp\"orer's catalogue: 
two observations for 1704 Jan 7 \& 8 with heliographic latitudes 
of $-7^\circ$ and $-8.2^\circ$, respectively, by Maraldi and La Hire; 
furthermore 1704 Jan 15-Feb 5 ($-9^\circ$), 
Feb 2-7 ($-8^\circ$), Feb 9 \& 10 ($-13^\circ$), 
and Mar 19-21 ($-10^\circ$). 
The diagram by RNR93 shows no points in 1704 Jan,
then three in March with $-15.6^\circ$ and $-10.2^\circ$. 

HS98 provides a large number of observers for this period of time
(the number in brackets is the number of observed sunspot groups according to HS98): \\
La Hire: Jan 7, 10, 11, 16, 17, 28, 29 (1), Feb 1-5 (1), Feb 10 (2), Mar 21 \& 24 (1), Apr 22 \& 23 (1) \\
Manfredi: Jan 12, 16, 18, 22, 25, 26, 30 (1), Feb 1, 2, 9, 12-15 (1), Mar 9, 10, 19, 21 (1), Apr 5, 11, 22, 23 (1) \\
Derham: Jan 16-19, 21-23, 30 (1), Feb 23 \& 25 (1), Mar 7-11, \& 13 (1), Apr 11-13 (1) \\
Maraldi: Jan 7 \& 8 (2), Jan 9-18 and 25-28 (1), Jan 29-Feb 5 (2), Feb 9 \& 10 (1), Mar 19-21, \& 24 (1) \\
Gray: Jan 17-19, 21-23, 25 (1), Mar 8-11 (1) \\
Plantade, De la Val, Salvago, De Clapier, Fulchiron, and Thyoli: each Feb 10 only (1) \\
R{\o}mer: Jan 3 (1), Feb 26 (1), Mar 21 (1) and Apr 5 (1), and finally \\
G. Kirch: Jan 9 \& 12 (2), Jan 13, 15, 16, and Jan 25-Feb 4 (1), Feb 11, 12, 15 (1), 
and Mar 18 \& 20 (1).
Again, one can conclude from the date ranges (March and April) given for the English observers,
Derham and Gray, that they are probably Julian dates not corrected to Gregorian by HS98.
Wurzelbaur (1 group each for Mar 21-24 and Apr 23 \& 24) is missing in HS98: 
while his detections on Mar 21, 23, \& 24 and Apr 23 are consistent
with other observers in HS98, his spots for Mar 22 and Apr 24 are completely new.
For the dates of Wurzelbaur's detections, HS98 list some observers who would have reported
a spotless Sun: Eimmart Mar 21-24, Manfredi Apr 24, La Hire: Apr 24;
for the last two dates, Apr 23 \& 24, Wurzelbaur mentioned that the spot was weak,
so that it may not have been detectable anymore for Manfredi and La Hire Apr 24.

\subsection{1705 Oct 16}
\label{16.10.1705}

In the postscript of letter 820 in Herbst's collection (2006), 
G. Kirch writes to Gottfried Wilhelm Leibniz on 1705 Oct 16:
\begin{quote}
P.S. Heute um 2,5 nach Mittage, hatten wir 2 feine Maculen in der Sonnen zu sehen. 
Etwan 3 Zoll vom Austrits-Rande.
\end{quote}

\begin{quote}
P.S. This afternoon [1705 Oct 16] around 2:30h, we 
[G. Kirch and his wife] 
got to see two fine spots in the Sun, about 3 inches from the egress edge.
\end{quote}

\begin{figure}
\begin{center}
\includegraphics[width=0.485\textwidth]{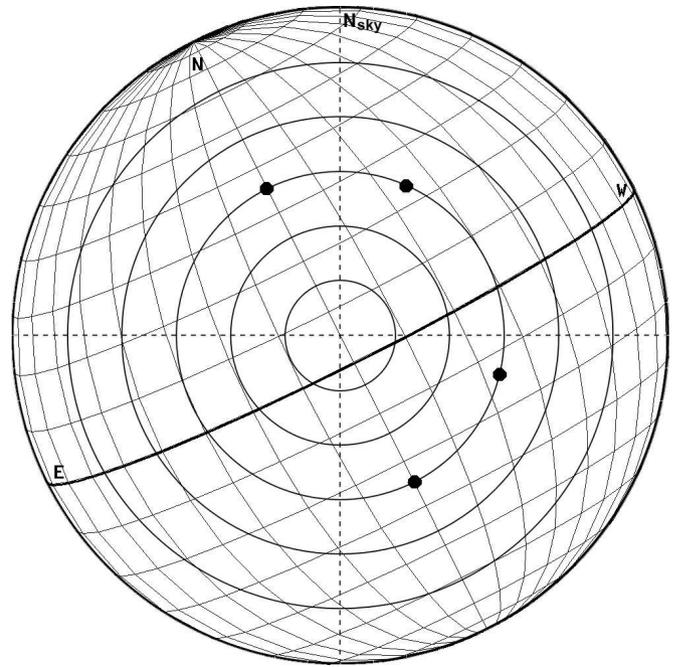}
\end{center}
\caption{Extreme locations for the sunspot seen on 1705 Oct~16 by
Kirch. The tilt of the Sun was $B_0=5.46^{\circ}$.}
\label{1705_10_16_1430_tip}
\end{figure}

The formal separation of 3~inches to the limb results in
a possible latitude range from $-24.4^{\circ}$ to $+35.2^{\circ}$ as shown
in Fig.~\ref{1705_10_16_1430_tip}. We may restrict the limb
to the range where typical latitudes between $-40^{\circ}$ and $+40^{\circ}$
exit the solar disk. We may assume that the observers of the
time called only this section of the limb the ``egress limb''.
Then we obtain a latitude range for the spots of $-13.9^{\circ}$ to
$+23.3^{\circ}$.

Given that our observers always saw spots on the southern hemisphere of the Sun,
they might have mentioned explicitly that the spot was on the northern hemisphere,
if this would have been the case -- as exception. In such a case, one could
limit the solutions to southern values. Alternatively, we could assume that the spot seen
here on 1705 Oct 16 was the same as observed on Nov 5-12, for which we deduce
the latitude in the next section to $4.2 \pm 17.1^{\circ}$.
However, we would like to refrain from both such assumptions.

To complete the description another mention by J.P. Wurzelbaur is added. 
In letter no. 829 he writes on 1706 Feb 2 (Herbst 2006):
\begin{quote}
Maculae Solares haben sich zwar offt sehen lassen; 
wie dann im Octobr. j\"ungsthin, da deren 2 noch nicht discum Solis quittiret hatten, 
albereit andere 2 mit einem sonderbahrem comitatu versehen hingegen einmachiret waren, 
und liesse sich von etlichen Jahren her ein ziemlicher Catalogus dergleichen observationum formiren.
\end{quote}

\begin{quote}
Sunspots were seen a lot; like in the last October [1705], 
when two did not have fully crossed the solar disc,
when two other spots with a accompaniment entered the disc. 
One could assemble a fair catalogue of similar observations for the last few years.
\end{quote}

Wurzelbaur describes not only the group of two spots of 1705 Oct, 
but also the entrance of another group with two spots (plus a {\it accompaniment},
original: {\it sonderbahrem comitatu}) during that time with an overlap in time.
He also mentions that many such instances were seen the last two years,
probably meaning that at least two sunspot groups were seen at once,
which was rare in earlier decades. 
He also seems to suggest to compile a sunspot data base.

Sp\"orer's catalogue does not contain any data for 1705 Oct,
but RNR93 do list five spots at $-0.2$, $-10.8$, $-9.8$, $-6.3$, and $-4.2^{\circ}$ for Oct.
HS98 do not contain any data for 1705 Oct 16.

Because G. Kirch describes the group to be close to the western limb, 
HS98 was searched for observations \textit{before} and around Oct 16: 
observations of 1 sunspot group each are listed for
La Hire (Oct 4, 5, 8-12, 14), Manfredi (Oct 3, 4), Derham (Oct 2, 3, 5-7), 
Plantade (Oct 4, 14, 19), Lalande (Oct 4), Muller (Oct 9, 14), and G. Kirch (Oct 14-16),
the dates given for G. Kirch were not mentioned in his letter from Oct 16.
HS98 also list observations of 2 groups on individual days, 
namely Manfredi (Oct 8, 10, 17, 18), Plantade (Oct 5, 9, 11-13, 15, 16),
Lalande (Oct 12), and Muller (Oct 10).
Wurzelbaur, who has observed one or two groups in October, is again not listed in HS98
(unfortunately, Wurzelbaur did not give the exact dates).
If the dates by Derham (Oct 2, 3, 5-7) are Julian, they would be Oct 13-18 Gregorian (11-day shift since 1700),
quite consistent with a spot been by Kirch on Oct 16 near the western edge.
While HS98 list 2 groups for Plantade for Oct 15 \& 16, Kirch specified that
he (and his wife) saw {\it two fine spots in the Sun, about 3 inched from the egreee edge},
so that they were probably in one group only.

\subsection{1705 Nov 5-12}
\label{12.11.1705}

G. Kirch describes in the 822nd letter from 1705 Nov 27 (Herbst 2006) 
to Gottfried Wilhelm Leibniz his next spot:
\begin{quote}
Wir haben eine Zeit her meist tr\"ube Wetter gehabt: 
Jedoch haben wir eine sch\"one gro\ss{}e Macul in der Sonnen observiret. 
Am 5 Nov. um 4 n. war sie 10 partes micrometri eines 7 sch\"uhigen Tubi vom Eintrits-Rande, 
ist 2'. 7''. da der Diameter $\odot$is 156 partes war, ist 33'. 4''. 
und den 12 Nov. zu Mittage war sie 40 partes micrometri ist 8'.29''. vom Austrits-Rande.
\end{quote}

\begin{quote}
For some time, we have had mostly bad weather, 
but we have observed a beautiful large spot in the Sun. 
On Nov 5, at 4h in the afternoon, it was 10 p.m. in a 7-foot telescope
from the ingress limb, i.e. 2' 7'' [arc min/sec].
while the Sun's diameter was 156 p.m., i.e. 33' 4''. 
On Nov 12 at noon it was 40 p.m., i.e. 8' 29'', from the egress limb.
\end{quote}

\begin{figure}
\begin{center}
\includegraphics[width=0.485\textwidth]{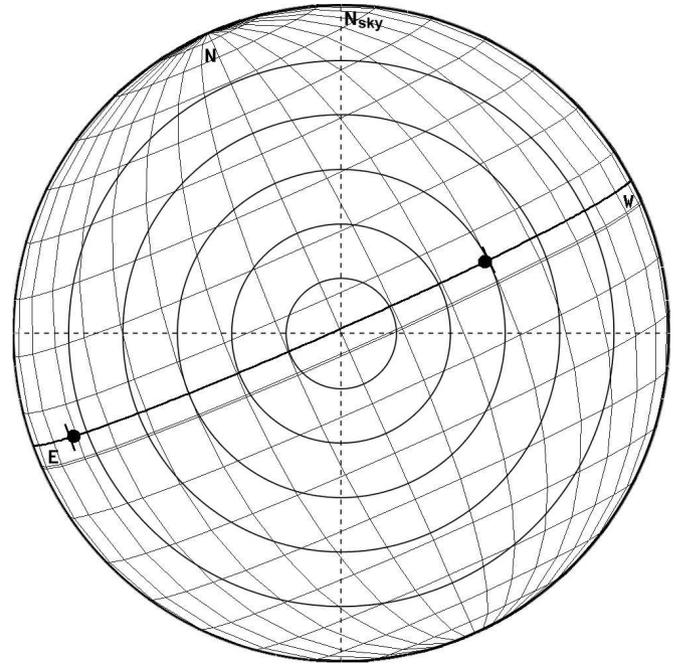}
\end{center}
\caption{Locations for the sunspot seen on 1705 Nov~5 and~12 by
Kirch. The tilt of the Sun on these dates was $B_0=3.49^{\circ}$
and $2.71^{\circ}$, respectively.}
\label{1705_11_05_1600_tip}
\end{figure}

We adopt the positions
\begin{itemize}
\item 1705 Nov 5, 4 pm: 68 p.m. from centre ($=0.8718R_\odot$)
\item 1705 Nov 12, noon: 38 p.m. from centre ($=0.4872R_\odot$)
\end{itemize}
This results in a latitude of $+4.2\pm17.1^{\circ}$ (Fig.~\ref{1705_11_05_1600_tip}). The large
error margin results from the fact that the
freedom in latitude is relatively large for
near-equator spots, since the separations from the
centre do not vary strongly when varying the
latitude. When going away from the equator, the
inch-rings move the possible locations closer in
heliographic longitude, but that is compensated
by the slower rotation rate.

Sp\"orer mentions an observation Nov 4-15 with a heliographic latitude of $-3^\circ$
from Derham, which could be the same spot as observed by G. Kirch.
The RNR93 diagram does not contain any data for this time,
but several for 1705 Oct (see previous section), the last at $-4.2^{\circ}$,
which could be the one observed also by Kirch and listed by Sp\"orer for Derham. 

The observers listed in HS98 saw 1 sunspot group each: 
La Hire (Nov 8, 10, 11, but spotless Nov 6 \& 7), Manfredi (Nov 4, 10, 14-17), 
Derham (Oct 25-Nov 4), Plantade (Nov 4-13), Cassini (Nov 10-13),
Lalande (Nov 4-15),
and G. Kirch (Nov 5, 7, 11, 12, but spotless Nov 4).
Again, the data for G. Kirch are not identical to those given
in his own letter.
While most observations according to HS98 lie from Nov 5 to 17,
which is just possible for one particular spot crossing the Sun,
the dates given for Derham (Oct 25-Nov 4) strongly indicate that they
were not transformed from the Julian to the Gregorian calendar.

\subsection{1706 Dec 11-16}
\label{13.12.1706}

\begin{figure}[h]
\begin{center}
\includegraphics[width=40mm,height=20mm]{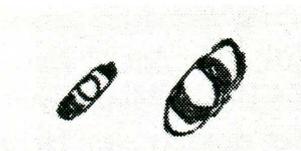}
\end{center}
\caption{Sketch by M.M. Kirch, G. Kirch's second wife, showing how she and their son 
Christfried Kirch saw the group on 1706 Dec 11.}
\label{1706_12_13_kirch_own}
\end{figure}

This letter to Leonhard Christoph Sturm (no. 845, Herbst 2006) from 1706 Dec 13,
written by G. Kirch's second wife Maria Margaretha Kirch (1670-1720),
contains a small drawing (Fig. \ref{1706_12_13_kirch_own}). 
M.M. Kirch writes to Sturm:
\begin{quote}
Am vergangenem Sonnabend haben wir sie auffm Observatorio 
durch einen 15 schuhigen Tubum besehen \dots 
Mein lieber Mann siehet ihn zwar nur l\"anglicht, und als ein paar schwartze L\"ochlein, allein ich und mein Sohn,
sehen ihn etwan auff diese Art oder gleich einer Kugel in einer breit r\"andrichten Sch\"u\ss{}el liegend \dots 
Itzt vormittags um 11,5 Uhr hat Er geme\ss{}en den Diametrum Solis 226 Viertel Micrometri die gr\"o\ss{}este 
Macul vom n\"achsten Rande, ist der Austrits-Rand, 26 partes, vom weitesten oder Eintrits-Rande 162 die Breite 
aller Maculn zu sammen 26 partes. Nun wird ihre Stelle Morgen freilich merklich ver\"andert seyn. wie sie den 
so wol gestern viel anders sahen als am Sonnabend, und heut anders als beyde Tage. Die gro\ss{}e ist sch\"on 
schwartz und Kernhafft, und hat einen getuschten Rand um sich.
\end{quote}

\begin{quote}
Last Saturday [Dec 11] we saw them at the observatory through a 15-foot telescope \dots 
My dear husband only sees it longish and like a few black holes, 
but my son and I see it like this [sketch] or like a sphere lying in a bowl with a broad rim. 
\dots
Now [Dec 13], before noon, around 11:30h, he [G. Kirch] measured the Sun's diameter to be 226 quarter micrometri [p.m.]. 
The largest spot is 26 p.m. from the nearest limb, which is the egress limb, 
and 162 p.m. from the furthest or the ingress limb. 
The width of all spots together is 26 p.m. 
While tomorrow, their position will be considerably different than today 
like it changed very much between yesterday and Saturday 
and today different from those two days. 
The large one is nicely black with a core and surrounded by a painted edge.
\end{quote}

Note that M.M. Kirch mentioned explicitly their {\it observatory}
and that the vision of G. Kirch was not as good as the vision of herself
and her son -- G. Kirch being almost 67 years old, his second wife only 36 years,
his son Christfried nearly 12 years young.

\begin{figure}
\begin{center}
\includegraphics[width=0.485\textwidth]{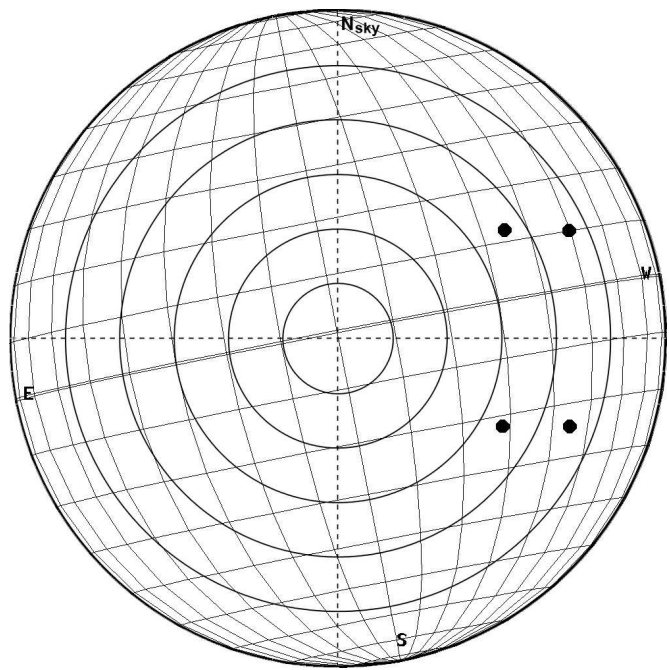}
\end{center}
\caption{Extreme locations for the sunspot group seen on 1706 Dec~13 by
M.M. and C.~Kirch, $B_0=-1.15^{\circ}$.}
\label{1706_12_13_1130_tip}
\end{figure}

Using the sequence of measurements from the eastern to the western
limb, we may assume the following quantities for 1706 Dec~13, 11:30:
\begin{itemize}
\item $0.717 R_\odot$ from the eastern limb to eastern spot
\item $0.097 R_\odot$ for the extent of the group
\item $0.097 R_\odot$ from the western spot to the western limb
\end{itemize}
The separation measurements were assumed to lie on a line parallel
to the celestial equator, as it would appear when measuring through
an equatorially mounted telescope. The position angle of the
heliographic north pole was $11.0^{\circ}$ towards the east, measured
from the celestial north pole. These quantities and assumptions
lead to the reconstruction shown in Fig.~\ref{1706_12_13_1130_tip}.
The midpoints of the possible group locations are $-23^{\circ}$ and $+11^{\circ}$.

In letter no. 848 to Hans Christian von Wolffsburg (1706 Dec 16, Herbst 2006)
from 1706 Dec 16, G. Kirch also writes in the postscriptum:
\begin{quote}
Heute vormittags war die gro\ss{}e Macul nur noch 12 p.m. vom Austrits-Rande,
ist 1/2 Zoll. Die n\"achste so ihr folget, sahen wir auch noch fein deutlich.
\end{quote}

\begin{quote}
Today before noon [1706 Dec 16] the large spot was only 12 p.m. [partes micrometri]
from the egress limb, means 1/2 inch. The following [spot] we could see clearly.
\end{quote}

If the latitude
was indeed $-23^{\circ}$, the synodic rotation rate would be $13^\circ$/d.
Within three days, the spot should have moved by $39^\circ$ in longitude. 

HS98 list in their table that La Hire observed on Dec 7 \& 17 one sunspot group and on Dec 11 two groups. 
With one group observed, HS98 list Manfredi (Dec 3, 5-8, 15),
Derham (Nov 28-Dec 5), 
Plantade (Dec 6-18), Cassini (Dec 7-15), Muller (Dec 15), Lalande (Dec 7), and
G. Kirch (Dec 11-14, 16, 17);
as usual Derham's dates are probably Julian.
Scheuchzer's observation (Z\"urich, Switzerland) of four sunspot groups on Nov 30 and Dec 1 and 
three groups on Dec 2 stands out (HS98);
La Hire would have reported a spotless Sun for Dec 1 and 2;
a spotless Sun on May 12 (or 22, if not shifted) as also reported by Scheuchzer (HS98)
would be consistent with a few other observers.

The observation by the Kirchs for Dec 11 (two spots close to each other in one group)
could be listed as two different groups seen by La Hire in HS98,
Plantade and Cassini agree that there was only one group on Dec 11.

\subsection{1707 Jan 27}
\label{27.01.1707}

In letter 851 on 1707 Jan 27 (Herbst 2006),
G. Kirch writes again to Gott\-fried Wilhelm Leibniz, 
where he only mentions the observed sunspot and the expected crescent-shaped Venus.
Regarding the spot G. Kirch's letter reads as follows:
\begin{quote}
Nach dem es etliche Tage her gantz tr\"ube gewesen, also, 
da\ss{} man die Sonne nicht hat besehen k\"onnen, 
und sich nun heute dieselbe gezeiget, 
habe ich durch einen 10~sch\"uhigen Tubum eine kleine Macul in derselben gefunden, 
deren Diameter wol kaum 1/8 einer Minute seyn mag. Sie war um 11 Uhr 36 Min. 64 Viertel Gewinde, 
oder partes micrometri vom Eintrits Rande, ist 9'. 18''. 
und bald darauff, nemlich um 11 Uhr 45 Min. 162 partes micrometri vom Austrits-Rande. 
Der Diameter Solis war 226 partes micrometri ist 32'. 49''.
\end{quote}

\begin{quote}
After many dull days, where the Sun could not be observed, 
I found a small spot with a 10-foot telescope. 
Its diameter is hardly an eighth of a minute. 
At 11:36h a.m. it was 64 quarter turns, means p.m., from the ingress limb, 
which is 9' 18'' [arc min/sec]. Shortly afterwards, namely at 11:45h a.m., it was 162 p.m. from the 
egress limb. The Sun's diameter was 226 p.m., means 32' 49''.
\end{quote}

\begin{figure}
\begin{center}
\includegraphics[width=0.485\textwidth]{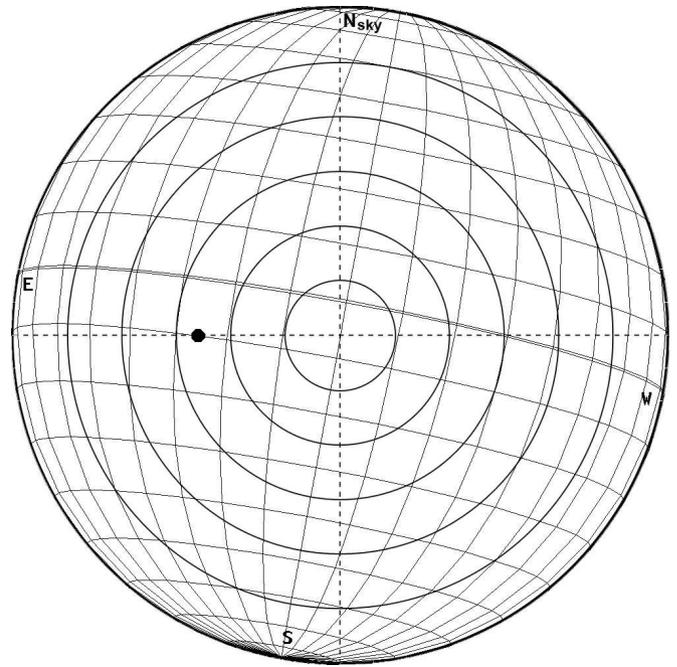}
\end{center}
\caption{Most likely location for the sunspot seen on 1707 Jan~27 by
G.~Kirch. The tilt of the Sun was $B_0=-5.93^{\circ}$.}
\label{1707_01_27_1140_tip}
\end{figure}

The exact middle of the Sun corresponds to a heliographic
latitude of the spot of $-5.3^{\circ}$ without using the position angle
of the Sun, given the measurements above and the
heliographic latitude of the Sun's midpoint of $-5.93^{\circ}$ for 1707 Jan~27, 11:40.
Using the position angle of the Sun of $10.5^{\circ}$ towards the west,
we obtain a spot latitude of $-10.0^{\circ}$. The corresponding
spot location is illustrated in Fig.~\ref{1707_01_27_1140_tip}.
(The above given diameter of the Sun, 32' 49'', is some $2.5\%$ too large,
possibly indicating their measurement precision.)

RNR93 plot a spot for 1707 end of Jan to early Feb with $-10.1^{\circ}$,
which is quite certainly the same spot as observed here by Kirch.
HS98 list only two observers for one sunspot group each around that time,
namely Manfredi (Jan 29) and G. Kirch (Jan 27 \& 28).
The extra date (Jan 28) is not given in the letter cited above,
so that HS98 had an extra source.
Surprisingly, according to HS98, La Hire would have reported a spotless Sun for Jan 27
and Kirch for Jan 29 \& 30 -- while Manfredi saw a spot on Jan 29.

\subsection{Sunspots 1707 March 5-8 (and an aurora March 6 observed by G. Kirch)}
\label{rest1707}

\begin{figure}
\begin{center}
\includegraphics[width=0.485\textwidth]{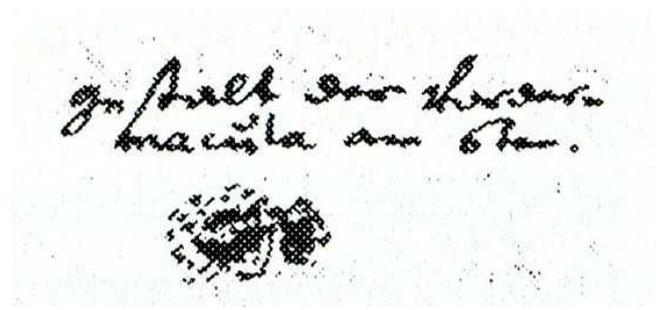}
\end{center}
\caption{Sketch of the spot for 1707 Mar 6 
attached by Sturm in his letter to Kirch.
His captions reads in German {\it gestalt der vordern macula am 6ten},
i.e. {\it shape of the foremost spot on the 6th} (Mar 6) .}
\label{maerz_sturm}
\end{figure}

On 1707 March 11, Leonhard Christoph Sturm wrote to G. Kirch (letter no. 853, Herbst 2006) 
about an observation of a sunspot group. He also attaches a small sketch (Fig. \ref{maerz_sturm}).
\begin{quote}
Am empfang meines zwey fachen berichts ... Nach demselben [bericht]
haben wir hier stets unruhiges wetter gehabt, ... Deme ungeachtet habe ich
doch durch die wolcken durch mit meinem 10. schuhigen Tubo feliciter bi\ss~ans
Ende observiret.
Den f\"unften habe die vordere macul noch schwartz und sch\"on gefunden, 
wie sie auch bi\ss{} an den rand hin geblieben, sie war aber etwas schmahler. 
Sie stunde 16/120. theil vom rande. die andere war sehr bla\ss{} und recht in 
zwey maculas zertheilet, wie ich bey einigem vorblicken der Sonne deutlich 
gesehen \dots den sechsten sahe ich die vordere noch viel schmahler 9/120. 
theil vom rande, die andere nur ein fach und gar nahe dabey. den 7ten. 
stunde die macul 2/120. vom rande und die bla\ss{}e  hart daran. 
den 8ten waren sie beide aus der disco hinweg.
Solchergestalt bringe ich heraus da\ss~sie durch den gantzen discum in 13 1/2. tag circiter
m\"ussen gegangen seyn.
\end{quote}

\begin{quote}
For the receipt of my double report ... After that [report], we had unstable weather, ...
Nevertheless, I have observed with my 10-feet tube luckily through the clouds until the end.
On the fifth [1707 March 5] I found the foremost spot to be black and nice, 
like it was near the limb, where it was a bit smaller. 
It was 16/120 parts from the limb. 
The other one was very faint and divided into two spots as I saw during the observation \dots 
On the sixth [March 6] the foremost spot was much smaller, 9/120 parts from the limb, 
the other one was single and nearby. 
On the seventh [March 7] the spot was 2/120 parts from the limb 
and the faint one nearby. 
The eighth [March 8] they were both vanished from the solar disc.
Therefore I deduce that it [the spot] trasversed the whole disc for 13.5 days.
\end{quote}

\begin{figure}
\begin{center}
\includegraphics[width=0.485\textwidth]{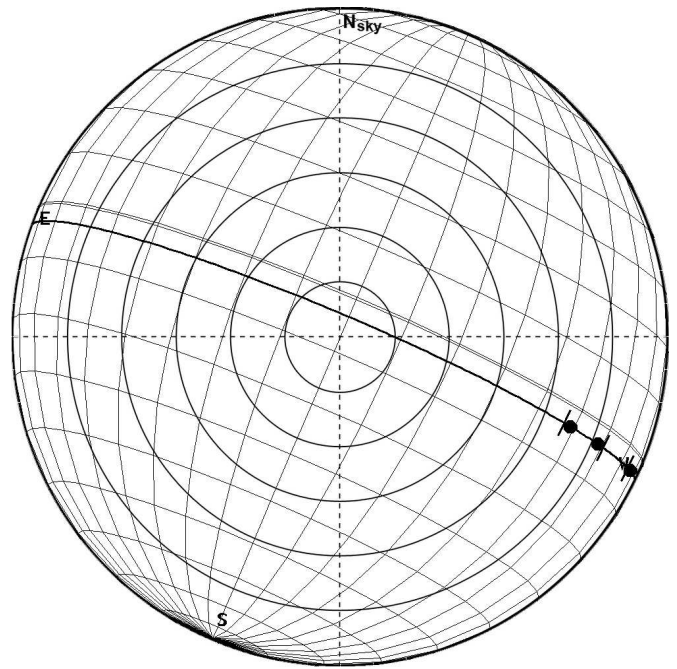}
\end{center}
\caption{Most likely location for the sunspot seen on 1707 Mar~5--7 by
Sturm. The tilt of the Sun on these days was $B_0=-7.24^{\circ}$, $-7.23^{\circ}$,
and $-7.22^{\circ}$.}
\label{1707_03_05_1200_tip}
\end{figure}

Assuming that the fractions given by Sturm are fractions
of the solar diameter, we have the following details:
\begin{itemize}
\item 1707 Mar 5 (assuming noon): $0.7333 R_\odot$
\item 1707 Mar 6 (assuming noon): $0.8500 R_\odot$
\item 1707 Mar 7 (assuming noon): $0.9667 R_\odot$
\end{itemize}
We have to assume the clock time during the day to
be 12h (noon) local time. The most likely solution is a spot
near the solar diameter at a latitude of $-3.5\pm27.0^{\circ}$.
Since the measurements indicate a constant, unphysical spot motion
as if the solar disk was flat, the inferred latitude
is highly uncertain. A real spot should have moved less
between Mar~6 and~7 than between Mar~5 and~6.
The corresponding reconstruction is illustrated in
Fig.~\ref{1707_03_05_1200_tip}

Since Sturm describes the last visibility of the spots, 
it is more likely that other observers saw the group already in 1707 Feb. 
A matching entry in Sp\"orer's catalogue (Feb 25-Mar 1) gives 
a heliographic latitude of $-6^\circ$ for Derham. 
It is quite well possible that this is the group seen by Sturm, namely two to three
spots close to each other in one group (Mar 5-7 in the western hemisphere).
RNR93 give two data points at the end of 1707 Feb with $-13.2$ and $-8.7^\circ$. 

Other observers in HS98 are: 
La Hire spotless on Mar 8,
G. Kirch is listed with the observation of two groups for Mar 1, 2, 4, 6, 7,
while the drawing by Sturm clearly shows one group with two spots.
HS98 list two groups for Sturm for Feb 28 and Mar 1, but nothing for Mar 5-8.
The date range for Derham (Feb 14-16, 18, 21, 24, Mar 6, 9, 11, 12 one group each) 
could well be Julian again.
Plantade observed one group on Feb 25 and 26, and two groups Feb 28-Mar 6,
then spotless on Mar 9; 
lastly HS98 list Muller (Feb 26-Mar 4, 6, 7) and Lalande (Feb 25-Mar 1) 
with the observation of one group each.

Given that the data in HS98 for Sturm (and G. Kirch) are different from what
we read in the letter cited above, HS98 used a different source, namely a book 
with sunspot observations by Sturm and Hertel (see HS98).
Sturm is listed by HS98 only 
for the two dates 1707 Feb 28 and Mar 1, not anywhere else.
Note also that Sturm, in his letter to Kirch,
gives 
1707 Mar 8 as spotless, which is not listed
in HS98, but consistent with spot detections since Feb 25 and also with
the spotlessness reported by La Hire for Mar 8 (HS98).
The two groups listed in HS98 for some observers may well be the two spots 
described by Sturm and by others, and being grouped into one group by most observers.

In his hand-written aurora catalog (see Schr\"oder 1996 for an edition), 
Gottfried Kirch's son Christfried Kirch (AD 1694-1740) listed
four entries for 1707, namely for Mar 6, Oct 21, Oct 29, Nov 27 (probably Gregorian dates);
for all four, the observer name is given (always Kirch); 
for Mar 6 and Nov 27, also Copenhagen is given as additional location;
for Mar 6 and Oct 29, an additional observer is given (hand-written in old Suterlin German),
probably {\it Seibl} which could be Seidel (see below).
No further details are given, so that it is hard to judge, whether it was a true aurora.  

Fritz (1873) lists a few aurorae for both Mar 1 and 6 (Gregorian) for different places,
also including Berlin (citing {\it G. Ki} for Kirch), Copenhagen (O. R\"omer),
and the location {\it Sch\"onberg in d. alt. Mark} citing
{\it Heusen, Beschreib. des Nordsch. Miscell.}, i.e. {\it Description of northern shine};
the given location is the region Altmark $\sim 100$ km west of Berlin, Germany.
Also de Mairan (1754, p. 186, 204-5) lists G. Kirch for Berlin,
O. R\"omer for Copenhagen, and {\it Christ. Mat. Seidelius} for 
{\it Schornberg dans la Vieille Marche} ({\it Altmark}) for both March 1 and 6
(citing {\it Employee Sup. pp. 141 \& 152}).

We did not find any mentioning of aurorae in the letter
dated 1707 March 11 from Leonhard Christoph Sturm to G. Kirch (letter no. 853, Herbst 2006),
in which he describes the sunspots in 1707 March.
As mentioned by Herbst (2006, p. 505) in relation to letter no. 853 from Sturm
to G. Kirch of 1707 Mar 11, there were four more letters from Sturm to G. Kirch's
wife Maria M. Kirch dated 1706 Dec 19, 1707 Feb 28, Mar 4 and Mar (?) 21, which are
not quoted in Herbst (2006). We obtained a copy of the (handwritten) letter 
dated {\it 21 M. 1707} (i.e. March or May) from K.D. Herbst 
(priv. comm., it is Bl. 230r-v from UB Basel Ms L Ia 724);
that the latter letter was sent in March (not May) becomes clear from
letter no. 863 dated 1707 Mai 9 from Hertel to Kirch (see Herbst 2006).
The aurora is also not mentioned in this letter.

However, in an extract of a letter from G. Kirch probably sent 
from Berlin to Dresden before 1707 Mar 14 (and after Mar 6) to
Ehrenfried Walther von Tschirnhaus (1651-1708) in Dresden, Germany 
(the only known letter partner of Kirch in the area of Dresden), 
Kirch's observation of the night of AD 1707 Mar 6 is described
in detail, we quote here in full letter no. 855 (Herbst 2006): 
\begin{quote}
Extract aus einem Schreiben, welches nach Dresden an einen Vornehmen
und seiner Gelehrsamkeit wegen sehr ber\"uhmten Manne gesandt 
einen gewi\ss en n\"achtlichen Nord-schein betreffend, der d. 6ten Martij allhir in Berlin
\"uber 2 Stunde observirt worden. \\
Er. Gn. ppp. Anbey benachrichtige, da\ss~den 6ten huius, war
am vergangenen Sonntage, Vormitternacht um 8 Uhr ein sehr ungew\"ohnlich heller Schein
gegen Norden am Himmel observirt worden, welcher seine Zuschauer in ziemliche Verwunderung
sezte, zumalen selbige \"uberzeuget waren, wie dieser helle Himmel weder vom Mond,
noch von der Sonn seine iezige Licht-Gestalt in der Nord-Seite entlehnen k\"onne.
Es war dieser N\"ordliche falsche Licht-Schein einem gedoppelten Regenbogen,
\"uber einanderstehend, nicht un\"ahnlich nur da\ss~die B\"ogen von considerabler Breite
erschienen. Zwischen den beiden B\"ogen war ein schwartzer Streiff, der aber keine
Wolcke seyn konte; ma\ss en in demselben die Sternen ohne Hinterung, wie sonst bey
hellem Himmel, gesehen wurden, unter denen war Lucida Lyrae und Westlich der gro\ss e
in cauda Cygni. Der \"oberste Bogen reichte bi\ss~\"uber den Kopff des Draconis hinaus
gegen den Polar-Stern zu; der unterste etwa bi\ss~an die Lyram. Zur rechten Ost-warts,
wie auch zur Lincken gegen Westen war der Schein, besonders da sichs zum Ende neigte,
am st\"arcksten. Man sahe unten anderm helle Strahlen, die von der Erden auffzusteigen
das Ansehen hatten, nicht anders als Ragetten, doch ohne Bogen, ad perpendiculum;
am Licht aber nur nicht so penetrant, sondern etwas neblichter, iedennoch in
S\"aulen-Form: Dahero ich sie dem Priester, dem ich sie zeigete, columnas 
igneas nigredine seu umbrosa intercapedine interstinctas, benannte. Und diese
bemerckte um den Asterismus Lyrae et Herculis herum und Ostwarts am meisten.
Gegen Westen konte ich dergleichen nicht wahrnehmen den kleinen Prospects wegen.
Von diesem falschen Lichte wurde der gantze N\"ordliche Horizont als von einem
schwachen Monden Schein mercklich erleuchtet. Dieser lichte Nord-Schein,
berichtet mich gleich izo ein Prediger in der Alt-Marck, 15 Meilen von hier
wohnend, habe sich um 7 Uhr angefangen, und um 10 geendigt, welche
W\"ahrung mit der hiesigen \"ubereinkomt. Es meldet dieser Prediger unter
andern auch, da\ss~der ganze Horizont, den er um und um haben sehen k\"onnen,
so helle werden, ob fiel des Monden Schein darauff: Denn der Erdboden
fein sein k\"onnen, und die H\"auser derer um ihn herum liegenden D\"orffer. 
Um 9 Uhr schwand der
Bogen, wo er am h\"ochsten allm\"ahlich und der Schein zog sich gegen Osten
und Westen. Ich erinner mich hiebey vor etlichen Jahren einen Bogen 
fr\"uh im Nebel gesehn zu haben. Er. Gn. Gedancken hier\"uber ppp.    
\end{quote}
We translate this to English as follows:
\begin{quote}
Extract from a letter, which was sent to Dresden to a highly ranked
man, very famous for his scholarship, about a certain northern glow at night-time,
which was observed here in Berlin on 6th March [1707] for more than two hours \\
greetings etc. pp Herewith, I inform you that on the 6th of this month, it was last Sunday
[indeed 1707 Mar 6 Gregorian was a Sunday; hence, 1707 Mar 13 as latest possible date for the letter], 
before midnight at 8 o'clock a very unusual bright glow was observed towards the
north on sky, about which the observers were rather surprised, because they were convinced,
that this bright sky could neither get its present light form on the northern side 
from the moon nor from the sun. This northern unusual [lit.: wrong] light form did not appear
dislike to a doubled rainbow, standing upon each other, but with the bow having
a considerble width. Between the two bows, there was a black strip, which could not
be a cloud; the brightness of the stars [in this area] were measured/seen without
reduction, as otherwise with a bright sky, among them $\alpha$ Lyr [lit.: Lucida Lyrae]
and $\alpha$ Cyg [lit.: in cauda Cygni]. The uppermost bow reached to beyond the head of
Draco towards the polar star; the lower bow up to about Lyra. Towards the right, east-wards,
as well as towards the left, west-wards, the glow was, in particular towards its ends,
strongest. Among others, bright rays were seen, which rose up from Earth,
having the appearance like rockets, but without bows, but perpendicular;
the light but not that bright, but a bit nebular, and in form of a column:
Therefore, I called it {\it fiery columns, which are being extinguished by
blackness or with dark interruptions}, 
when I showed it to a [certain] priest. And these were noticed most
around the constellation of Lyra and Hercules and towards the East.
Towards the West, I could not see it that way, because of the smaller
viewing angle. The whole northern horizon was brigthened by this wrong
light as from a weak moon. This bright northern glow was also reported to me
by a preacher from the Alt-Mark [region in East-central Germany $\sim 100$ km 
west of Berlin], living 15 [German] miles from here,
it would have started around 7 o'clock and ended at around 10 o'clock,
which is roughly consistent with here. This preacher reports among other
matters, that the whole sky, which he could see all around, 
became so bright, as if lunar light would fall onto them:
because he could see the ground well, and the houses of the villages around it.
Around 9 o'clock, the bow disappeared slowly where it was highest,
and the glow moved towards East and West. I remember now, many years ago,
to have seen a bow early [in the morning] in fog. 
Your thoughts about it, greetings etc. pp.
\end{quote}

Three to four of five aurora criteria in Neuh\"auser \& Neuh\"auser (2015a)
are fulfilled: 
(i) night-time (on March 6, it was dark by 7 or 8h p.m. in Berlin
or Altmark, respectively, whether or not they used sun-dial hours),
(ii) northern direction,
(iii) color ({\it fiery columns}), the mentioning of a {\it rainbow} does not
neccessarily means colors, but he meant form and width of the bow, and maybe
(iv) motion and dynamics ({\it bright rays were seen, which rose up from Earth,
having the appearance like rockets}) -- it is not clear whether Kirch describes
moving rays or non-moving rays or narrow columns
(the fifth criterion would have been repetition within a few nights).
Even though it is otherwise unusual or even problematic to describe an aurora
to be not unlike a rainbow (here: {\it not ... dislike a doubled rainbow}),
see Neuh\"auser \& Neuh\"auser (2018), 
and even though Kirch compares the observed bow at the end of the letter with an early morning fog bow,
the phenomenon observed by G. Kirch was very probably a true aurora
(instead of, e.g., a lunar halo display with arcs called {\it rainbow} and a {\it fiery} light pillar).
1707 Mar 6 was also close to new moon on Mar 4.
(Furthermore, G. Kirch knew the phenomenon of lunar halo displays,
as is evident, e.g., from his drawing of an elaborated display on 1684 Jan 24 (Julian) in
a letter to Johannes Hevelius, Gdansk, Poland (no. 256 in Herbst 2006) --
with the drawings now located in the archive at Paris Observatory (BO Paris, C.1.16,
no 29, folio 2299r-v and no 30, folio 2300) and reprinted in
Neuh\"auser \& Neuh\"auser (2015b),their figures 4 and 5.) 

G. and C. Kirch (as well as Seidel) observed an auroral arc from the west
through high up in the north (between the head of Draco and the polar star) to the east --
with the stars {\it Lucida Lyra} (i.e. the brightest in Lyra, i.e. $\alpha$ Lyr = Vega)
and {\it in cauda Cygni} (i.e. {\it in the tail of Cygnus}, i.e. $\alpha$ Cyg = Deneb, where the
name {\it Deneb} came from the Arabic Dhanab (ad-Daj\={a}ja) meaning {\it tail (of the hen)})
seen in the (lower) arc or between the two arcs -- moving at low altitude from the NNW to the NNE in the evening
(Herbst (2006) incorrectly gave Albireo for the star {\it in cauda Cygni}).

Similar auroral acrs were drawn by C. Kirch (1716) based on later observations on 1716 Mar 17,
including two arcs with a black (night-sky) gap between them and rays (rockets)
perpendicular to the arc -- see Fig. 24 for his sketch.

\begin{figure}
\begin{center}
\includegraphics[width=0.485\textwidth]{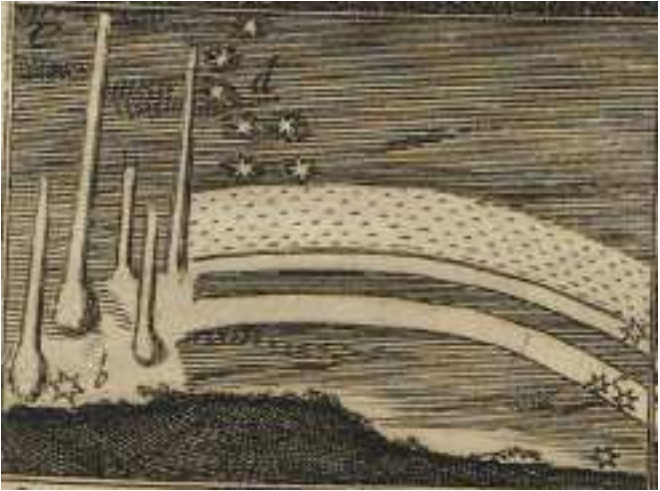}
\end{center}
\caption{Aurora observation by C. Kirch (1716) for 1716 Mar 17,
which is compared to the one on 1707 Mar 6 (in the letter of 1716, footnote 4). 
For 1716 Mar 17, C. Kirch writes here: 
{\it Um 10. Uhr 20. Minuten war dieser Bogen doppelt, wie Fig. C. andeutet. 
Mitten zwischen beyden / war ein schmaller tunckler Streiff oder Bogen ...
Es fingen aber bald am West-Ende dieser Bogen / die Strahlen an / in die H\"ohe zu fahren ...
mit auffahrenden Strahlen und S\"aulen}, which is
{\it At 10 hours 20 minutes, this bow was double, as shown in figure C.
In the middle between the two, there was a narrow dark strip or bow ...
At the western end of the bow, rays started to go up high ...
with rising rays and columns.} This discription for 1716 Mar 17 is quite
similar to the one for 1707 Mar 6.}
\end{figure}

Given that 1707 Mar 6 on the Gregorian calendar was indeed a Sunday, as given above in the letter,
the dating is clear, even though the year is not given in the letter extract.
The date 1707 Mar 6 was also given by C. Kirch for an aurora (see below)
and in a later letter concept of the Kirchs:
{\it der merckliche Nordschein 1707 den 6 Mart.},
i.e. {\it the remarkable northern glow of 1707 Mar 6}, 
from letter no. 858 in Herbst (2006), see below footnote 4.

The aurora was observed not only in Berlin by G. Kirch, but also in the
Altmark region by an unnamed priest, another letter partner of G. Kirch;
the Altmark observation was most certainly performed at the town
{\it Sch\"onberg in d. alt. Mark} as given in Fritz (1873) for an aurora observation on Mar 1 and 6 
(citing {\it Heusen, Beschreib. des Nordsch. Miscell.}, i.e. {\it Description of northern shine});
also, for 1707 Oct 19 \& 20, Fritz (1873) gives aurorae from Sch\"onberg citing G. Kirch after
{\it Pastor Seidel},\footnote{A Christoph Matth\"aus Seidel (1668–1723) was a pietistic (like
G. Kirch) protestant pastor at Sch\"onberg AD 1700-1708 (Schicketanz 2005).} who 
is most certainly the priest/preacher mentioned by Kirch above;
this letter partner was not identified by Herbst (2006);
In C. Kirch's hand-written aurora catalog, in addition to 1707 Mar 6,
the three entries for 1707 are Oct 21 (Kirch), Oct 29 (Kirch, Seibl/Seidl),
and Nov 27 (Kirch, Copenhagen), which is slightly different from Fritz (1873): Oct 19 \& 20.
The town of Sch\"onberg is $\sim 100$ km west of Berlin, i.e. some 15 German miles (given
by Kirch), where one such mile was $\sim 7.5$ km. 
The aurora was also observed by G. Kirch's son C. Kirch (born 1694), 
probably still with his parents in Berlin at that time, see below.
Ehrenfried Walther von Tschirnhaus (1651-1708) was a scholar of natural
science and philosophy, he lived and worked in Dresden, Germany,
towards the end of his life.

Frobesius (1739) also lists the aurora observations of 1707 Mar 6 by Seidel from Sch\"onberg 
({\it C.M. Seidelii ... Schoenbergae in veteri Marchia}) and Kirch from Berlin 
({\it G. Kirchii, berol. astronomi}) in his aurora catalog, but he gives {\it horam 7. inter et 10 vespertinam}
(i.e. {\it 7th to 10th hour in the evening}) for Kirch and {\it Hora circiter VIII.}
(i.e. {\it at around the 8th hour}) for Seidel, i.e. the other way around compared to
the Kirch letter above.

This is most certainly the only true aurora observed by G. Kirch.
Kirch also wrote a short treatise about this observation (Kirch 1710).
G. Kirch (born 1639), like most adults living around 1707
(also e.g. Sturm, born 1669, or von Tschirnhaus, born 1651),
lived mostly throughout the Maunder Minimum with no or very few aurorae --
hence, no to little experience.

C. Kirch (AD 1694-1740) wrote in his paper about the strong aurora of 1716 Mar 17 (Kirch 1716)
also briefly about the aurora on 1706 Mar 6: \\
{\it auch Anno 1707 eben dergleichen Phenomenon, wie wohl lange nicht so starck
als dieses itzige, mit angesehen} \\ 
which we translate to English as follows: \\
{\it also in the year 1707 just the same phenomenon seen, 
even if not as strong as the current one.} \\
Also in his later paper about the strong aurora of AD 1729 Nov 16 \& 17 (Kirch 1729),
he again briefly mentioned his own observations in AD 1707 March: \\
{\it Noch vorher habe Anno 1707 im Merz einen mercklichen
Nord-Schein mir angesehen, da dieses Phaenomenon als etwas ganz neues und unbekanntes 
angesehen ward, aber gegen den letzt erschienenen nicht zu vergleichen war},\\
which we translate to English as follows: \\
{\it Earlier, in the year 1707 in March, I have also seen an unusual northern glow,
when this phenomenon was seen as completly new and unknown, but it could not
be compared to the one which appeared lately [1729].} \\
This quotation confirms that the aurora phenomenon was {\it completely new and unknown}
to even professional astronomers in the early 18th century.

According to his letter to G. Kirch, 
Sturm observed two sunspots AD 1707 Mar 5-7, one of them approaching the western limb.
According to HS98, Sturm saw one spot on Feb 28 and Mar 1,\footnote{HS98 gave as source:
{\it Sturm, L. C., 1707. Sonnenflecken Buch, v. 2, no. 4.
Manuscript at Akademie fur Wissenschaften zu Berlin},
which seems to be a manuscript (not a book), which we could not find.}
Kirch on Mar 1, 2, 4, 6, and 7, and others from Feb 25 to Mar 1.
From Sturm's statement quoted above that he estimated that the spot was on the Sun for
13.5 days until Mar 8, it would have appeared in the sun on Feb 23.
Hence, it would be possible that the active region
associated with this spot or group was also responsible for the release
of the energetic particles towards the end of Feb of early March 
producing an aurora on Earth on Mar 6.

Given bad weather before and around Mar 5 (mentioned in the letter by Sturm
quoted above), Sturm could not observe the aurora on Mar 6.
As seen in the next section (3.22), it was probably the same spot group
that was again observed Mar 23 (two weeks after Mar 8, when it disappeared
at the western limb), so that it was long-lived and, hence, probably large
(for Mar 23, Hertel mentioned and drew at least six spots (labeled until f),
see next section, but the drawing is lost).

According to Herbst (2006), there would be more aurora observations
by the Kirchs described in another undated letter (no. 858 in Herbst 2006)
for Mar 17/18 and Apr 12/13 and 13/14 (1707 according to Herbst), but they
are misdated and were observed in AD 1716.\footnote{Letter no. 858
in Herbst (2006) is a concept for (or copy of) a letter written by the
hand of Christfried Kirch; Herbst (2006) assumed that Christfried wrote it
on a request by and for his father Gottfried, so that he included it in his
edition of the letter exchange of G. Kirch. 
The addressee is again not mentioned,
as in letter no. 855 (see above); because the text mentions  
{\it Nord-schein ... hier in der Nacht zwischen den 17 und 18 Mart} 
(i.e. {\it northern glow ... here in the night between 17 and 18 March})
and compares it to the {\it Nord-schein 1707 den 6 Mart.}
(i.e. {\it northern glow 1707 Mar 6}),
Herbst (2006) assumed that this letter would again 
by directed to von Tschirnhaus (as no. 855 above).
After briefly mentioning the aurora of Mar 17/18, the letter then 
describes in detail more aurorae on Apr 12 and 13: 
{\it Am heiligen Oster-tage den 12 Apr. wurde ich wiederum 
eines Nord-Scheins gewar, als ich die beyden Planeten Jupiter and Venus
abends observirte ...}
(i.e. {\it On the holy Easter day on 12 Apr, I noticed again a northern glow
when I observed the planets Jupiter and Venus in the evening ...}),
observed at least from 10:45h to 12h pm.
Later, the letter continues {\it N\"achst folgenden tag, den 13 Apr.
als am Ostermontage ... gegen Norden wieder heller was als gew\"ohnlich ...}
(i.e. {\it On the next following day, 13 Apr, when on Easter monday
... towards north it was brighter than usual ...}).
We summarize that we have aurorae for Mar 17/18 and Apr 12/13 \& 13/14,
for which the years are not explicitely mentioned, and the former is
compared to the aurora {\it 1707 den 6 Mart} (year mentioned).
The problem with the dating of the aurorae Mar 17 to Apr 13 to 1707
is mainly that in 1707, Easter was not on Apr 12, but on Apr 24 (e.g. Grotefend 2007).
(By 1707, the difference between the Julian and Gregorian calendars was
11 days, but it is unlikely to explain the offset by assuming that Kirch would have
given the Julian date here, because he consequently always gave the 
Gregorian date since 1700 Mar 1; also, the offset between Apr 12 and 24
is 12 days (not 11) -- and it is also unlikely that Kirch meant the
previous evenings.)
From 1707 (year mentioned in that letter) to 1740 (death of C. Kirch),
the only year in which Easter fell on Apr 12 (Gregorian) was 1716.
In the hand-written aurora catalog of C. Kirch, aurorae are listed  
for 1716 Mar 16 ({\it Kirch zu Danzig}, Gdansk, now Poland) and twice for April 
(without days nor other details like observer name or location),
but none for 1707 Mar 16 nor Apr.
It is therefore very likely that C. Kirch was the official author
of this letter -- and not G. Kirch, who died in 1710.
Furthermore, C. Kirch wrote a short paper about his aurora observations
of 1716 Mar 17 (Kirch 1716), where he specified that he observed it in
Danzig (now Gdansk, Poland) and where he attatched drawings of auroral arcs as observed.
(According to Herbst (2006), the fact that the text author writes in this letter
{\it da\ss~ich meine schuldige Pflicht bey der hochansehlichen Societ\"at
der Wissenschafften, in acht nehmen} (i.e. {\it that I observe my duty
at the highly-ranked society of sciences}), would be additional evidence that G. Kirch
would be the author; however, C. Kirch became director of the Berlin observatory
some time later in 1716, so that the above text is not an argument against C. Kirch as 
official sender of this letter in 1716.)
Since Ehrenfried Walther von Tschirnhaus had already died in 1708 in Dresden, Germany,
he was then also not the addressee of the letter we date to 1716 Apr (or later).
The planets Jupiter and Venus were well visible during the close conjunction
around 1716 Apr 12 and 13 in the evenings, while Venus was not visible 1706 Apr 12 and 13.
New moons were 1716 Mar 23 and Apr 22, so that it was indeed dark enough for
aurora observations in the nights 1716 Mar 17 and Apr 12 \& 13.
Fritz (1873) also does not list any aurorae for 1707 Mar 17 nor April,
but he does list the strong aurora of 1716 Mar 15-17, seen on Mar 17
as far south as Cadiz, Spain, and Lisbon, Portugal, and also at 49 other places;
Fritz (1873) gives {\it Mar 17 ... Berlin, d. g. Nacht ungem. gross
(Chr. Kirch, G. Ki.)} (i.e. {\it the whole night unusualy large});
that he cites both Christfried and Gottfried Kirch could be due to the
fact that he assumed that the hand-written aurora catalog was not by
C. Kirch, but by G. Kirch or both. 
The aurora of 1716 Mar 17 was the largest display since the start of the
Maunder Minimum.
Also de Mairan (1754, p. 186, 206) listed aurorae for 
1716 Mar 15, 17 and Apr 11, 12, and 13; in his listing
of auroral observations by C. Kirch (p. 499/500), 
there are three entries for 1716: Mar 17, Apr 11, 12.
Halley wrote about it {\it Nothing of this kind has occured in England for more
than 80 years, nor of the same magnitude since 1574} 
(Halley 1716, dated 1716 Mar 6 Julian, i.e. 1716 Mar 17 Gregorian).}

In letter no. 859 (Herbst 2006) from Constantin Gabriel Hecker 
(1670-1721, astronomer in Gdansk, Poland) to G. Kirch,
written after 1707 Apr 17, the lunar occultation of Apr 16/17 is described
in detail, but it is also reported a {\it fire} that might be an aurora:
{\it ... having climbed the observatory of Hevelius already at 9h pm
and having directed the tube, I saw in the neighbourhood of the town
a light or rather a fire, looking like a burning house; ... see in a
village next to the city something similar, in one word:
the terrible display of houses and warehouses 
set of fire by the Moscowers/Moscowits [meaning Russians in general] and the whole night fires}
(our translation from Herbst's German translation of Hecker's Latin text).
While it might be dubious anyway, whether to be able to observe an
aurora in a full moon night (before the lunar occultation),
Gdansk (Poland) is directly at the south coast of the Baltic sea, 
so that none of the surrounding villages are north of it;
and there were indeed a lot of hostilities between Sweden, Poland, and
Russia during the Great Northern War 1700-1721, e.g. in
1716 all of Sweden's Baltic and German possessions were lost.

\subsection{On and around 1707 March 23}
\label{march1707}

Christian Gottlieb Hertel wrote on 1707 Apr 25 (letter no. 861, Herbst 2006) 
to G. Kirch referring also to two sketches, which were not found:
\begin{quote}
Die erste Figur stellet diejenigen maculn, wovon MHH. 
Kirch dem Herrn Professori in dem Letzteren geschrieben, 
in ihrer eigentlichen Gestalt, proportionirten Gr\"o\ss{}e und Schw\"artze, 
mit flei\ss{} gezeichnet vor, so offt ich selbe wegen Wetters sehen k\"onnen. 
Die andere Figur aber die wiedergekommene vorige maculen 
welche sich wieder bi\ss{} an den Rande des Au\ss{}trits sehen la\ss{}en, 
doch so, da\ss{} sie fr\"uh in loco m noch ziemlich deutlich, 
nachmittage in n al\ss{} einen bla\ss{}en Nebel, 
weiter gegen Abend aber gar nicht mehr habe erblicken k\"onnen, 
wie genau und lange ich auch darnach gesucht. 
Doch konte Sie noch nicht v\"ollig herum seyn weil mir der Raum, 
so sie noch zu gehen hatte, fast zugro\ss{} vorkommet, mutma\ss{}e dannenhero, 
sie mu\ss{} damals verschwunden sein. 
Diejenigen aber, so zu Ende des Monats Martii erschienen, sind nicht wiederkommen.
\end{quote}

\begin{quote}
The first sketch shows the spots my esteemed colleague wrote about to the Professor 
in the last letter [unrecorded] in their real shape, proportional size, and blackness. 
I diligently drew it, as often as the weather allowed me to see it. 
The other sketch shows the reappearing previous spot, 
which was observable until the egress limb. 
In the morning, as seen in position m, it was nice and plain, 
in the afternoon in position n it was like a faint nebula, 
later in the evening it was not visible any more, even though of a detailed and long search. 
But it could not be completely through [the Sun's disc] for the space
to the limb, which it would still have to go, was almost too large. 
That is why I assume, it vanished that time. 
Those [spots], which appeared this way at the end of the month of March, did not come back.
\end{quote}

Since Hertel already knew that the spots did not reappear by the time 
he wrote
the letter (1707 Apr 25), the observations of the spots mentioned must have ended
around April 10 or earlier.

In the postscriptum of this letter, Hertel gives a date:
\begin{quote}
gar artig war es anzusehen, al\ss{} dem 23 Martii die macula c herunterw\"arts gegen d wich, 
stieg auff der andern Seite der Scheibe die neu wiedergekommene e 
umb eben so viel gegen f in die h\"ohe, und zwar zu einer zeit, wie die erstere Figur andeutet.
\end{quote}

\begin{quote}
It was fine to observe, as on March 23 the spot c went downwards to d and 
on the other side of the disc the reappearing [spot] e 
went just as much high against f, during the time the first figure shows.
\end{quote} 

The HS98 tables list the following observers, the numbers in brackets are the 
number of sunspot groups: \\
La Hire: Mar 20 \& 22 (2), 23, 24, 26, 28, and 31 (1), \\
Manfredi: Mar 24, 26, 27, 29, 30 (1) \\
G. Kirch: Mar 18-22, 24, 29, 30 (2) \\
Derham: Mar 18, 21 (1), spotless Mar 30 \\
Plantade: Mar 19, 20, 22 (2), Mar 23, 28-30 (3) \\
Lalande: Mar 20-23, 25-28 (1), Mar 24 (2), and  \\
Hertel: Mar 19 (2). \\
Since the date given by HS98 for Hertel is different from our
citation of Hertel's letter above (Mar 23), HS98 must have used a
different source, namely a book with sunspot observations from Hertel and Sturm (see HS98). 
Hertel's letter is not inconsistent with an additional observation on Mar 19.
The above quotation from Hertel's letter is consistent with two sunspot groups.

Derham reported a spot from Mar 6 to 21 with breaks (HS98), which could be Julian.
According to HS98, the spot was observed (otherwise, except for Derham's dates) since Mar 18,
so that these new sightings could indeed (as claimed by Hertel) be the reappearance  
of the spot seen by several observers until Mar 7 (previous section).
Hertel probably concluded about the reappearance both from the time span
since the last sighting (disappearance at the egress limb) and/or
the heliographic latitude, maybe also from the spot size or form,
which was, however, known to vary much.
A similar heliographic latitude is consistent with Sp\"orer giving
$-9^\circ$ for Mar 20-28 and $-6^\circ$ for Feb 25-Mar 1,
both for Derham, the latter date range is close to Sturm's observation
Mar 5-7.

\subsection{1708 Aug 8-18 and Sep 2-13}
\label{1708Aug}

J.P. Wurzelbaur writes in letter no. 889 from 1708 Oct (Herbst 2006) 
about further observations:
\begin{quote}
\dots \"Uber die von 11 Augusti in der Sonne entstandene 
und bis 18 ejusdem darinn observirte auch am 3:$^{ten}$ Septembr: 
widerherumbkommene und bis 14 ejusdem die Sonne durchwanderte maculas solares \dots
\end{quote}

\begin{quote}
\dots About the sunspots formed on [1708] Aug 11 and observed until [Aug] 18 
and anew entering on Sept 3 and wandered through the Sun's disc until [Sept] 14 \dots
\end{quote}

The lastly mentioned observation on 1708 Sept 14 is discussed in Sect. \ref{14.09.1708},
together with the solar eclipse observed then. 

These spots are also described by G. Kirch in letter 892 from 1708 Dec 19 to Wurzelbaur:
\begin{quote}
Die $\odot$ Maculen habe ich folgender Gestalt observiret: 
Am 8 Aug. um 4 1/4 nachmittags betrachtete ich die $\odot$ 
durch einen 10 schuhigen Tubum und fand sie rein. 
Am 10 Aug \dots und um halb 5 nachmittags die $\odot$ durch einen 10 schuhigen Tubum betrachtete, 
fand ich eine sehr schwache Macul in derselben, die war doppelt, 
um 4 Uhr 42' war die gr\"o\ss{}este 8.5' vom Ost-Rande. 
Die kleine aber war um eine Min von der gr\"osesten gegen den Ost-Rand. 
Weil aber die Maculen \"ubel zusehen waren, 
so konte man nicht so gar genau me\ss{}en. 
Den 11 Aug um 9 Uhr vormittags besahen wir die Maculen, 
die waren nun fein deutlich zu sehen. 
und es lie\ss{} sich zwischen beyden die dritte fast erblicken. 
Um 9 Uhr 10 Min. vormittags war die gr\"o\ss{}ere Macul 12' vom Ost-Rande der Sonnen. 
Den 18 Augusti Um 8 Uhr 34 Min vormittags war die gro\ss{}e Macul, 
sehr nahe am west-Rande der Sonnen und um 9 Uhr 43 klebte sie gantz am Sonnen-Rande. 
Den 2 Septemb um 3 nachmittags bekamen wir die Sonne zu sehen, 
und fanden die Macul wieder. Sie war sehr klein und nahe am Rande, 
iedoch klebte sie nicht am Sonnen Rande, sondern man konte noch deutlich 
R\"aumchen darzwischen erkennen, etwan so gro\ss{} als die Macul selbst breit 
ist und zwar kaum. Sonst war sie kernhafftig. 
Den 3 Sept um 8 vormittags war die Macul deutlich und braun. 
Um 9 vormittags war es wieder etwas hell, und die Macul fast 1 Min vom Ost-Rande. 
Um 2 nachmittags war sie \"uber 1 min davon. Als wir sie zum ersten mal sahen, 
war sie gelb, mit einen langen schwachen Nebel \dots
\end{quote}

\begin{quote}
The solar spots I observed as follows: 
On [1708] Aug 8 around 4:15h p.m. I observed the Sun through a 10-foot telescope 
and found it clear. On Aug 10 \dots around 5 o'clock p.m. I observed the Sun through 
a 10-foot telescope and found a faint spot in it. 
This spot was doubled and at 04:42h p.m. the largest was 8.5' from the eastern limb. 
The smaller one was about a minute from the larger, against the eastern limb. 
Because the spots were hard to see, one could barely measure. 
On Aug 11 around 9 o'clock a.m. we observed the spots, which were fine and clear now. 
Between the two there was a third one nearly visible. At 9:10h a.m. 
the large spot was 12' from the eastern limb. On Aug 18 at 8:34h a.m. 
the large spot was near the western limb and at 9:43h a.m. 
it stuck on the limb. On Sep 2 around 3 o'clock p.m. we saw the Sun 
and found the spot again. It was small and near the limb, but did not stick to it. 
We saw clearly some space between it, nearly as big as the spot itself, 
which was not very big. Apart from that it had a core. 
On Sep 3 around 8 o'clock a.m. the spot was clear and brown. 
At 9 o'clock a.m. it was bright again and the spot was nearly 1 minute 
from the eastern limb. At 2 o'clock p.m. it was more than 1 minute from it. 
As we observed it, it was of yellow colour and with a faint nebula \dots
\end{quote}

Note that the authors now use arc min (') as unit for measurements on the solar disk.
(Minute as unit of time was also abbreviated with the same sign (').)

\begin{figure}
\begin{center}
\includegraphics[width=0.485\textwidth]{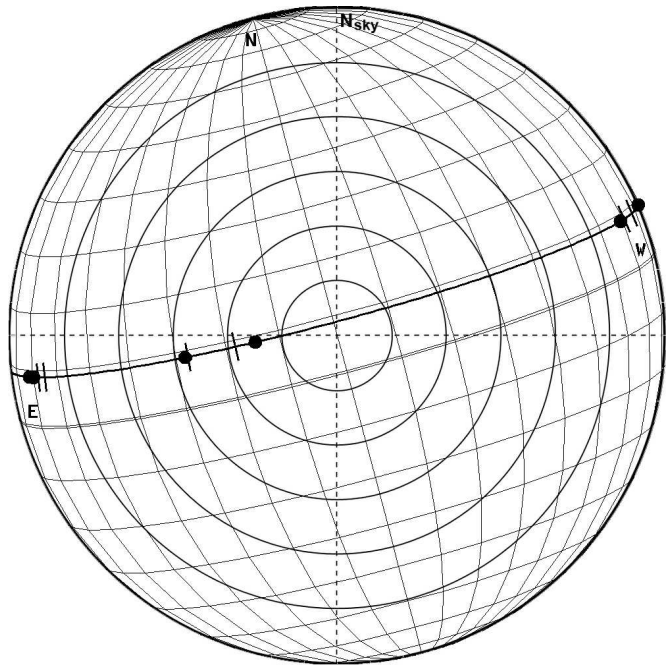}
\end{center}
\caption{Reconstruction of the sunspot location seen on 1708 Aug~10--Sep~13 by
G. Kirch. The tilt of the Sun was $B_0=6.61^{\circ}$, $6.64^{\circ}$, $6.94^{\circ}$, $7.24^{\circ}$,
and $7.16^{\circ}$ on those days.}
\label{1708_08_10_1642_tip}
\end{figure}

From this description as well as the description in the following subsection,
we deduce the positions:
\begin{itemize}
\item 1708 Aug 10, 4:42 pm: 8.5' from E, i.e. $0.4688 R_\odot$
\item 1708 Aug 11, 9:10 am: 12' from E, i.e. $0.25 R_\odot$
\item 1708 Aug 18, 9:43 am: 0.05' from W, i.e. $0.9968 R_\odot$
\item 1708 Sep 3, 9 am: 0.9' from E, i.e. $0.9437 R_\odot$
\item 1708 Sep 3, 2 pm: 1.1' from E, i.e. $0.9312 R_\odot$
\item 1708 Sep 13, 3:25 pm: 1.15' from W, i.e. $0.9281 R_\odot$
\end{itemize}
We need to assume that the locations of the second revolution
on Sep~3--13 refer to the same spot as in August. The solution
is a latitude of $+8.9\pm10.5^{\circ}$. The corresponding spot locations
are shown in Fig.~\ref{1708_08_10_1642_tip}.

With an observation between 1708 Aug 11-18,
Sp\"orer's catalogue gives a heliographic latitude of $-7^\circ$
for Wideburg (or Wiedenburg as in HS98, or Wideburgio as on his books, or Wiedeburg).
RNR93 however did not contain any data for 1708 Aug. 

Because the spot on 1708 Sept 13-14 is discussed in the next section, 
only the entries of HS98 of 1708 Aug are listed here: 
La Hire is listed with an observation of one group on Aug 11-16 (spotless Aug 8 \& 10),
Manfredi from Aug 11, 12, 16, 17, and spotless on Aug 8, 10, 18,
Muller is listed for one group Aug 12-17.
Wideburg is also listed for the observation of one group on Aug 11 and 12
and 2 groups Aug 13-17 (spotless until Aug 10 and on Aug 18).
The dates given for Derham (July 31, Aug 1, 5, 6, 22, 24, 28) and Gray (July 30-Aug 11),
one group each, could be Julian dates.
Kirch and Wurzelbaur are not listed by HS98 for August.

\subsection{1708 Sept 13-14}
\label{14.09.1708}

This spot is a special one in the G. Kirch correspondence, 
for it is observed during the solar eclipse of 1708 Sept 14,
shortly before the egress of the spot.
In the previous subsection, it was mentioned that this spot was
also observed Sep 3-14 by Wurzelbaur as well as by Kirch on Sep 2 and 3.

As mentioned in Neuh\"auser et al. (2015) to compare with P. Becker's observations
on that date, G. Kirch describes in letter 892 from 1708 Dec 19 (Herbst 2006) 
to Johann Philipp Wurzelbaur the sunspot's movement since its first discovery on Aug 8
(previous section).

In the letter, there are three important sections describing 
the spot on Sep 13, the solar eclipse on Sep 14, 
and then the position of the spot during the eclipse.

\begin{quote}
Den 13 Sept. um 3 Uhr 25' nachmittags war die Macul 8 partes micrometri 
10 schuhigen Tubi vom West-Rande, ist 1'. 9'' Diameter Solis war 224 partes micrometri 
ist 32'. 0'' \dots 
Das Ende war um 9 Uhr 42'. 37''. 
Wann ich nun aus vielen andern Observationibus nach rechne, 
und den Anfang um 7 Uhr 31 Min 7'' setze, 
so hoffe ich und bin es ziemlich versichert, da\ss{} ich um keine Minute fehlen kan. \dots
Das beste bey der Sonnen-Finsterni\ss{} h\"atte ich schier verge\ss{}en: 
nemlich die sehr kleine Macul, welche nahe am West-Rande der Sonnen noch zu erblicken war. 
Diese ward nicht vom Mond bedeckt, sondern da man sie am n\"achsten beym Mond zu seyn sch\"atzete, 
war sie etwan 6 partes micrometri 10 schuhigen Tubi von ihm, w\"are 51''. 
und dieses war um 8 Uhr 8'. 41''. \dots
\end{quote}

Translated, preserving the original wording, G. Kirch describes (text also in Neuh\"auser et al. 2015):
\begin{quote}
[Sept 13, 3:25 p.m.] in the 10-foot tube, the spot was 8 micrometer parts [p.m.] 
away from the western edge, which is 1' 9'' [arc min/sec], the diameter of the Sun was 224 p.m., which is 32'.
\dots The end [of the eclipse, Sept 14] was at 9 o'clock, 42' 37'' a.m. If I take other observations into account, 
I'm able to recalculate its beginning to 7 o'clock 37' 7'' a.m., what is certainly correct. \dots
\dots The best during the solar eclipse was the very small spot, which was visible 
close to the western edge of the Sun. This part was not eclipsed by the moon, 
it was also quite close to the lunar limb, so it could be estimated to lie about 6 p.m. 
[micrometer parts in the] 10-foot tube from it [the lunar limb], i.e. 51''. 
And this was at 8 o'clock 8 min 41 sec. One could not always see it, 
because it was very weak and small, and the clearness of the sky was variable.
\end{quote}

\begin{figure}
\begin{center}
\includegraphics[width=0.485\textwidth]{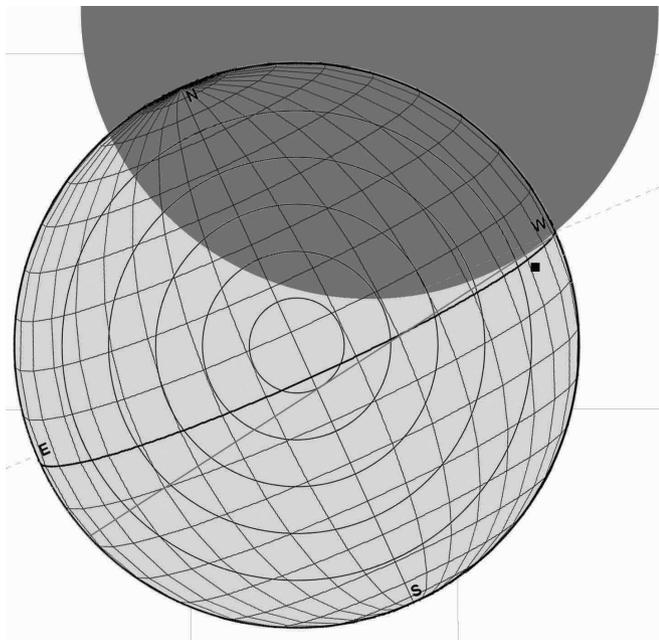}
\end{center}
\caption{Reconstruction of the sunspot location seen on 1708 Sep~14 by
G. Kirch during a partial solar eclipse. The dark grey circle is the
Moon at 08:08:41 (Kirch's time), the grey, straight line is the path
of the lower limb of the Moon, the dashed line is the ecliptic.}
\label{1708_09_14_kirch_eclipse_grid}
\end{figure}

We compare the end time of the eclipse determined to be
9:42:37 by G.~Kirch, with the ephemeris of the solar eclipse
and find 8:43:25~UT (same result from StarCalc~5.73 as well as
NASA eclipse web page\footnote{eclipse.gsfc.nasa.gov/SEsearch/SEsearchmap.php?Ecl=17080914}).
As it seems that Kirch did not observe the beginning of the eclipse,
we did not used his (inferred) first contact time.
If we apply the resulting clock system
difference of 00:59:12 to the closest approach of the Moon
to the spot, we obtain 7:09:29~UT. For this time, we compute
the location of the Moon as well as the path of the southern
limb of the Moon across the solar disk with StarCalc~5.73 and
show it in Fig.~\ref{1708_09_14_kirch_eclipse_grid}. Since
Kirch mentioned that this was the closest approach of the Moon
to the spot and that it never covered it, the spot location
can only be near the indicated square with a latitude of
$-2^{\circ}$. It is difficult to assess the uncertainty, but we may
conclude that the location is -- within the errors margins --
compatible with the above estimate of $+8.9\pm10.5^{\circ}$
(previous subsection). Since the
path of the Moon gives us a small uncertainty on the
observation of Sep~14, we conclude that the spot was on the
solar equator or slightly below it. Everything north of the equator would have been
covered by the Moon during the eclipse.

The heliographic latitude of this very spot, as observed by G. Kirch,
based on the letter quoted above,
was obtained to be $-4.5 \pm 1.5^\circ$ by Neuh\"auser et al. (2015) --
discussed there in comparison with observations by P. Becker.
In comparison, Sp\"orer's catalogue gives a latitude of $-5^\circ$ (for Wideburg) --
RNR93 list two spots in Oct with $-6.6$ and $-5.3^\circ$.

We also list the entries in HS98 for 1708 Sep:
La Hire (Sep 3-5, 7, 8, 10-12, 14-18);
Manfredi (Sep 2, 3, 6-8, 10, 15-17);
Wideburg on Sep 2 one sunspot group and on Sept 3-11 two groups (spotless Sept 14); 
G. Kirch (Sep 11);
Derham, (Aug 22-24, 28, \& Sep 1); 
Gray (Aug 27-Sep 8);
Blanchini (Sep 11, 12), 
and M\"uller (Sep 4, 5, 14).

It may appear surprising that all other observers (except La Hire and M\"uller)
did not report about the spot being visible during the solar eclipse (Sep 14);
according to HS98, Wideburg even reported explicitly a spotless Sun for Sep 14.
The date ranges for Derham and Gray are to be shifted by 11 days (Julian);
the remaining date range in HS98, Sep 1 to 18, is fully consistent with
G. Kirch describing it for Sep 2 to be shortly after ingress (previous section),
as well as with Wurzelbaur for Sep 3-14, the latter is not listed in HS98.
We do not know the source of HS98 for Kirch's observation on Sep 11.

In addition to the observations listed in HS98,
Neuh\"auser et al. (2015) present the observations by Peter Becker in
Rostock, Germany, for 1708 Sep 10 and 11 for $b \simeq (0 \pm 5)^{\circ}$,
and that he had bad weather on Sep 12-14, i.e. during the eclipse.
Becker also reported that {\it it was ensured from Berlin 
that the spot was still there on the 14th [of Sep],
the day of the eclipse, on the Sun and it was noticed},
probably referring to G. Kirch's observation in Berlin
(there is no letter exchanged between Kirch and Becker found, Herbst 2006).
The spot seen by Becker on Sep 10 \& 11 being 2 and 1.5 twelfth, respectively, from the
western limb could well be the spot seen by Kirch about 1' from the limb on Sep 14.

\subsection{1708 Nov 24-Dec 2}
\label{30.11.1708}

\begin{figure}[h]
\begin{center}
\includegraphics[width=70mm,height=55mm]{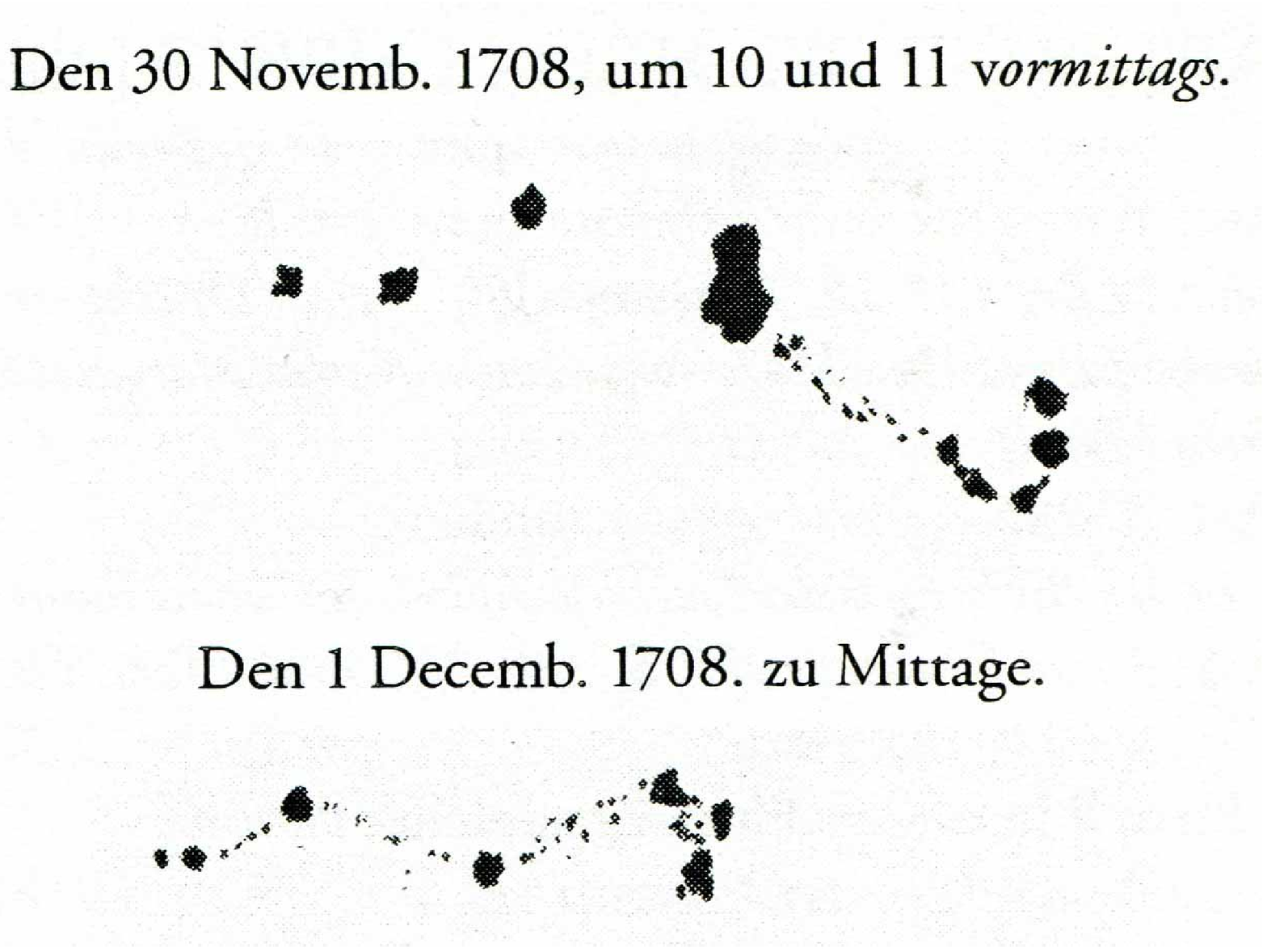}
\end{center}
\caption{Drawing by G. Kirch, captions translate as follows: 
{\it 1708 Nov 30 around 10h and 11h before noon (top), 1708 Dec 1 at noon (bottom),} 
both drawings are captioned with (translated): 
{\it rough arrangement of the many sunspots, 
through a 10-foot telescope, if the Sun's disc was nicely large, 
ca. 5-foot long, projected from the ocular down to the floor.}
According to his letter to Wurzelbaur, G. Kirch has 
labeled the right-most part as {\it h} 
(10' from the western limb), the central double spot as {\it a b}
(11'26'' from the western limb), and the left-most part as {\it e g f}
(12'51'' from the western limb); Kirch also specified that spot {\it f}
was 3'26'' from group {\it h}, so that spot {\it f} is the
very left-most spot in this drawing located $\sim$ 13'26'' from the western limb,
while spots {\it e g} are the two other spots in that group,
whose separation to the limb may have been measured from its central or western
spot; all separations for Nov 30. The central part ({\it a b}) on Nov 30
is probably identical to the double spot mentioned for Nov 24 to be 
1'26'' from the eastern limb.}
\label{1708_11_30_kirch_own}
\end{figure}

G. Kirch describes this observation in letter no. 891 to G.W. Leibniz on 1708 Dec 1 (Herbst 2006),
also discussing his consideration about the nature of sunspots.
Attached to the letter is a drawing (Fig. \ref{1708_11_30_kirch_own})
showing the spot group on Nov 30 and Dec 1:
\begin{quote}
Am 24 Nov. sahen wir etwas davon zum ersten mahl, 
am Ost- oder Eintrits-Rande der Sonnen, 
dabey war auch eine helle Facula, 
dergleichen ich an Helligkeit noch nie gesehen \dots 
Nun die Sonne sich gestern wieder sehen lie\ss{}, 
fand man anstatt der zweyen \dots eine gantze Menge Maculen \dots
Jedoch auch ziemlich deutlich, und fein weit von ein ander \dots 
Weil nun die Maculen, wie ich g\"antzlich dar vor halte, 
ein Rauch seyn, von einem neuen, in der Sonnen entstandenen, Brande; 
Wie solches auch die Faculen, (welche gemeiniglich nur am Rande der Sonnen 
zu sehen seyn) bezeugen: Als k\"onte man solches billich ein Himmlisches Feuer-Werck nennen.
\end{quote}

\begin{quote}
On [1708] Nov 24 we saw it for the first time, near the eastern ingress limb. 
It also had a bright facula, which was as bright as I have never seen before \dots 
Yesterday [Nov 30] the Sun was visible again, 
one found not two \dots but a whole lot of spots \dots
But they were very clear and nicely separated from each other \dots 
Because the spots, as I fully support, are a smoke from a new fire forming in the Sun, 
as also the faculae attest it (which are commonly seen only on the Sun's limb); 
one could well call this a heavenly firework.
\end{quote}

In letter no. 892 (Herbst 2006) from 1708 Dec 19, G. Kirch also mentions the spot 
he saw on 1708 Nov 24. 
After some overcast days, he was able to observe that spot again and wrote to Wurzelbaur:
\begin{quote}
Den 24 Novemb. um 12 zu Mittage, fand ich durch einen 7 sch\"uhigen Tubum, 
am Ost-Rande der Sonnen, eine kleine Macul, die war doppelt, 
und hatte eine sch\"one helle facul. 
Durch den 10 schuhigen Tubum war sie 10 partes micrometri vom Ost-Rande, 
ist 1'. 26''. oder 0 Dig. 32'. 
Hier auff war in etlichen Tagen kein Wetter zum observiren.  
Am 30 Novemb. aber funden wir die Maculen in gro\ss{}er Menge, 
wie beyliegende Figur ausweiset. 
h war vom West-Rande der Sonnen 70 partes micrometri ist 10'. 0''. 
h von f 24 partes micrometri ist 3'.26''. 
a b vom West-Rande 80 partes micrometri ist 11'. 26''. 
e g f vom West-Rande 90 partes micrometri ist 12'. 51''. 
Der Diameter Solis war 230 partes micrometri ist 32'. 51''. 
Den 1 Decemb. waren die Maculen, wie mich deuchtet, 
kleiner worden, und den 2 Dec. sahe man, bey st\"urmischer und unreiner Lufft 
die 2 Haubt-Maculen, auch bisweilen noch die dritte zwischen den beyden, oben, aber man konte nichts me\ss{}en.
\end{quote}

\begin{quote}
On Nov 24 at 12 o'clock at noon I found one [spot] through a 7-foot telescope, 
near the eastern limb of the Sun. It was a small, doubled spot with a nice and bright facula. 
With a 10-foot tube it was 10 p.m. from the eastern limb, means 1' 26'' [arc min/sec] or 0 inches 32'. 
The following days the weather was dull and not good enough for observing. 
On Nov 30 we found a large number of sunspots, as the attached figure shows 
[unrecorded]. 
h was 70 p.m., means 10' from the western limb. h from f 24 p.m., means 3' 26''. 
a b from the western limb 80 p.m., means 11' 26''. e g f from the western limb 90 p.m., 
means 12' 51''. The Sun's diameter was 230 p.m., means 32' 51''. 
On Dec 1 it was, as it seemed, smaller and on Dec 2 one could see, with the stormy 
and unclean air, the two main spots and occasionally the third above between them, but could not measure.
\end{quote}
G. Kirch  then proceeds to describe a variable star in Cetus, probably Mira itself.

\begin{figure}
\begin{center}
\includegraphics[width=0.485\textwidth]{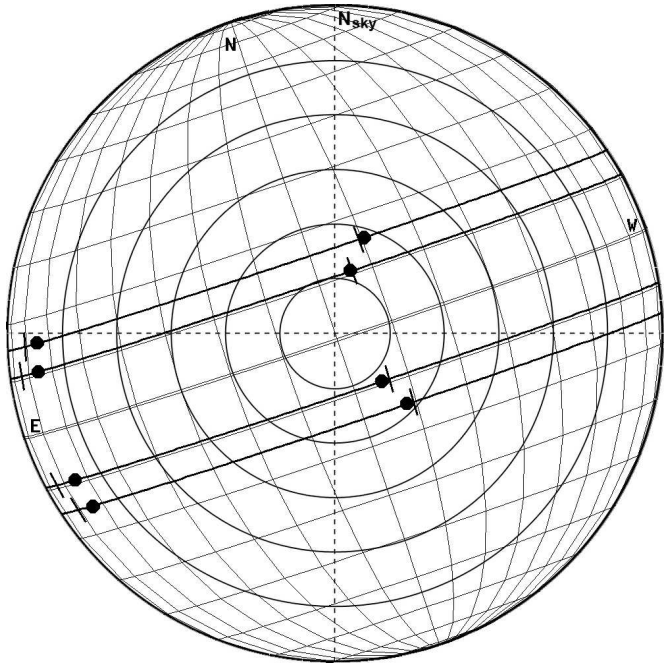}
\end{center}
\caption{Reconstruction of the sunspot location seen on 1708 Nov~24 and~30 by
G. Kirch. The reconstruction is for spots h and a+b.
The tilt angle of the Sun was $B_0=1.21^{\circ}$ and $0.45^{\circ}$, respectively.}
\label{1708_11_24_1203_tip}
\end{figure}

We find the following information in the letter by G.~Kirch:
\begin{itemize}
\item 1708 Nov~24, noon: $r=0.9130$ to the east
\item 1708 Nov~30 (assuming noon): $r=0.2174$, spot g 
\item 1708 Nov~30 (assuming noon): $r=0.3043$, spot ab 
\item 1708 Nov~30 (assuming noon): $r=0.3913$, spot h
\end{itemize}
The solution for combining Nov~24 with spot g on Nov~30 delivers
the two latitudes $-9.4\pm8.2^{\circ}$ and $10.6\pm8.2^{\circ}$, the combination
with (double) spot a+b delivers $-14.7\pm9.4^{\circ}$ and $15.5\pm9.4^{\circ}$,
and the combination with spot h delivers $-20.0\pm7.9^{\circ}$ and
$20.8\pm7.1^{\circ}$. The reconstruction with spot g, which is the
westernmost and the one giving the lowest latitudes, is shown
in Fig.~\ref{1708_11_24_1203_tip}.

In Neuh\"auser et al. (2015), another observation is given for three spots (or groups) on 1708 Dec 1,
namely by Peter Becker from Rostock, Germany, which is not listed in HS98.
Becker specified: 
\begin{quote}
\dots on the 1st of December of last year [1708] at 12 o'clock, 
there were -- again on the southern side close to the ecliptic
between the three and four twelfths from the edge -- three small spots seen, two of them on a straight line,
separated by about one twelfth in solar latitude, / the third and smallest, however, was a bit closer and
above the spot, / which stood closer to the edge.
\end{quote}
From this description, Neuh\"auser et al. (2015) derived the heliographic latitudes
to be $b = -10.6 \pm 3.4^{\circ}$, $-11.9 \pm 3.9^{\circ}$, and $-8.4 \pm 3.7^{\circ}$, see their figure 5 and table 2.
These three values for Becker are consistent with our constraint for Kirch in Fig. \ref{1708_11_24_1203_tip},
but definitely on the southern hemisphere according to Becker.

It is quite well possible that those three {\it spots} seen by Becker were part
of the large group seen and drawn by G. Kirch, namely the left-most and right-most parts in
Fig. \ref{1708_11_30_kirch_own} (bottom for Dec 1 like Becker): 
in the left-most part, Kirch saw three spots, but Becker only the largest one,
and for the right-most part in the drawing by Kirch, Becker saw the larger spot towards the south-west
and the smaller one to its upper left; then, it is possible that the small, central spot in
the drawing by Kirch (for Dec 1) was not visible to Becker;
Kirch used a 10-foot telescope tube (Dec 1708), Becker 4- and 7-foot telescopes (given for Jan 1709).
Kirch gave 10' to 12'55'' as separation range of those two parts
from the western limb for Nov 30, which is well consistent with
Becker giving $\sim$ 8' to 10.5' ({\it three and four twelfth from the edge}) 
for Dec 1.\footnote{While Neuh\"auser et al. (2015) assumed that Becker was referring
to the western edge here, as a few sentences earlier in his text for 1708 Sep,
we can confirm this assumption here by comparison with Kirch. This would then also
confirm that Kirch indeed has drawn the spots with west to the right and south to the bottom.
If the smallest spot seen by Becker would be the central spot seen on
Dec 1 by Kirch, then it would not be to the upper left of the right-most one,
as given by Becker.}

The smallest spot drawn by Becker for an observation (of new spots) one month later has 
a size of 25 millionth of a solar hemisphere (28 MSH after correction of foreshortening,
Neuh\"auser et al. 2015), so that we can assume this value
as upper limit for spots seen by Kirch, but not by Becker
(Becker for Jan 1709; {\it can pretty easy and clearly recognize them},
while for Dec 1708: {\it three small spots}).

For an observation on Nov 30-Dec 1, Sp\"orer's catalogue gives a heliographic latitude 
of $-10^\circ$ observed by Wideburg, probably one of the spots seen by G. Kirch and Becker,
whereas the RNR93 diagram do provide two data for Nov 1708 at $-8.2$ and $-4.3^{\circ}$.

In HS98 matching observations are: 
La Hire (Nov 14, 15, 17, 18 and Dec 1) one group (spotless Nov 24),
Wideburg (Nov 14-21, 26, 30, Dec 1) one group and Nov 24 two groups,
Cassini (Nov 12-18) and Muller (Nov 17, 18, 24-30) one group each, and
G. Kirch (Nov 24, Nov 29-Dec 2) also one group. 
Interestingly, the dates given in HS98 for G. Kirch are almost as found by us
in his two letters, the only additional day in HS98 is Nov 29.
When Wideburg reported two groups on Nov 24, while all other report one group
(but consisting of many spots as seen in the letter and drawing by Kirch),
it may well be that Wideburg reported two spots for Nov 24 (but in one group).
The total range from Nov 12 (Cassini) to Dec 2 (Kirch) is too long for any one spot:
the spot seen by La Hire, Wideburg, and Cassini (on Nov 12 and 14) cannot
have been seen any more by Kirch on Nov 30 and Dec 1. 

The HS98 data base includes
errors in dates and group numbers as shown not only in this work
(see also Svalgaard \& Schatten 2016, Neuh\"auser et al. 2015, and
Neuh\"auser \& Neuh\"auser 2016).

\subsection{1709 Jan 6-11}
\label{1709jan}

The first sunspot in 1709 is described by J.P. Wurzelbaur 
in his letter from 1709 Apr 
(no day given, but Kirch notes to have received the letter on Apr 15)
to G. Kirch (No. 893, Herbst 2006):
\begin{quote}
Die am 6 Januarii nach etlichen tr\"uben und regenerischen tagen 
albereit \"uber die helffte disci solaris auancirte Sonnen Maculam 
so am 7:$^{\rm ten}$ etwan noch 3 digitos a Limbo occidentali abgestanden, 
folgenden tage aber wegen eingefallener strengen K\"alte 
und dahero entstandener dicken D\"unste nicht hat betrachtet werden k\"onnen, 
am 11:$^{\rm ten}$ ejusdem nachmittag aber in der Sonne nicht ferner anzutreffen gewesen \dots
\end{quote} 

\begin{quote}
After many dull and rainy days a spot was seen on [1709] Jan 6
already past the centre of the solar disc. 
On Jan 7 it stood ca. 3 inches from the western limb, 
the following days it got bitterly cold and thick air 
prohibiting an observation. On Jan 11 in the afternoon it was no longer visible \dots
\end{quote}

G. Kirch also mentioned those spots in a short note 
he wrote under Wurzelbaur's letter 
\begin{quote}
Den 10 Jul habe geantwortet und Maculen von 6 Jan bis 6 Febr. geschickt \dots
\end{quote}

\begin{quote}
I responded on [1709] July 10 and sent sunspots from Jan 6 until Feb 6 \dots
\end{quote}
When G. Kirch writes {\it from Jan 6 until Feb 6}, he does not necessarily
mean {\it all} days in this period, but at least the first and last date.

\begin{figure}
\begin{center}
\includegraphics[width=0.485\textwidth]{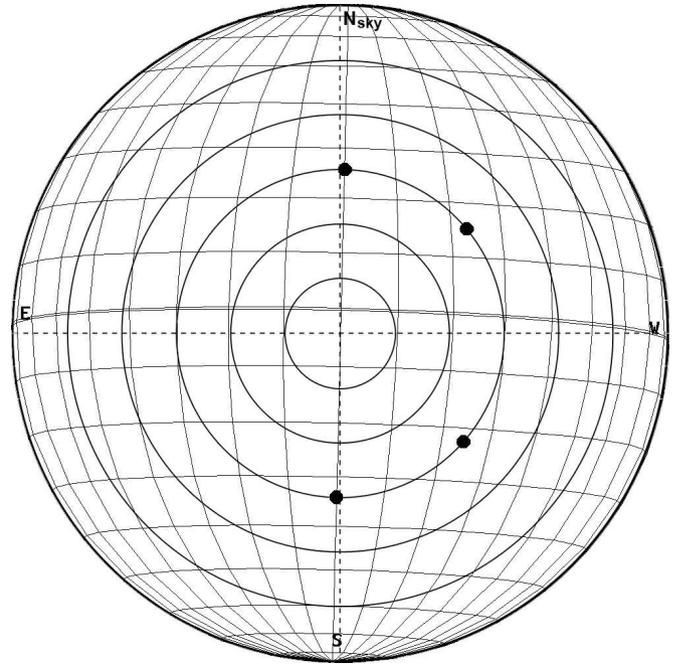}
\end{center}
\caption{Reconstruction of the possible sunspot location seen on 1709 Jan~7
according to the single given separation from the western limb. The
tilt angle of the Sun was $B_0=-4.16^{\circ}$.}
\label{1709_01_07_1200_tip}
\end{figure}

Three inches from any limb on 1709 Jan~7 results in a heliographic
latitude of $-34.0^{\circ} < b < +25.6^{\circ}$. The mentioned proximity from the
western limb indicates a much lower latitude though. Within the
sector on the Sun, defined by the points where the $\pm40^{\circ}$-latitude
lines touch the solar limb, the limiting latitudes are
$-22.5^{\circ} < b < +15.1^{\circ}$ (see Fig.~\ref{1709_01_07_1200_tip}).

Nearly the same date range is given in the Sp\"orer cataloge (Jan 6-10) 
with a heliographic latitude of $-16^\circ$ observed by G. Kirch and Wideburg. 
The diagram by RNR93 gives two data points in 1709 Jan with latitudes of $-6.4$ and $-5.3^\circ$.
HS98 list the observation of one group for La Hire (Jan 6, 7, 10), 
Wideburg (Jan 6-10), 
Wolf is listed for two groups on Jan 6,
Muller (Jan 6-9), and
G. Kirch (Jan 6, 7, 9, 10, then spotless on Jan 11); 
The {\it two groups} listed for Wolf for Jan 6 by HS98 are most certainly
the two spots drawn by Becker for that day, forming only one group.
Wurzelbaur is missing in HS98.
Derham is listed (HS98) for Jan 15, 21, and 22, probably still Julian,
as he is also listed for one group for 1708 Dec 26, which could be 1709 Jan 6 Gregorian.

In Neuh\"auser et al. (2015) the observations by Becker with drawings for 1709 Jan 5-9
are presented, which are missing in HS98. For the two spots as drawn by Becker
(on the western hemisphere near but past the disk centre, as described by Wurzelbaur), 
they derive heliographic latitudes from $-9^{\circ}$ to $-15^{\circ}$, consistent with Sp\"orer
and RNR93, and also with our constraint from Kirch.
Becker saw two spots in one group Jan 5-7 and then one of the two spots only on Jan 8 \& 9;
he also reported a spotless Sun for Jan 3 and 10.
From the fact that Kirch and La Hire saw one spot on Jan 10, while Becker reported a spotless Sun,
we can conclude that the former observers had better equipment -- and that the spot on Jan 10
had an (uncorrected) area smaller than 25 MSH 
(by comparison to Becker's spot drawing in Neuh\"auser et al. 2015).
We also note that 1709 Jan 11 was explicitly given as spotless by Wurzelbaur
(the spots could have rotated out of view).

\subsection{1709 Jan 29-Feb 6}
\label{last_two}

In letter no. 893 dated (early) April 1709 (see previous subsection for the dating), 
J.P. Wurzelbaur
lists all the sunspots, he had observed throughout the year 1709,
including the spot seen until early Feb which was already mentioned in the previous subsection:

\begin{quote}
am 29:$^{ten}$ aber nach etlichen tr\"uben tagen 
annoch allerdings in voriger gestalt und gr\"osse albereit 2 digitos redux widerumb einger\"ucket, 
und am 31:$^{ten}$ das mittel noch nicht erreicht den 1:$^{ten}$ Februarii 
aber nachmittags fast einen digitum \"uberschritten hatte, 
folgenden tages aber in unver\"anderter Gr\"o\ss{}e und soliditet forgefahren, 
zugleich auch eine neue macula welche einen rauchigen Anfang nach sich zoge, 
und gestern noch nicht zu sehen war, 3 digitos circiter von dem Ostlichen Rande entstanden, 
folgends aber wegen 14 t\"age eingefallenem tr\"uben Wetters 
bey uns weiter nich haben observirt werden k\"onnen \dots
\end{quote}

\begin{quote}
On [1709 Jan] 29 after many dull days it [spot] was already 2 inches in the Sun's disc. 
On [Jan] 31 it was not quite in the centre [of the Sun] 
and on Feb 1 in the afternoon it was almost 1 inch past the centre. 
The next day it proceeded with unchanged size 
and a new spot with a smoky beginning was behind it
that was not seen the day before, it formed 3 inches from the eastern limb, 
but could not be observed for the next 14 days because we got bad weather \dots 
\end{quote}

Some of the dates from Wurzelbaur are
within the date range given by G. Kirch in the short note 
partly cited in the previous section, namely letter no. 893 dated early Apr 1709:
\begin{quote}
I responded on July 10 and sent sunspots from Jan 6 until Feb 6,
the solar eclipse on 1709 March 11 [Greg.] as well.
\end{quote}
When G. Kirch writes {\it from Jan 6 until Feb 6}, he does not necessarily
mean {\it all} days in this period, only the first and last dates are certain active days.
He also mentioned the observation of the solar eclipse 1709 March 11, which can
most certainly be considered a spotless day, because no spots are mentioned
(according to HS98, La Hire and Feuillee reported a spotless day for Mar 11).

\begin{figure}
\begin{center}
\includegraphics[width=0.485\textwidth]{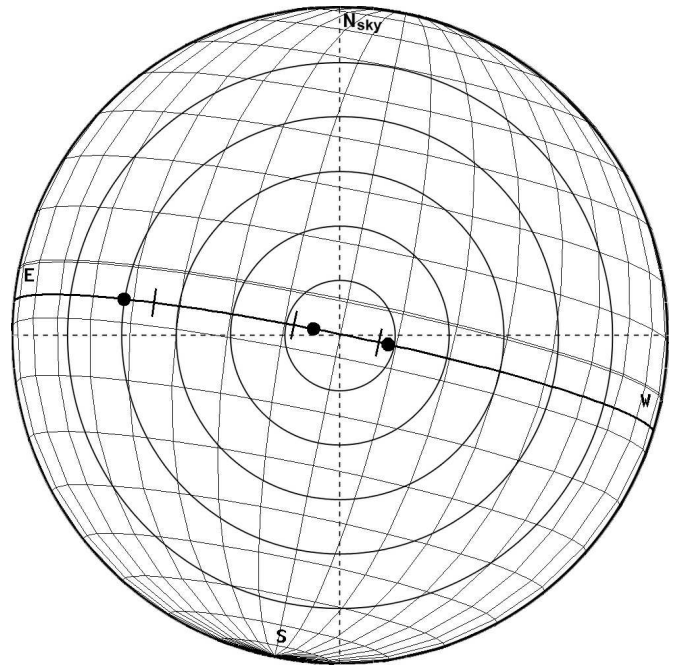}
\end{center}
\caption{Reconstruction of the sunspot location seen on 1709 Jan~29, 31, and Feb~1,
when $B_0=-6.11^{\circ}$, $-6.24^{\circ}$, and $-6.31^{\circ}$, respectively.}
\label{1709_01_29_1200_tip}
\end{figure}

We adopt the following positions for the reconstruction:
\begin{itemize}
\item 1709 Jan 29 (assuming noon): 4"
\item 1709 Jan 31 (assuming noon): 0.5"
\item 1709 Feb 1 (assuming 3 pm): 0.9"
\end{itemize}
The resulting latitude fitting those measurements best is
$-6.0\pm4.9^{\circ}$. The corresponding reconstruction is shown in
Fig.~\ref{1709_01_29_1200_tip}.

G. Sp\"orer's catalogue gives a heliographic latitude of $-11^\circ$ 
for an observation by Wideburg on Jan 26-Feb 6.
The diagram in RNR93 also contains one data point with  a heliographic latitude
of $-10.1^\circ$ at early Feb 1709.
These data are roughly consistent with our constraint for the first spot mentioned by Wurzelbaur.

HS98 list the observation of one sunspot group for La Hire (Jan 26, 28, 29, 30), 
Wideburg (Jan 26-Feb 2, 4-6), 
Derham (Jan 15, 21, 22, and Jan 26-Feb 6, maybe Julian), Feuillee (Jan 31, Feb 5), 
and G. Kirch (Jan 26-31, Feb 3, 5, 6). 
The observation of 2 groups is also given for La Hire and Wideburg for Feb 3, 
whereas Feuillee observed even 3 groups Feb 1-4 (according to HS98). 

When Wurzelbaur wrote {\it The next day [Feb 2] ... a new spot with a smoky beginning was 
behind it [the other spot] that was not seen the day before, it formed 3 inches from the eastern limb},
he must mean a 2nd group, because the other one was already past the solar centre.

\subsection{1709 Aug 23-29}
\label{very_last}

The very last observation is about several sunspots observed by J.P. Wurzelbaur 
since 1709 Aug 23. 
In the previously cited letter no. 895 (Herbst 2006) from Wurzelbaur
dated 1709 Sept, he writes to G. Kirch:
\begin{quote}
\dots und ist es seithero in der Sonne ruhig gewest bis in den abgewichenen Augustum, 
da ich am 23 ejusdem nach etlichen tr\"uben tagen 4 maculas worunter eine sonderlich schwarz, 
albereit 1 digitum \"uber das centrum der Sonne auancirt angetroffen, 
die sich folgende tag wenig ver\"andert, am 25:$^{ten}$ aber vormittag in 5, 
nachmittag in 6 stuck zerstreuet, den 26:$^{ten}$ aber fr\"uhe 
umb 6 Uhr widerumb in 2 gedoppelte conjungirt hatten, worvon am 27:$^{ten}$ 
nur Eine etwan 1 digitum vom Westrand abstehend, nach dem tr\"uben 28:$^{ten}$ aber, 
am 29:$^{ten}$ zwar durch den Nebel weiter nichts in der Sonne zu ersehen gewesen.
\end{quote}

\begin{quote}
\dots and it was quiet in the Sun since then until the last August, 
when I saw on the 23rd after many dull days 4 spots, 
one of which was remarkably black and nearly 1 inch past the centre of the Sun, 
also on the next day only slightly changed. 
On [Aug] 25 before noon [there were] five [spots], in the afternoon six pieces distributed, 
on [Aug] 26 at 6 o'clock in the morning they conjugated into two double [spots], 
one of them was nearly 1 inch from the western limb on Aug 27. 
After the dull [Aug] 28th, on [Aug] 29th, [observing] through fog, and the Sun was clear.
\end{quote}

In summary, there were spots 1709 Aug 23 to 27 and a spotless Sun on Aug 29.

\begin{figure}
\begin{center}
\includegraphics[width=0.485\textwidth]{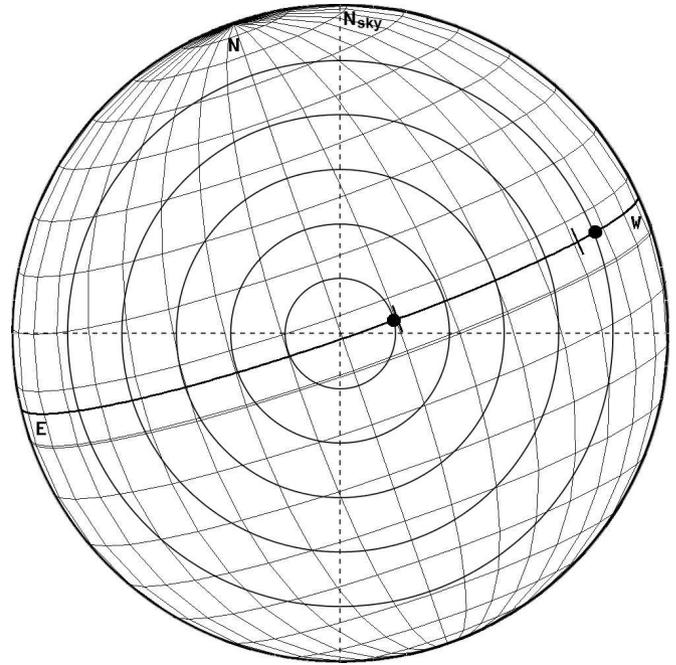}
\end{center}
\caption{Reconstruction of the sunspot location seen on 1709 Aug~23 and 26,
when $B_0=7.09^{\circ}$ and $7.15^{\circ}$, respectively.}
\label{1709_08_23_0600_tip}
\end{figure}

We interpret the measurements as follows:
\begin{itemize}
\item 1709 Aug 23 (assuming noon): 1"
\item 1709 Aug 26, 6 am: 5"
\end{itemize}
These result into a latitude of $6.0\pm6.8^{\circ}$. 
If we would also use 6 am for Aug 23, the solution would be $6.1 \pm 7.1^{\circ}$,
i.e. not really different;
Fig.~\ref{1709_08_23_0600_tip} shows the corresponding reconstruction.

G. Sp\"orer's catalogue lists an observation for Aug 25-27 for $b=-7^\circ$.
The RNR93 diagram does not give any point in 1709 Aug.

HS98 list one sunspot group for La Hire (Aug 25, 27, but spotless Aug 28-30), 
G. Kirch (Aug 21-27, then spotless Aug 28 to Sep 1),
Derham (Aug 13, probably Julian), Lalande (Aug 25-27),
and  Muller (Aug 24, 25, 27, then spotless Aug 30 and 31),

\onecolumn
\setcounter{table}{0}
\captionof{table}{All reconstructed sunspots from the Kirch letters (Herbst 2006)
with Gregorian dates (days for which we could reconstruct the latitude b), observer, our section, spot latitude b $[^\circ]$, 
number of groups (and number of spots in brackets if given) according to Kirch and his letter partners, 
other observers in Hoyt \& Schatten (1998, HS98), 
latitude b $[^\circ]$ in Ribes \& Nesme-Ribes (1993, RNR),
and in Sp\"orer (1889).}
\label{tab_reconstruct}

\begin{tabularx}{\textwidth}{|p{1.8cm}|p{1.7cm}|p{0.6cm}|p{1.9cm}|p{1.4cm}|p{4.4cm}|p{0.8cm}|p{1.8cm}|} \hline
obs. date & observer & Sec & latitude b {[}$^\circ${]} & \# grps (\# spts) &
\multirow{2}{*}{\begin{tabular}[c]{@{}l@{}}HS98 observers:\\ max. \# groups\end{tabular}}                  &
\multirow{2}{*}{\begin{tabular}[c]{@{}l@{}}RNR \\ b {[}$^\circ${]} \end{tabular}} &
\multirow{2}{*}{\begin{tabular}[c]{@{}l@{}}\normalsize Sp\"orer b {[}$^\circ${]}\\ \footnotesize(observer)\end{tabular}} \\ \hline

\begin{tabular}[c]{@{}l@{}}1680\\May 20+22\end{tabular} & \begin{tabular}[c]{@{}l@{}}G. Kirch\\Ihle\end{tabular}      & 3.1 & $-0.6 \pm 13.2$ & 1 (3) & G. Kirch: 1, Cassini: 1 & & \\ \hline
\begin{tabular}[c]{@{}l@{}}1680\\Jun 15-23\end{tabular} & \begin{tabular}[c]{@{}l@{}}G. Kirch\end{tabular} & 3.2 & \begin{tabular}[c]{@{}l@{}}$-8.8 \pm 7.0$\\$-9.6 \pm 6.8$\end{tabular} & 1-2 & G. Kirch: 1, Cassini: 1 & & \\ \hline
\begin{tabular}[c]{@{}l@{}}1684\\May 6\end{tabular}   & \begin{tabular}[c]{@{}l@{}}G. Kirch\\Ihle\end{tabular}      & 3.4 & $-20.6 \pm 3.5$ & 1  & \begin{tabular}[c]{@{}l@{}}La Hire: 1-2, Flamsteed \\ G. Kirch, Cassini, Clau-\\ sen, Cassini, Ettmuller: 1\end{tabular} &  & (a)    \\ \hline

\begin{tabular}[c]{@{}l@{}}1684\\Jul 5+6\end{tabular} & \begin{tabular}[c]{@{}l@{}}G. Kirch\\Schultz\end{tabular}   & 3.5 & $-20.4 \pm 13.6$ & 1 & \begin{tabular}[c]{@{}l@{}}La Hire, Flamsteed, Caswell,\\ Cassini, G. Kirch, Eimmart,\\ Hevelius, Glielmini: 1\end{tabular} &  & -10.8 (Cassini)         \\ \hline

\begin{tabular}[c]{@{}l@{}}1686 Apr 25\\ to May 1\end{tabular} & G. Kirch       & 3.6 & $-13 \pm 5$ or $4 \pm 5$ & 1 & \begin{tabular}[c]{@{}l@{}}La Hire: 3\\ Cassini, G. Kirch: 1\end{tabular} & & -15 (La Hire)                       \\ \hline

\begin{tabular}[c]{@{}l@{}}1688\\Dec 14+15\end{tabular}    & G. Kirch & 3.7 & $-10.5 \pm 6.0$ & 1 (2) &   &       &                                                 \\ \hline

\begin{tabular}[c]{@{}l@{}}1700\\Nov 7-13\end{tabular}     & Wurzelbaur   & 3.9 & $3.2 \pm 7.2$ & 1-2 (3) & \begin{tabular}[c]{@{}l@{}}La Hire: 2\\ Cassini, Wurzelbaur:                                                 1\end{tabular}  &     & \begin{tabular}[c]{@{}l@{}}-9.5 (La Hire,\\ Cassini,\\ Wurzelbaur)\end{tabular} \\ \hline

\begin{tabular}[c]{@{}l@{}}1701\\Nov 3-9\end{tabular}  & Ihle         & 3.11 & $-12.4 \pm 8$ & 1 & La Hire, Cassini, Jartoux: 1  &  & \begin{tabular}[c]{@{}                           l@{}}-12\\(Cassini)\end{tabular}                                                         \\ \hline

\begin{tabular}[c]{@{}l@{}}1702\\Dec 22-28\end{tabular}  & Wurzelbaur   & 3.12  &\begin{tabular}[c]{@{}l@{}} $-20.0 \pm 6.8$ \\or $13.9 \pm 6.4$\end{tabular}  & 1 & La Hire, Cassini, Manfredi: 1                                                                                      &                   & \begin{tabular}[c]{@{}l@{}}-11 (La Hire\\ and Cassini)\end{tabular}        \\ \hline

\begin{tabular}[c]{@{}l@{}}1703 May\\24 to Jun 2\end{tabular}    & \begin{tabular}[c]{@{}l@{}}G. Kirch\\Hoffmann\\Ihle\end{tabular}  & 3.13  & $-0.7 \pm 3.5$  & 1-2 (8) & \begin{tabular}[c]{@{}l@{}}La Hire: 2, Stannyan,\\Manfredi, Eimmart, Cassini,\\ G. Kirch, Hoffmann: 1\end{tabular} &      & \begin{tabular}[c]{@{}l@{}}-2 (La                                  Hire,\\ Cassini)\end{tabular}                                                                   \\ \hline

\begin{tabular}[c]{@{}l@{}}1703\\Jul 8-15\end{tabular}  & \begin{tabular}[c]{@{}l@{}}G. Kirch\\Hoffmann\end{tabular}    & 3.15 & $7 \pm 21$  & 1 (3) & \begin{tabular}[c]{@{}l@{}}G. Kirch: 3, La Hire: 2
                        \\Manfredi, Eimmart, Cassini,\\ Derham, R{\o}mer, Stannyan: 1\end{tabular}               &    & \begin{tabular}[c]{@{}l@{}}-19\\(La Hire)\end{tabular}                             \\ \hline

\begin{tabular}[c]{@{}l@{}}1705\\Nov 5-12\end{tabular}   & G. Kirch         & 3.18  & $4.2 \pm 17.1$      & 1 & \begin{tabular}[c]{@{}l@{}}La Hire, Manfredi,                                               Derham,\\ Plantade, Kirch, Cassini: 1\end{tabular} &
               -4.2         & -3 (Derham)                                                                                          \\ \hline

1707 Jan 27   & G. Kirch          & 3.20  & $\sim -10$         & 1 & Manfredi, G. Kirch: 1 &  -10.1     &   \\ \hline

\begin{tabular}[c]{@{}l@{}}1707\\Mar 5-7\end{tabular}   & Sturm          & 3.21  & $-3.5 \pm 27.0$      & 1 (3+) & \begin{tabular}[c]{@{}l@{}}La Hire, Sturm, G.                            Kirch,\\ Plantade: 2, Manfredi, \\Derham, Muller, Lalande: 1\end{tabular}     &  -13.2 -8.7                                                                                                                                                                    & \begin{tabular}[c]{@{}l@{}}-6\\(Derham)\end{tabular} \\ \hline
\begin{tabular}[c]{@{}l@{}}1708\\Aug 10-18\end{tabular}   & G. Kirch          & 3.23  & $8.9 \pm 10.5$ & 1 & \begin{tabular}[c]{@{}l@{}}Wideburg: 2, Muller,\\ la Hire, Manfredi, G. Kirch,\\ Derham, Blanchini, Gray: 1\end{tabular}  &                                                                                                                                                         & \begin{tabular}[c]{@{}l@{}}-7 (Wied-\\enburg)\end{tabular}                                                                             \\ \hline

\end{tabularx}

\newpage

\begin{tabularx}{\textwidth}{|p{1.8cm}|p{1.7cm}|p{0.6cm}|p{1.9cm}|p{1.4cm}|p{4.4cm}|p{0.8cm}|p{1.8cm}|} \hline
obs. date & observer & Sec & latitude b {[}$^\circ${]} & \# grps (\# spts) &

\multirow{2}{*}{\begin{tabular}[c]{@{}l@{}}HS98 observers:\\ max. \# groups\end{tabular}}                  &
\multirow{2}{*}{\begin{tabular}[c]{@{}l@{}}RNR \\ b {[}$^\circ${]} \end{tabular}} &
\multirow{2}{*}{\begin{tabular}[c]{@{}l@{}}\normalsize Sp\"orer b {[}$^\circ${]}\\ \footnotesize(observer)\end{tabular}} \\ \hline

\begin{tabular}[c]{@{}l@{}}1708\\ Sep 14\end{tabular}   & G. Kirch & 3.24 & $\sim -2$ (b) & 1 & La Hire, Muller: 1 & & \begin{tabular}[c]{@{}l@{}}-5 (Wied-\\enburg)\end{tabular} \\ \hline

\begin{tabular}[c]{@{}l@{}}1709 Jan 29\\to Feb 1\end{tabular}   & Wurzelbaur   & 3.27  & $-6.0 \pm 4.9$ & 1 & \begin{tabular}[c]{@{}l@{}}Feuillee: 3\\ La Hire,                                       Wideburg: 2\\ G. Kirch, Derham, Wolf: 1\end{tabular}     & -10.1 & \begin{tabular}[c]{@{}l@{}}-11 (Wied-\\enburg, Kirch\\Derham)\end{tabular}                                                          \\ \hline

\begin{tabular}[c]{@{}l@{}}1709 \\ Aug 23-26\end{tabular}     & Wurzelbaur   & 3.28 & $6.0 \pm 6.8$ & 1 (6)  & \begin{tabular}[c]{@{}l@{}}La Hire, G. Kirch, Derham,                               \\Lalande, Muller: 1\end{tabular}                 &     &  -7                                                            \\ \hline
\end{tabularx}

\smallskip

Remarks: (a) Different spot(s) listed in RNR93 and Sp\"orer ($-11^{\circ}$ for Cassini).
(b) A similar value ($-4.5 \pm 1.5^{\circ}$) was already given for this spot in Neuh\"auser et al. (2015).

\newpage

\begin{table*}
\caption{We list all differences between the information on number of sunspot groups
in the letters studied here and the tables of Hoyt \& Schatten (1998, HS98).
All dates given are Gregorian; we always mean Gottfied Kirch, if the observer is listed as Kirch
(unless otherwise specified);
{\it n/a} means {\it not available}, i.e. no observations known to HS98.
In addition to the differences listed here,
data of the British observers Caswell, Derham, Stannyan, Gray, and Sharp
as listed in HS98 are Julian dates,
not corrected to the new style (10 or 11 day shift), maybe also for Eimmart 1684 and Arnold 1688.
In addition, data on spotlessness from Flamsteed are probably not correct, see Sect. 3.2.
There are also incorrectly interpreted generic statements by Siverius, Wurzelbaur, Ihle, and Angerholm in HS98:
they probably did not observe on each and every day listed as {\it spotless}.}
\begin{tabular}{l|l|lll} \hline
Date       & Groups     & Groups in & Spots in & Sect. \\
dd:mm:yyyy & in HS98    & letters   & letters  & or else \\  \hline
20:05:1680 & Ihle: 2    & Kirch: 1  & Kirch: 2 & 3.1 (a) \\
28:05:1680 & Kirch: n/a & Kirch: 1  & Kirch: 1 & 3.1 \\ \hline
18:06:1680 & Kirch: 1   & Kirch: 2  & Kirch: 2 & 3.2 \\
19:06:1680 & Kirch: 1   & Kirch: 2  & Kirch: 2 & 3.2 \\
20:06:1680 & Kirch: 1   & Kirch: 2  & Kirch: 2 & 3.2 \\
23:06:1680 & Kirch: n/a & Kirch: 1  & Kirch: 1 & 3.2 \\  \hline
21:07:1681 & Kich: n/a  & Kirch: 0 & & Table 3 \\
23:07:1681 & Kirch: n/a & Kirch: 1  & Kirch: 1 & 3.3 \\
28:07:1681 & Kirch: n/a & Kirch: 1  & Kirch: 1 & 3.3 \\
31:07:1681 & Kirch: n/a & Kirch: 0  & & 3.3 \\ \hline
05:05:1684 & La Hire: 2 & Ihle: 0   & & 3.4 (b) \\
05:05:1684 & Ihle: n/a  & Ihle: 0   & & 3.4 (b) \\
06:05:1684 & Ihle: 1    & Ihle: 1   & & 3.4 (b) \\ 
06:05:1684 & Schultz: n/a & Schultz: 0 & & Table 4 \\ \hline
05:07:1684 & Schultz: n/a & Schultz: 1 & Schultz: 1 & 3.5 \\
06:07:1684 & Schultz: n/a & Schultz: 1 & Schultz: 1 & 3.5 \\
06:07:1684 & Kirch: 0     & Kirch: 1   & Kirch: 1   & 3.5 (d) \\
07:07:1684 & Schultz: n/a & Schultz: 1 & Schultz: 1 & 3.5 \\ \hline
14:12:1688 & Kirch: n/a & Kirch: 1   & Kirch: 2   & 3.7 \\
15:12:1688 & Kirch: n/a & Kirch: 1   & Kirch: 2   & 3.7 \\
18:12:1688 & Kirch: n/a & Kirch: 0   &   & 3.7 \\
19:12:1688 & Kirch: n/a & Kirch: 0   &   & 3.7 \\ \hline
%
%
10:06:1700 & Wurzelbaur: n/a & Wurzelbaur: 1 & Wurzelbaur: few & 3.8 \\
11:06:1700 & Wurzelbaur: n/a & Wurzelbaur: 1 & Wurzelbaur: few & 3.8 \\
12:06:1700 & Wurzelbaur: n/a & Wurzelbaur: 1 & Wurzelbaur: few & 3.8 \\
13:06:1700 & Wurzelbaur: n/a & Wurzelbaur: 1 & Wurzelbaur: few & 3.8 \\ \hline
06:11:1700 & Wurzelbaur: 1   & Wurzelbaur: 1 or 2 & Wurzelbaur: 2 & 3.9 \\
07:11:1700 & Wurzelbaur: 1   & Wurzelbaur: 1 or 2 & Wurzelbaur: 2 & 3.9 \\
09:11:1700 & Wurzelbaur: 1   & Wurzelbaur: 1 or 2 & Wurzelbaur: 3 & 3.9 \\
10:11:1700 & Wurzelbaur: 1   & Wurzelbaur: 1 or 2 & Wurzelbaur: 3 & 3.9 \\
11:11:1700 & Wurzelbaur: 1   & Wurzelbaur: 1 or 2 & Wurzelbaur: 2 & 3.9 \\
12:11:1700 & Wurzelbaur: 1   & \multicolumn{2}{l}{Wurzelbaur: bad weather} & 3.9 \\
13:11:1700 & Wurzelbaur: 1   & Wurzelbaur: 1      & Wurzelbaur: 2 & 3.9 \\ \hline
27:11:1700 & Wurzelbaur: n/a & Wurzelbaur: 0 & & Table 5 \\ \hline
\end{tabular}
\end{table*}

\newpage

\begin{table*}
\begin{tabular}{l|l|lll} \hline
Date       & Groups     & Groups in & Spots in & Sect. \\
dd:mm:yyyy & in HS98    & letters   & letters  & or else \\  \hline
31:12:1700 & Kirch: 2        & Kirch: 1 or 2      & Kirch: 2      & 3.10 \\ \hline
03:11:1701 & Ihle: n/a       & Ihle: 1      & Ihle: 1 & 3.11 \\
07:11:1701 & Ihle: n/a       & Ihle: 1      & Ihle: 1 & 3.11 \\
09:11:1701 & Ihle: n/a       & Ihle: 1      & Ihle: 1 & 3.11 \\
10:11:1701 & Ihle: n/a       & Ihle: 0      &         & 3.11 \\ \hline
17:12:1702 & Ihle: n/a       & Ihle: 0      &         & Table 6 \\ \hline
31:12:1702 & Wurzelbaur: n/a & Wurzelbaur: 0 & & Table 5 \\
01:01:1703 & Wurzelbaur: n/a & Wurzelbaur: 0 & & Table 5 \\ \hline
11:01:1709 & Wurzelbaur: n/a & Wurzelbaur: 0 & & Table 5 \\ \hline
29:08:1709 & Wurzelbaur: n/a & Wurzelbaur: 0 & & Table 5 \\ \hline
17:12:1702 & Ihle: n/a       & Ihle: 0      &         & 3.12 \\
22:12:1702 & Ihle: n/a       & Ihle: 1      & Ihle: 1 & 3.12 \\
22:12:1702 & Wurzelbaur: n/a & Wurzelbaur: 1 & Wurzelbaur: 1 & 3.12 \\
23:12:1702 & Wurzelbaur: n/a & Wurzelbaur: 1 & Wurzelbaur: 1 & 3.12 \\
24:12:1702 & Ihle: n/a       & Ihle: 1      & Ihle: 1 & 3.12 \\
28:12:1702 & Wurzelbaur: n/a & Wurzelbaur: 1 & Wurzelbaur: 1 & 3.12 \\
31:12:1702 & Wurzelbaur: n/a & Wurzelbaur: 1 & Wurzelbaur: 1 & 3.12 \\
01:01:1703 & Wurzelbaur: n/a & Wurzelbaur: 0 &      & 3.12 \\ \hline
26:05:1703 & Ihle: n/a       & Ihle: 1       & Ihle: 1 & 3.13 \\
27:05:1703 & Ihle: n/a       & Ihle: 1       & Ihle: 1 & 3.13 (e) \\
04:06:1703 & Ihle: n/a       & Ihle: 1       & Ihle: 1 & 3.13 \\ \hline
09:07:1703 & Kirch: 3        & \multicolumn{2}{l}{Kirch: 1-3 spots in 1 group (f)} & 3.15 \\
10:07:1703 & Kirch: n/a      & Kirch: 1      & Kirch: 3 & 3.15 \\
11:07:1703 & Kirch: n/a      & Kirch: 1      & Kirch: 3 & 3.15 \\ \hline
21:03:1704 & Wurzelbaur: n/a  & Wurzelbaur: 1 & Wurzelbaur: 1 & 3.16 \\
22:03:1704 & Wurzelbaur: n/a  & Wurzelbaur: 1 & Wurzelbaur: 1 & 3.16 \\
23:03:1704 & Wurzelbaur: n/a  & Wurzelbaur: 1 & Wurzelbaur: 1 & 3.16 \\
24:03:1704 & Wurzelbaur: n/a  & Wurzelbaur: 1 & Wurzelbaur: 1 & 3.16 \\
23:04:1704 & Wurzelbaur: n/a  & Wurzelbaur: 1 & Wurzelbaur: 1 & 3.16 \\
24:04:1704 & Wurzelbaur: n/a  & Wurzelbaur: 1 & Wurzelbaur: 1 & 3.16 \\ \hline
16:10:1705 & Plantade: 2      & G.+M.M. Kirch: 1    & G.+M.M. Kirch: 2    & 3.17 (g) \\ 
16:10:1705 & M.M. Kirch: n/a  & M.M. Kirch: 1       & M.M. Kirch: 2       & 3.17 (b,g) \\ \hline
11:12:1706 & La Hire: 2       & G.,M.M.,C. Kirch: 1 & G.,M.M.,C. Kirch: 2 & 3.17 (g,h) \\
11:12:1706 & M.M. Kirch: n/a  & M.M. Kirch: 1       & M.M. Kirch: 2       & 3.17 (g,h) \\
11:12:1706 & C. Kirch: n/a    & C. Kirch: 1         & C. Kirch: 2         & 3.17 (g,h) \\
12:12:1706 & M.M. Kirch: n/a  & M.M. Kirch: 1       & M.M. Kirch: 2       & 3.17 (g,h) \\
13:12:1706 & M.M. Kirch: n/a  & M.M. Kirch: 1       & M.M. Kirch: 2       & 3.17 (g,h) \\ \hline
05:03:1707 & Sturm: n/a       & Sturm: 1            & Sturm: 2            & 3.21 \\
06:03:1707 & Plantade: 2      & Sturm: 1            & Sturm: 2            & 3.21 \\
06:03:1707 & Sturm: n/a       & Sturm: 1            & Sturm: 2            & 3.21 \\
06:03:1707 & Kirch: 2         & Sturm: 1            & Sturm: 2            & 3.21 \\
06:03:1707 & Plantade: 2      & Sturm: 1            & Sturm: 2            & 3.21 \\
07:03:1707 & Sturm: n/a       & Sturm: 1            & Sturm: 2            & 3.21 \\
08:03:1707 & Sturm: n/a       & Sturm: 0            &                     & 3.21 \\ \hline
23:03:1707 & Hertel: n/a      & Hertel: 2           & Hertel: several & 3.22 \\ \hline
\end{tabular}
\end{table*}

\newpage

\begin{table*}
\begin{tabular}{l|l|lll} \hline
Date       & Groups     & Groups in & Spots in & Sect. \\
dd:mm:yyyy & in HS98    & letters   & letters  & or else \\  \hline
08:08:1708 & Kirch: n/a       & Kirch: 0            &                  & 3.23 \\
10:08:1708 & Kirch: n/a       & Kirch: 1            & Kirch: 2         & 3.23 \\
11:08:1708 & Kirch: n/a       & Kirch: 1            & Kirch: 3         & 3.23 \\
11:08:1708 & Wurzelbaur: n/a  & Wurzelbaur: 1       & Wurzelbaur: 1    & 3.23 \\
18:08:1708 & Kirch: n/a       & Kirch: 1            & Kirch: 1         & 3.23 \\
11:08:1708 & Wurzelbaur: n/a  & Wurzelbaur: 1       & Wurzelbaur: 1    & 3.23 \\
02:09:1708 & Kirch: n/a       & Kirch: 1            & Kirch: 1         & 3.23 \\
03:09:1708 & Kirch: n/a       & Kirch: 1            & Kirch: 1         & 3.23 \\ 
03:09:1708 & Wurzelbaur: n/a  & Wurzelbaur: 1       & Wurzelbaur: 1    & 3.23 \\ 
03:09:1708 & Wideburg: 2      & Wurzelbaur/Kirch: 1 & Wurzelbaur/Kirch: 1 & 3.23 \& 3.24 \\ 
10:09:1708 & -                & Becker: 1           & Becker: 1        & 3.24 (c,i) \\
11:09:1708 & -                & Becker: 1           & Becker: 1        & 3.24 (c,i) \\
13:09:1708 & Kirch: n/a       & Kirch: 1            & Kirch: 1         & 3.24 \\
14:09:1708 & Kirch: n/a       & Kirch: 1            & Kirch: 1         & 3.24 (j) \\
14:09:1708 & Wurzelbaur: n/a  & Wurzelbaur: 1       & Wurzelbaur: 1    & 3.23 \& 3.24 (j) \\ \hline
%
%
06:01:1709 & -                & Becker: 1            & Becker: 2        & 3.26 (i) \\
06:01:1709 & Wolf: 2          & Becker: 1            & Becker: 2        & 3.26 (c,i) \\
06:01:1709 & Wurzelbaur: n/a  & Wurzelbaur: 1        & Wurzelbaur: 1    & 3.26     \\
07:01:1709 & Wurzelbaur: n/a  & Wurzelbaur: 1        & Wurzelbaur: 1    & 3.26     \\
11:01:1709 & Wurzelbaur: n/a  & Wurzelbaur: 0        &                  & 3.26     \\ \hline
29:01:1709 & Wurzelbaur: n/a  & Wurzelbaur: 1        & Wurzelbaur: 1    & 3.27     \\ 
31:01:1709 & Wurzelbaur: n/a  & Wurzelbaur: 1        & Wurzelbaur: 1    & 3.27     \\
01:02:1709 & Wurzelbaur: n/a  & Wurzelbaur: 1        & Wurzelbaur: 1    & 3.27     \\
02:02:1709 & Wurzelbaur: n/a  & Wurzelbaur: 2        & Wurzelbaur: 1+1  & 3.27     \\ \hline
23:08:1709 & Wurzelbaur: n/a  & Wurzelbaur: 1        & Wurzelbaur: 4    & 3.28     \\
24:08:1709 & Wurzelbaur: n/a  & Wurzelbaur: 1        & Wurzelbaur: 4    & 3.28     \\
25:08:1709 & Wurzelbaur: n/a  & Wurzelbaur: 1        & Wurzelbaur: 5-6  & 3.28     \\
26:08:1709 & Wurzelbaur: n/a  & Wurzelbaur: 1        & Wurzelbaur: 4    & 3.28     \\
28:08:1709 & Wurzelbaur: n/a  & Wurzelbaur: 0        &                  & 3.28     \\ \hline
\end{tabular}

\smallskip

Remarks: (a) We have no direct information from the observer listed here
for HS98, so that it is uncertain whether there is a discrepancy to the Kirch letters:
the observer listed here for HS98 could have seen a different number of groups
(but this in unlikely). 
(b) According to G. Kirch.
(c) Observer P. Becker is not listed in HS98.
(d) The detection of one group from 1684 June 18-31 as listed by HS98 for Kirch
is probably based on Julian dates, not corrected to the new style (see Sect. 3.5),
similar in 1684 in HS98 for Eimmart and Caswell.
(e) 8 spots in one group on this date according to Hoffmann (Sect. 3.13, Fig. 15).
(f) Kirch: 1 spot on July 8, 3 spots July 10 \& 11, one spot July 15, probably all in one group.
(g) M.M. Kirch is Gottfried Kirch's second wife, Maria Margaretha,
and C. Kirch is their son Christfried Kirch.
(h) According to the letter written by M.M. Kirch, Sect. 3.19.
(i) Neuh\"auser et al. (2015).
(j) Observation during solar eclipse.
\end{table*}

\twocolumn

\section{Conclusion and summary}
\label{summary}

We summarize our findings as follows:
\begin{itemize}
\item With excerpts from letters to and from G. Kirch (edited by Herbst 2006), we investigated 28 periods
of sunspot observations from Kirch and his letter partners in the last few
decades of the Maunder Minimum (1680 to 1709) -- most data are found in the letters,
but two in his {\it Himmels-Zeitung} (Sect. 3.1 and 3.2) and one in his {\it Ephemeries} (Sect. 3.7).
\item The letters and observations from Kirch are mostly by Gottfried Kirch, 
but some also by his 2nd wife Maria M. Kirch and their son Christfried Kirch.
\item The partners of the letter exchanges studied here are G. Schultz, J. Hevelius, G.S. D\"orfel, G. Teubner, G.C. Eimmart,
J.A. Ihle, J.P. Wurzelbaur, S. Reyher, J.H. Hoffmann, G.W. Leibniz, O. R\o mer, L.C. Sturm, H.C. von Wolffsburg,
and C.G. Hertel.
\item We found information on some 35 spot groups, often with several spots per group.
\item We present 11 original drawings of spots from the letters, mostly one spot or group.
\item One of the spot observations was performed during the solar eclipse of 1708 Sep 14 (Sect. 3.24).
\item We could constrain the heliographic latitude for 19 groups (Table 1), five of them are new values: 
1680 May and June (2 groups), 1684 May, and 1688 Dec -- all consistent with a southern hemisphere.
\item We added the five new latitudes into the butterfly diagram for the Maunder Minimum (Figs. 31 \& 32).
\item We found 17 explicite spotless days, some of them as yet unknown (e.g. Tables 2-6).
\item There may be one case of a (quasi-) simultaneous observation of sunspots and an aurora:
Sturm wrote to Kirch about the spots 1707 Mar 6-8 and Kirch mentioned to have observed himself 
an aurora on Mar 6 in his letter to Tschirnhaus (Sect. 3.21); the aurora was also observed by
G. Kirch's son C. Kirch and a pastor named Seidel, another letter partner of G. Kirch.
The spot(s) were already present on the Sun since about Feb 23, so that some interaction
could have released energetic particles responsible for the aurora.
\item We noticed a number of differences by comparison of the tables in Hoyt \& Schatten (1998)
and the information in the letters (see Table 2); however, we should keep in mind here that
the HS98 data base includes errors in dates and group numbers as shown not only in this work
(see also Svalgaard \& Schatten 2016, Neuh\"auser et al. 2015, Neuh\"auser \& Neuh\"auser 2016, Vaquero et al. 2016),
so that a comparison of Kirch's letters with other observers (based on HS98) is preliminary, as long as the
original sources have not been revisited.
\item As was noticed before, several generic statements about spotlessness were overinterpreted by
Hoyt \& Schatten (1998) assuming that the observer would have observed each and every day during a long period;
we found discrepancies in this regard here for the observers Siverius, Wurzelbaur, Ihle, and Angerholm.
\item Some observations by Flamsteed and Stancarius were misinterpreted by Hoyt \& Schatten (1998) as
to mean spotlessness, but the observers did not report on spots nor spotlessness.
\item Also, the dates given in Hoyt \& Schatten (1998) for the British observers Caswell, Derham, Stannyan, Gray, and Sharp 
during the Maunder Minimum are their original Julian dates, not shifted to the new style (10 to 11 day shift),
partly also to be applied to Flamsteed (e.g. 1684 May). Furthermore, also for some observations in protestant parts of Germany
before 1700 (when they moved to the new calendar style), Hoyt \& Schatten (1998) did not yet convert their 
Julian dates to Gregorian (e.g. Kirch for 1684 Apr 26-May 7).
\end{itemize}

In Tables 2-6 in the appendix, we also present solar observations by the astronomers 
from Sect. \ref{spots}, i.e. from letters to or from Kirch,
that did not result in a sighting of new sunspots; we list
\begin{itemize}
\item dates given for spotlessness,
\item additional quotations on other solar observations, and
\item generic statements about spotlessness.
\end{itemize}

We have compiled our data on latitudes in Table 1, also in comparison to HS98, RNR93, and Sp\"orer (1889).
We present an updated butterfly diagram for the Maunder Minimum:
we plot the spot latitudes from Sp\"orer (1889) and Ribes \& Nesme-Ribes (1993,
as listed by Vaquero et al. 2015) as well as a few additional ones from Neuh\"auser et al. (2015) for the observer P. Becker.
We also indicate all other spot groups for this time period from Hoyt \& Schatten (1998).

\begin{figure*}
\centering
\includegraphics[width=0.675\textwidth,angle=270]{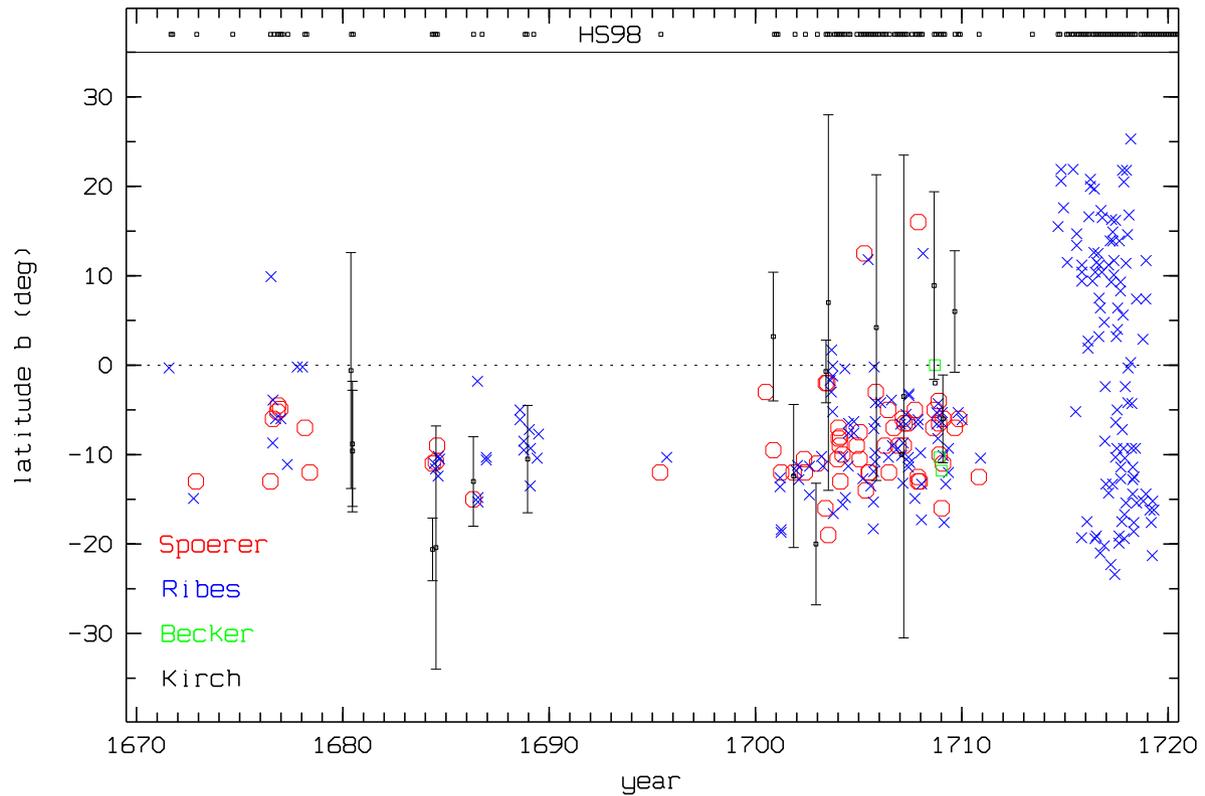}
\caption{New butterfly diagram for the period AD 1670-1720
(with the last decades of the Maunder Minimum ending around 1715)
with previously known spot latiudes from Sp\"orer (1889) as red circles 
and from Ribes \& Nesme-Ribes (1993) as blue crosses (taken from Vaquero et al. 2015), 
as well as from Neuh\"auser et al. (2015) for the observer P. Becker as green squares,
plus all those estimated here from the letters to and from Kirch 
(including those from Kirch's {\it Himmelszeitung} for 1680 May and June as well as from Kirch's ephemeries for 1688 Dec) 
as black dots -- most with error bars
(for the two cases in 1686 Apr/May and 1702 Dec 24, where we found two solutions (Table 1), we plot here the southernmost ones);
small dots in the upper panel show all spot groups listed by Hoyt \& Schatten (1998, HS98).}
\label{butt_new1}
\end{figure*}

\begin{figure*}
\centering
\includegraphics[width=0.675\textwidth,angle=270]{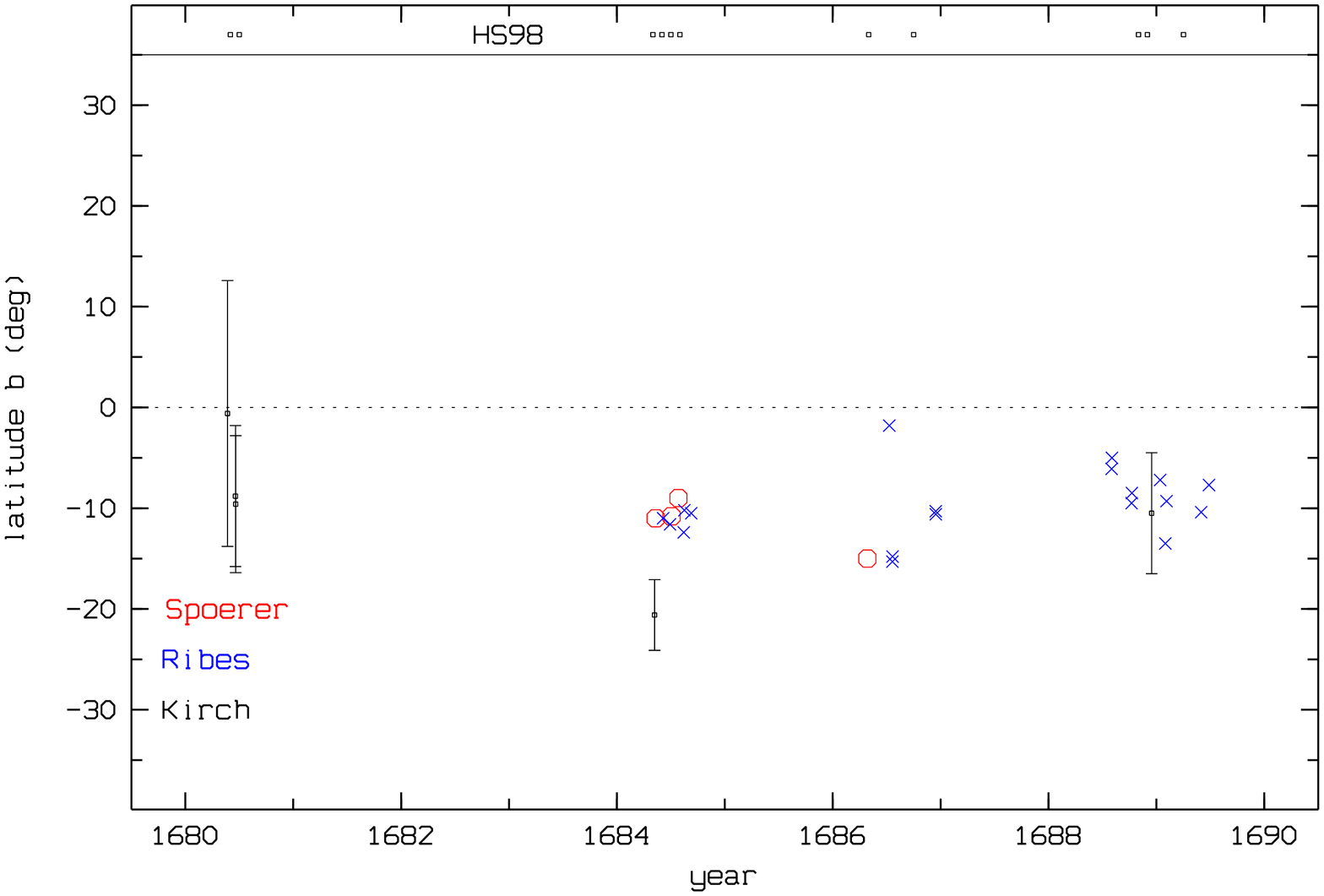}
\caption{New butterfly diagram for the period AD 1680-1690 only (part of the Maunder Minimum,
where the five new spot positions lie) with previously known spot latitudes from Sp\"orer (1889) as red circles
and Ribes \& Nesme-Ribes (1993) as blue crosses (taken from Vaquero et al. 2015), 
plus the five new ones with latitudes as estimated here from Kirch's {\it Himmelszeitung} for 1680 May (Sect. 3.1)
and 1680 June (Sect. 3.2),
from Ihle's (and Kirch's re-drawn) drawing for 1684 May 6 found in a letter from Kirch to Hevelius (Sect. 3.4),
and from Kirch's ephemeries for 1688 Dec (Sect. 3.7)
as black dots with error bars; small dots in the upper panel show all spot groups listed by Hoyt \& Schatten (1998, HS98).}
\label{butt_new2}
\end{figure*}

Our new butterfly diagram for the Maunder Minimum (Figs. \ref{butt_new1} and \ref{butt_new2}) 
shows five new spots with latitudes determined for the first time
(all our other latitudes are consistent with previously published values for those days within $2 \sigma$).
The latitudes found for the spots described in the letters to and from Kirch are all consistent -- within
their error bars -- with locations near the equator or on the southern hemisphere.

\bigskip

{\bf Acknowledgements.}
We consulted the data base of Hoyt \& Schatten (1998) at
ftp://ftp.ngdc.noaa.gov/STP/space-weather/solar-data/solar-indices/sunspot-numbers/group.
We acknowledge eclipse predictions by Fred Espenak, NASA GSFC Emeritus,
obtained from eclipse.gsfc.nasa.gov/solar.html.
We acknowledge the JPL Horizons service for providing ephemeris for the solar P angle
(ssd.jpl.nasa.gov/horizons.cgi). We also used StarCalc~5.73.
RN would also like to thank Dagmar L. Neuh\"auser for valuable discussion,
in particular she noticed that the Kirch letter so far dated 1707 April
(with three auroral dates) was actually written in 1716.
We thank Mark Booth (U Jena) for proof-reading an earlier version of the manuscript.
We acknowledge Hans Gaab (Nuremberg) for his help in reading the 1707 letter
from Sturm to Maria M. Kirch.
We also acknowledge very good suggestions by our referee, Dr. Jose Vaquero.

\section{Appendix}

\onecolumn
\setcounter{table}{2}
\captionof{table}{Remaining solar observations by G. Kirch (date format: dd:mm:yyyy); 
first datable spotless days, then 
reports about other solar events, where spots may have been mentioned if seen,
and finally generic statements about spotless periods.}
\begin{tabularx}{\textwidth}{l|X|X|l} \hline
Greg. date & text in Herbst (2006) & our translation & letter \\ \hline
\multicolumn{4}{l}{Dated reports about spotlessness} \\ \hline
29.05.1680 & Den 19. (29.) May Vormittags um 9. Uhr fand ich die Sonne wiederum gantz rein. & 
On May 19 (29), I found the Sun yet again entirely spotless. & (*) \\
29.06.1680 & Den 19. (29.) Junii war die Sonne wiederum gantz rein ohne Maculen. &
  On Jun 19 (29), the Sun was yet again entirely spotless. & (*) \\
21.07.1681 & Den 11/21 Jul. ist sie noch nicht drinnen gewesen.&
	On July 11/21 {[}Jul./Greg.{]} it {[}spot{]} was not yet on it.&
	94\\ 
31.07.1681&
	Donnerstags war sie auch durch den 13 sch\"uhigen Tubum nicht mehr zu finden.&
	On Thursday {[}July 31, Greg.{]} it was not visible any more, even through a 13-foot telescope.&
	96\\ 
18.+19.12.1688&
	Die sequente, coelo defaecatissimo, Sol purus cernebatur.&
	The following day {[}Dec 19, Greg.{]}, as the sky was clear, and a spotless Sun was seen.&
	594\\ 
08.08.1708&
	Am 8 Aug. um 4 1/4 nachmittags betrachtete ich die $\odot$ durch einen 10 schuhigen Tubum und fand sie rein.&
	On Aug 08 {[}Greg.{]} around 04:15h p.m. I observed the Sun through a 10-foot telescope and found it clear.&
	892 \\ \hline
\multicolumn{4}{l}{Other solar observations} \\ \hline
%
%
12.07.1684&
	Die neulichste Sonnenfinsterni\ss{} haben wir allhier zwar observiret, von Anfang bi\ss{} zu Ende\dots &
	Although we observed the recent solar eclipse from beginning to end\dots &
	276\\ \cline{1-4}
11.05.1687&
	Das neueste was ich am Himmel observiret, ist die kleine Sonnenfinsterni\ss{}, aber immer Schade, da\ss{} das tr\"ube Wetter uns den Anfang zu sehen versagte.&
	My newest observation was the small solar eclipse. Too bad, the dull weather did not allow us to observe its 			beginning.&
	356\\ \cline{1-4}
13.09.1689&
	Die Sonnenfinsternis habe ich zwar meinen M\"oglichkeiten entsprechend beobachtet, aber weil mir ein geeigneter 		Ort und die notwendingen Instrumente fehlen, halte ich diese Beobachtung mit Recht nur f\"ur mittelm\"a\ss{}ig.&
	As far as possible I observed the solar eclipse. But considering the shortage of a suitable location and the 	essential instruments, I consider this observation to be mediocre.&
	409\\ \cline{1-4}
10.11.1690&
	Gott hat auch Gnade gegeben, da\ss{} ich in Erfurt erlanget was ich gew\"unschet, sintemahl ich Mercurium beynahe 		3 Viertel St. lang gesehen, bi\ss{} er an den $\odot$en rand r\"uhrete.&
	God had mercy since I had the opportunity in Erfurt, as I wished, to observe Mercury nearly 3 quarter hours in the Sun, until it touched 		the solar limb.&
	448\\ \cline{1-4}
22.06.1694&
	Die neulichste Sonnenfinsterni\ss{}, am 22 Jun. Dienstags, welche (laut Rechnung) nicht hier her gelangen sollte, haben wir doch allhier gesehen, wie wol sehr klein.&
	The latest solar eclipse on June 22 {[}Greg.{]}, Tuesday, which should not be visible (according to calculations), we saw nonetheless, but very small.&
	547 (**)\\ \cline{1-4}
23.-25.09.1699&
        Wir hatten im Kloster Zelle sch\"on Wetter, den Montag nach Mittage, den Dienstag und Mittwoch Tag und Nacht;&
        In abbey Neuenzelle we had good weather on Monday afternoon, Tuesday and Wednesday by day and night; (solar                     eclipse: 23.09.1699)&
        683\\ \cline{1-4}
\end{tabularx}

\newpage

\begin{tabularx}{\textwidth}{l|X|X|l}
Greg. date & text in Herbst (2006) & our translation & letter \\ \hline
12.05.1706&
        Dem erstlich wollte ich meine Observation gedachter Sonnen-Finsterni\ss{} gerne zuvor ausarbeiten\dots &
        I wanted to edit my observation of the solar eclipse {[}May 12, Greg.{]} first, because the circumstances were                  very miserable\dots &
        848\\ \cline{1-4}
11.03.1709&
        \dots Auch die Sonnenfinsternis den 11 Mart 1709.&
        \dots the solar eclipse on 1709 March 11 {[}Greg.{]} as well.&
        893\\ \hline
\multicolumn{4}{l}{Generic statements about spotlessness} \\ \hline
%
25./27.04.1686&
        Es ist nun fast 2 Jahre, da\ss{} wir keine Macul in der Sonnen gehabt haben.&
        It has been almost 2 years since we had spots in the Sun.&
        320 \\ \hline
\end{tabularx}

\smallskip

Remarks: (*) from Kirch's {\it Himmelszeitung} (Sect. 3.1 \& 3.2).
(**) The letter was written by Kirch in Guben, Germany, east of Berlin;
maximal obscuration in both Guben and Berlin was $\sim 1\%$ only (www.eclipse.nasa.gov).
The text given above is from one of the two versions (concepts) of this letter,
both cited in Herbst (2006, p. 190-194); it may be surprising that Kirch dated the
eclipse with the Gregorian date, but since he wrote to a Jesuit (Adam Adamandus Kochanski 
in Warsaw, Poland), he may have chosen the Gregorian date to avoid confusion u
(Kocjanski's reply is in Latin and dated Gregorian, Herbst 2006, p. 199-200); in the 
other letter version, he dated the eclipse Julian; in the other letter concept, he also wrote 
{\it Ich h\"atte zwar wol eher wieder geschrieben,
aber ich wartete erstlich auff die kleine Sonnenfinsterni\ss ...}, i.e.
{\it I would have written earlier, but I waited for the small solar eclipse ...},
showing that he did expect the eclipse to happen, but being small
(a lengthy calculation about it by Kirch is found in his letter to Kochanski).

\newpage

\setcounter{table}{3}
\captionof{table}{Remaining solar observations by Schultz (date format: dd:mm:yyyy)}
\begin{tabularx}{\textwidth}{l|X|X|l} \hline
Greg. date & text in Herbst (2006) & our translation & letter no. \\ \hline
\multicolumn{4}{l}{Dated reports about spotlessness} \\ \hline
06.05.1684&
	F\"ur die nachricht von der macula solari sage Ich schuldigen danck; Ich habe bald selbigen tages, so gutt ich gek\"ont, nach geschawet, aber nichts mehr sp\"uhren k\"onnen; vermuthlich wird sie bereits au\ss{} der Disco solaris 	herau\ss{} seyn;&
	I owe you a debt of gratitude for the report on the spot in the Sun; on that same day {[}May 06, Greg.{]} I looked 	for it as good as I could but could not find the spot; probably it has already left the Sun's disc.&
	271\\ \cline{1-4}
\multicolumn{4}{l}{Other solar observations} \\ \hline
12.07.1684&
	Was die Eclipsin Solis anbetrifft, habe Ich dieselbte auch observiret\dots &
	I was able to observe the solar eclipse [July 12, Greg.] as well\dots &
	277\\ \cline{1-4}
11.05.1687&
	Da\ss{} Ich auff desselbten letzteres vom 9/19 April: itzo erst antwortet hat die am 1/11 May eingefallen Sonnenf\"unstern\"uss{}, derer Successionem Ich zu gleich mit \"uberschreiben wollen, veruhrsachet.&
	The delayed reply to your letter from April 09/19 [Jul./Greg.] is due to the solar eclipse on May 01/11, which I 		successfully could observe.&
	358\\ \cline{1-4}
13.09.1689&
	Nach dem sich nun j\"ungsthin das gl\"ucke gef\"uget, da\ss{} Ich die neulichste Eclipsin Solis observiren k\"onnen, habe Ich dannenhero anla\ss{} genommen, dem Herren mit diesem Brieflein auffzuwarten und gedachte 				Observation beyliegend zu communicieren.&
	After I was lucky enough to observe the latest solar eclipse, I decided to report you my observation in this 			letter.&
	401\\ \cline{1-4}
10.11.1690&
	\dots da Ich denn eine geraume halbe virtel stunde zeit hatte, die Sonne in meiner Camera obscura zu betrachten, 		und mich zu versichern, da\ss{} vor di\ss{}mal vom Merkur nicht die geringste spuhr vorhanden sey\dots &
	\dots I had half a quarter hour to observe the Sun through my camera obscura and to make sure that Mercury was not visible in it\dots &
	450\\ \cline{1-4}
\multicolumn{4}{l}{Generic statements about spotlessness} \\ \hline
Summer 1685&
	Sonst habe ich den gantzen Sommer nichts besonderes observiret, auch in sole keine maculam sp\"uhren k\"onnen, ob 		Ich gleich wochentlich mehr al\ss{} ein mal darnach geschauet.&
	Apart from that I observed {[}the Sun{]} during the whole summer, but could not find any sunspots, although I 			looked for them more than once a week.&
	304\\ \cline{1-4}
Summer 1686&
	Ich habe diesen gantzen Sommer \"uber, wochentlich ein mal oder zwey, wann es das wetter und andere Gescheffte 			verg\"onnet, die Sonne per tubum in camera obscura observiret, niemals aber einige spuhr einer maculae seindt.&
	During the whole summer I observed {[}the Sun{]} once or twice a week, if the weather or other businesses allowed 		it, through a telescope in the camera obscura but there was no trace of any spot\dots &
	338\\ \cline{1-4}
Aug-03.10.1689&
	Ich habe vom Medio Augusti bi\ss{}her wochentlich etliche mal versucht, ob Ich irgend eine maculam in sole 				antreffen m\"ochte, habe aber nie keine gefunden\dots &
	Since the middle of August {[}1689{]} I looked for sunspots many times a week but could not find any of them.&
	401 \\ \cline{1-4}
\end{tabularx}

\newpage

\setcounter{table}{4}
\captionof{table}{Remaining solar observations by Wurzelbaur (date format: dd:mm:yyyy)}
\begin{tabularx}{\textwidth}{l|X|X|l} \hline
Greg. date & text in Herbst (2006) & our translation & letter no. \\ \hline
\multicolumn{4}{l}{Dated reports about spotlessness} \\ \hline
27.11.1700&
	\dots daher ich dann den 27:$^{ten}$ und etliche folgende tage mich \"offters darnach umbgesehen aber weiter 			nichts von einigen flecken erblicken k\"onnen.&
	That is why I looked for them on the {[}Nov{]} 27th {[}Greg.{]} and the following days but could not find any 			spots.&
	750\\ \cline{1-4}
\begin{tabular}[c]{@{}l@{}}31.12.1702\\ 01.01.1703\end{tabular}&
	\dots den 31:$^{ten}$ Decembr: aber und primo Januarii 1703 nichtes mehr davon angetroffen&
	\dots on Dec 31 {[}1702, Greg.{]} and on Jan 1 {[}1703, Greg.{]} nothing of it {[}spot{]} was seen &
	786\\ \cline{1-4}
11.01.1709&
	\dots am 11:$^{ten}$ ejusdem nachmittag aber in der Sonne nicht ferner anzutreffen\dots &
	\dots On Jan 11 [Greg.] in the afternoon it [spot] was no longer visible \dots &
	893\\ \cline{1-4}
29.08.1709&
	\dots am 29:$^{ten}$ zwar durch Nebel weiter nichts in der Sonne zu ersehen\dots &
	\dots on {[}Aug{]} 29th {[}Greg.{]} through fog, the Sun was clear \dots &
	895\\ \cline{1-4}
\multicolumn{4}{l}{Other solar observations} \\ \hline
13.09.1689&
	\dots Eclipsi Solis d. 3. Septembris hh. pomeridiem observanda nos admonuissent.&
	\dots and the observation of the solar eclipse on Sept 03 {[}Sept 13, Greg.{]} in the afternoon hours \dots The 			desirable clarity of the sky invited to observe it, and allowed to watch it joyfully from beginning to end.&
	408\\ \cline{1-4}
10.11.1690&
	Nachdem ich den Tubum auf den jenigen Ort/ wo der Austritt der Sonnen aus den Wolcken zu erwarten/ gerichtet/ und 		diesselbe auf die/ dem Tubo entgegen gestellte Tafel gefallen/ habe ich in deroselben obern Theil ein mittelm			\"assiges Flecklein gemercket/ wie in Fig. II zu sehen/ welches ich auch f\"ur eine maculam solarem w\"urde 			gehalten haben/ woferne nicht theils die vermuthete Gegenwart de\ss{} Mercurii, theils die geschwinde Bewegung 			dieses Fleckleins mich eines andern beredet h\"atte.&
	After I aimed the telescope at the location, where I expected the Sun to come out of the clouds, and it was 			projected to the board across the telescope, I saw in its upper part a medium large spot, as shown in Fig. II {[}		not presented{]}. I could have thought, that it was a sunspot but the expected presence of Mercury and its quick 		movement proofed me wrong.&
	446\\ \cline{1-4}
22.06.1694&
	\dots al\ss{} habe mit diesem hiebeygehenden gehaltene observationibus der beeden j\"ungsten Eclipsium aufwarten		\dots communicieren wollen.&
	\dots and would like to know how the attached observations of the two eclipses {[}solar: June 22, Greg.; lunar: 		July 7, Greg.{]} match with your observations.&
	548\\ \cline{1-4}
03.11.1697&
	MERCURIUS prope limbum Solis iterato observatus NORINBERGAE a Joh: Philippo Wurzelbaur&
	MERCURY near the Sun's limb, observed again in NUREMBERG by Joh. Philipp Wurzelbaur (caption)&
	591\\ \cline{1-4}
23.09.1699&
	H. Wurzelbaur l\"a\ss{}t Seine observationem Eclipseos $\odot$ w\"urcklich in N\"urnberg drucken.&
	Mr. Wurzelbaur is going to print his observation of the solar eclipse {[}Sept 23, Greg.{]} in Nuremberg. (original 	text by U. Junius)&
	688\\ \cline{1-4}
12.05.1706&
	\dots da\ss{} Er nemlich bis dahin keine Nachricht erhalten, wie man die Sonnenfinsternus den 12 Maji in N				\"urnberg gesehen habe zu berichten: Da\ss{} die v\"ollige beschreibung selbiger observation albereit auf Primo 		Juni\dots berichtet habe.&
	\dots since he {[}G. Kirch{]} did not get any report on the observation of the solar eclipse on May 12 {[}Greg.{]} 	in Nuremberg, even though I reported on it on June 01.&
	836\\ \cline{1-4}
\end{tabularx}
\newpage
\begin{tabularx}{\textwidth}{l|X|X|l} \hline
Greg. date & text in Herbst (2006) & our translation & letter no. \\ \hline
05./06.05.1707&
	\dots und darauf gestern und heute erwarteten Mercurii in der Sonne\dots dahero ich heute vor tages auf deren 			Aufgang desto eyferiger gewartet\dots aber es war weder Mercurius noch die geringste macula darinne nicht zu 			erblicken\dots &
	\dots and the expected appearance of Mercury in the Sun, yesterday and today {[}May 05, 06, Greg.{]}\dots That is 		why I awaited its {[}Sun's{]} rise all the more today\dots but it was neither Mercury nor the smallest spot in 			it.&
	862\\ \cline{1-4}
14.09.1708&
	Nechstverschienen 14 Septembr: erwartete ich mit verlangen auch die Sonnenfinsternis zu observiren\dots jedoch 			sahe ich zwischen 8 und 9 Uhr durch vom wind f\"ur\"uber getriebene etwas d\"unne W\"olcklein mit blosen augen 			zweymal, wie die Sonne \"uber die helffte und wohl etwan 7 digg: vom Mond bedeckt gewesen.&
	Last Sept 14 {[}Greg.{]} I awaited to observe the solar eclipse\dots but between 8 and 9 o'clock I saw the Sun through the 		clouds, driven by the wind, and appr. half of it, 7 inches, covered by the moon.&
	889\\ \cline{1-4}
\multicolumn{4}{l}{Generic statements about spotlessness} \\ \hline
$\sim$ 1684 & Die maculae solares welche vor 20 Jahren wohl rare gewesen & Sunspots, which apparently were rare 20 years ago & 810 \\ \cline{1-4}
6/1695-5/1700 & auch nach A$^\circ$: 1695 im Ma\"yen,
unerachtet fast t\"aglicher durchsuchung disci Solaris keine ersehen k\"onnen,
bi\ss{} im Juni des j\"ungstabgewichenen Jahres
& not even after 1695 May, in spite of nearly daily examination of the Sun's disc,
until June last year & 750 \\ \cline{1-4} 
summer 1709 & und ist es seithero in der Sonne ruhig gewest bis in den abgewichenen Augustum,
da ich am 23 ejusdem nach etlichen tr\"uben tagen ... & and it was quiet in the Sun since then until the last August,
when I saw on the 23rd after many dull days ... & 895 \\ \hline
\end{tabularx}

\newpage

\setcounter{table}{5}
\captionof{table}{Remaining solar observations by Ihle (date format: dd:mm:yyyy)}
\begin{tabularx}{\textwidth}{l|X|X|l} \hline
Greg. date & text in Herbst (2006) & our translation & letter no. \\ \hline
\multicolumn{4}{l}{Dated reports about spotlessness} \\ \hline
05.05.1684&
	Itzt da ich schlie\ss{}en will, k\"ombt H. Ihle und berichtet, da\ss{} eine Macula in der Sonnen, welche Er 			gestern nicht gesehen.&
	Now, coming to an end, Mr. Ihle is coming and reports, that a spot [is] in the Sun, which he has not seen 				yesterday [May 5, Greg.].&
	275\\ \cline{1-4}
1688/9 or 1697/8 & \dots aber gantz ledig befunden am 20. und 21. Decemb: 6. 7. 19 Januar \dots &
    \dots but fully clear on [1688 or 1697] Dec 20 and 21 [Jul.] and [1689 or 1698] Jan 6, 7, 19 \dots 
    (see Sect. 3.7 for discussion of the dating) \\ \cline{1-4}  
10.11.1701&
	\dots folgenden tag konte ich davon gar nichts mehr sp\"uhren, da doch die Sonne klar genug war, doch eine kurtze 		zeit\dots &
	\dots the next day [Nov 10, Gerg.] I could not see it [spot] any more, even though the Sun was clear at least for a short while			\dots &
	761\\ \cline{1-4}
17.12.1702&
	Am 17 Decembr: Mittags befunde ich die Sonne gantz rein\dots &
	On Dec 17 [Greg.] around noon I found the Sun fully clear\dots &
	781\\ \cline{1-4}
\multicolumn{4}{l}{Other solar observations} \\ \hline
07.11.1677&
	Allhier in Leipzig war H. Ihle, der damals Merkur einen einzigen Blick in der Sonnen ertappete. (report by G. Kirch)&
	Here in Leipsic Mr. Ihle was the only one who saw the Mercury in the Sun. (report by G. Kirch)&
	448\\ \cline{1-4}
03.07.1693&
	\dots am 23 Junii, hier zu Leipzig hat un\ss{} das Gew\"olcke sehr geplagt, ich h\"atte gern vor und nach Mittage 		etliche Altitudines Solis\dots genommen\dots bi\ss{} 1. uhr, XI', da brachte sie einen gantz kleinen Anfang zur 		Finsterni\ss{} mit\dots &
	\dots on June 23 [Jul.] the clouds bothered us here in Leipsic. I would have liked to measure the Sun's altitude before and after noon		\dots until 01:11h p.m., when I could see a small part of the solar eclipse\dots &
	536\\ \cline{1-4}
\multicolumn{4}{l}{Generic statements about spotlessness} \\ \hline
1684 Jan-Apr & die 26 April: s.v. 1684 [Julian] ...  Die Praecedente, et anno amplius, Solem plane vacuum offendimus & ... 1684 May 6 [Greg.] ... On the day before as in the whole year so far, the Sun was without spots. & 267 \\ \cline{1-4}
1701&
	Nach dem ich nun \"uber 20 Jahr sehr wenig maculas solares vermercket\dots Alle tage habe ich in gedachter langen 		zeit achtung drauff gegeben, so offt ich von Wolcken, oder unau\ss{}etzlich verrichtungen nicht gehindert worden		\dots &
	Since I did observe only very few sunspots in the last 20 years\dots During that time I paid attention to it every 	day, as long as the weather or unavoidable businesses allowed me to.&
	761\\ \cline{1-4}
Oct 1701&
	\dots an j\"ungst verwichenem Octobris war die Sonne noch gantz rein\dots &
	\dots in the recent October the Sun was still completely clear\dots &
	761\\ \cline{1-4}
\end{tabularx}

\newpage

\setcounter{table}{5}
\captionof{table}{Remaining solar observations by Sturm (date format: dd:mm:yyyy)}
\begin{tabularx}{\textwidth}{l|X|X|l} \hline
Greg. date & text in Herbst (2006) & our translation & letter no. \\ \hline
\multicolumn{4}{l}{Dated reports about spotlessness} \\ \hline
08.03.1707&
	\dots den 8ten waren sie beide an der disco hinweg\dots &
	\dots on [March] 8th [Greg.] they were both vanished from the solar disc\dots &
	853\\ \cline{1-4}
\multicolumn{4}{l}{Other solar observations} \\ \hline
04.-06.05.1707&
	\dots habe darauff den Vierten, 5$^{ten}$ und 6$^{ten}$ bi\ss{} mittage nach der Sonne gesehen, sie in camera 			vereobscura expiciret, welches an einem freyen morgen horizont von auffgang der Sonnen an bi\ss{} Abends um 7. Uhr 	continuirlich geschehen k\"onnen\dots bin ich versichert, da\ss{} Mercuri in die Sonne nicht gekommen.&
	\dots I then observed the Sun on the [May] 4th, 5th and 6th [Greg.] until noon through my camera obscura, which was possible 		from the sunrise on a clear morning until the evening around 7 o'clock\dots I am sure now that Mercury was not 			visible in the Sun.&
	863\\ \cline{1-4}
\end{tabularx}

\end{document}